\documentclass{jfm}
\usepackage{graphicx}
\usepackage{subfigure}
\usepackage{epstopdf, epsfig}
\usepackage{amsmath}
\usepackage{overpic}
\usepackage[colorlinks, citecolor=blue]{hyperref}
\usepackage{upgreek}
\usepackage{lineno}
\usepackage{rotating}
\usepackage{multicol}
\usepackage{multirow}
\usepackage[mathscr]{eucal}

\newcommand{\ri}{\mathop{\rm i}\nolimits}
\newcommand{\re}{\mathop{\rm e}\nolimits}

\shorttitle{Principle of  fundamental resonance }
\shortauthor{R. Song, M. Dong and L. Zhao}

\title{Principle of  fundamental resonance  in hypersonic boundary layers: an asymptotic viewpoint }

\author{
  Runjie Song\aff{1},
  Ming Dong\aff{1}\corresp{\email{dongming@imech.ac.cn}}
 \and Lei Zhao \aff{2} \corresp{\email{lei\_zhao@tju.edu.cn}}}

\affiliation{\aff{1}State Key laboratory of Nonlinear Mechanics, Institute of Mechanics, Chinese Academy of Sciences, Beijing 100190, China
\aff{2}Department of Mechanics, Tianjin University, Tianjin, 300072, China
}

\begin{document}

\maketitle

\begin{abstract}
The fundamental resonance (FR) in the nonlinear phase of the boundary-layer transition to turbulence appears when a dominant planar instability mode reaches a finite amplitude, and the low-amplitude oblique traveling modes with the same frequency \textcolor{magenta}{as} the dominant mode, together with the stationary streak modes, undergo the \textcolor{magenta}{strongest} amplification among all the Fourier components. This regime may be the most efficient means to trigger the natural transition in hypersonic boundary layers. In this paper, we aim to reveal the intrinsic mechanism of the FR in the weakly nonlinear framework based on the large-Reynolds-number asymptotic technique. It is found that the FR is in principle a triad resonance among a dominant planar fundamental mode, a streak mode and an oblique mode. In the \textcolor{magenta}{major part} of the boundary layer, the nonlinear interaction of the fundamental mode and the streak mode seeds for the growth of the oblique mode, whereas the interaction of the oblique mode and the fundamental mode drives the roll components (transverse and lateral velocity) of the streak mode, which leads to a stronger amplification of the streamwise component of the streak mode due to the lift-up mechanism. {This asymptotic analysis clearly shows that the dimensionless growth rates of the streak and oblique modes are the same order of magnitude as the dimensionless amplitude of the fundamental mode $(\bar \epsilon_{10})$,} and the amplitude of the streak mode is $O(\bar \epsilon_{10}^{-1})$ greater than that of the oblique mode. The main-layer solution of the streamwise velocity, spanwise velocity and temperature of both the streak  and  the oblique modes become singular as the wall is approached, and so a viscous wall layer  appears underneath. The wall layer produces an outflux velocity to the main-layer solution, inclusion of which leads to an improved asymptotic theory, whose accuracy is confirmed by comparing with the calculations of the nonlinear parabolised stability equations (NPSE) at moderate Reynolds numbers and the secondary instability analysis (SIA) at sufficiently high Reynolds numbers.
\end{abstract}


\section{Introduction}
In the development of hypersonic vehicles, to calculate accurately the drag and heat flux  is a task of the first priority, which requires an  accurate prediction of the laminar-turbulent transition. For high-altitude flight conditions, transition is often triggered by a natural route, for which four phases, including the receptivity, linear instability, nonlinear resonance, and turbulence, appear in sequence. For supersonic or hypersonic boundary layers, there may exist more than one discrete instability modes, which are referred to as the Mack first, second, ... modes, according to the ascending order of their frequencies \citep{Mack_1987}. It was revealed by the asymptotic analysis that only the Mack first mode with $\Theta > \tan^{-1} \sqrt{M^2-1}$ (where $\Theta$ and $M$ denote the wave angle and Mach number, respectively) \textcolor{magenta}{has} the viscous nature \citep{smith_1989,liu_dong_wu_2020}, while the quasi-two-dimensional Mack first and all the higher-order modes are inviscid \citep{smith_brown_1990,cowley_hall_1990,dong_liu_wu_2020,zhao_he_dong_2023}. Usually, the Mack second mode appears when the Mach number is approximately over 4, and its growth rate peaks when it is planar (two-dimensional). In contrast, the Mack first mode appears in all supersonic boundary layers, which is more unstable when it is oblique (three-dimensional).
The linear evolution of these modes was confirmed by quite a few  numerical works, such as \cite{Fedorov_2011} and \cite{Zhong_Wang_2012}. When the unstable modes are accumulated to finite amplitudes, the nonlinear interaction among different Fourier components becomes the leading-order impact, \textcolor{red}{showing three major nonlinear regimes,} including the oblique-mode breakdown \citep{Thumm_1991,Fasel_1993,chang_malik_1994,leib_lee_1995,mayer_vonterzi_fasel_2011,mayer_wernz_fasel_2011}, the subharmonic resonance \citep{SARIC_1984,Herber1988} and the fundamental resonance \citep{sivasubramanian_fasel_2015,hader_fasel_2019}.

\begin{figure}
\centering
\begin{overpic}[width=0.98 \textwidth]{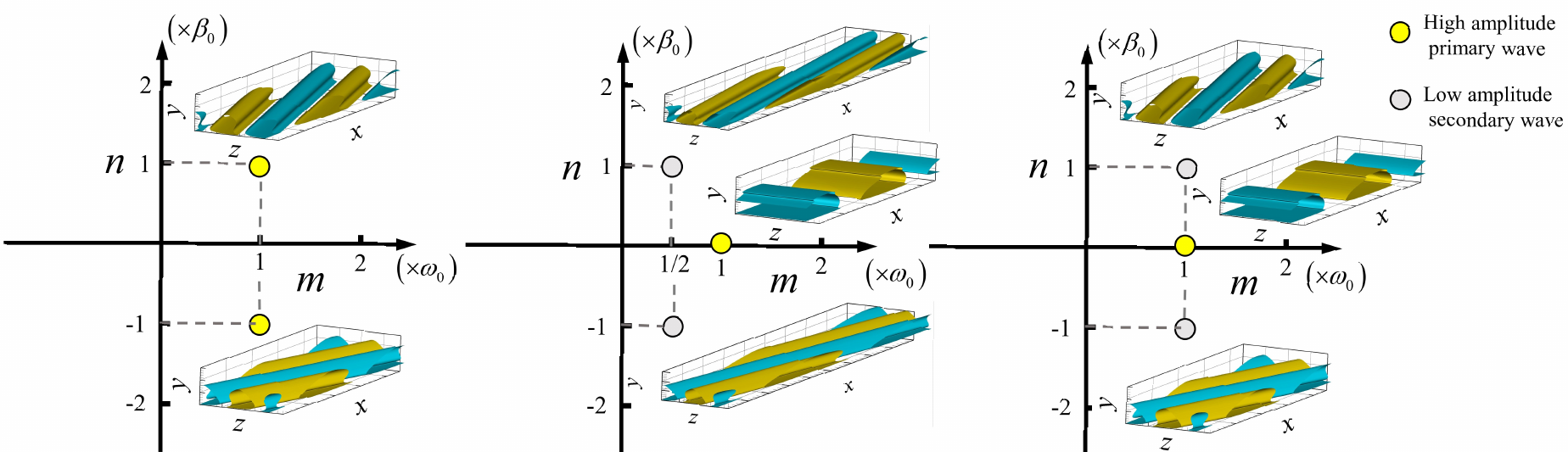}
\put(-1,25){(a)}
\put(30,25){(b)}
\put(60,25){(c)}
\end{overpic}
\caption{Characteristic parameters in the frequency–spanwise-wavenumber space. (a): the oblique-mode breakdown regime; (b): the subharmonic resonance regime; (c) the fundamental resonance regime. $\omega_0$ and $\beta_0$ denote the fundamental frequency and spanwise wavenumber, respectively. \textcolor{cyan}{This sketch is a modified version of figure 10 in \citet{hader_fasel_2019}.}}
\label{fig:pinpu_sketch}
\end{figure}



The oblique-mode breakdown appears when the dominant perturbations in the early nonlinear phase are a pair of three-dimensional (3D) travelling waves with the same frequency but opposite spanwsie wavenumbers, as sketched in figure~\ref{fig:pinpu_sketch}-(a). Such a regime was \textcolor{magenta}{pioneered} by \cite{Thumm_1991} and \cite{Fasel_1993} using the direct numerical simulation (DNS) approach and subsequent studied by \cite{chang_malik_1994} using the nonlinear parabolised stability equations (NPSE) approach. 
They all found that the growth rates of the second-order harmonics and the stationary streak mode are equal to twice of those of the oblique modes, because they are driven by the self or mutual interaction of the oblique modes. Particularly, the streak mode shows a much greater amplitude than the second-order harmonics. Such a phenomenon was recently explained by \cite{Song_2022} using the weakly nonlinear analysis based on the large-Reynolds-number asymptotic technique. Considering the growth rate of the travelling Mack mode to be much smaller than its wavenumber, the transverse and lateral perturbation velocities of the streak mode, showing a roll structure, are \textcolor{magenta}{primarily} amplified due to the mutual interaction of the oblique modes, but its streamwise velocity perturbation undergoes a further amplification due to the lift-up mechanism induced by the roll structure. For a low-Mach-number supersonic boundary layer, the most unstable linear perturbation is usually the oblique first mode, which ensures the dominant perturbation in the early nonlinear phase to be three-dimensional, indicating that  the oblique-mode breakdown  regime is  likely to be triggered in this configuration.

The subharmonic resonance (SR) appears when the dominant perturbation in the early nonlinear phase is planar, or two-dimensional (2D), \textcolor{magenta}{and the frequency of the most amplified 3D perturbations is half of that of the 2D mode; see the sketchmatic in figure~\ref{fig:pinpu_sketch}-(b).} Such a regime usually appears in  a subsonic or an incompressible boundary layer, as observed numerically \citep{Herber1988} and experimentally  \citep{SARIC_1984}, and the rapid amplification of the 3D modes is attributed to the secondary instability (SI) based on the Floquet theory \citep{Herber1988}. For supersonic boundary layers, \cite{kosinov_1990,Kosinov_1994} and \cite{Kosinov_1996} reported a generalised subharmonic regime, for which the dominant perturbation is a 3D Mack first mode, and the two most promoted SI modes are subharmonic in frequency. This scenario was later confirmed by the numerical simulations in \cite{mayer_wernz_fasel_2011}.


For a hypersonic boundary layer, the 2D Mack second mode is the most linearly amplified perturbation, which could be the dominant perturbation in the early nonlinear phase. \textcolor{red}{Using the DNS approach, the most amplified 3D modes are found to be those with the same frequency as the 2D second mode \citep{sivasubramanian_fasel_2015,hader_fasel_2019}, which was also confirmed by the secondary instability analysis (SIA) in \cite{chen_zhu_lee_2017}.} This scenario is referred to as the fundamental resonance (FR) and a sketchmatic for the FR in the spectrum space is shown in figure~\ref{fig:pinpu_sketch}-(c). Based on the critical-layer theory \citep{Wu_2004,Wu_zhang_2022}, \cite{Wu_Luo_2016} deduced the evolution equations for the oblique modes and claimed that the 2D mode acts as a catalyst to promote the growth of the oblique modes. The 2D dominant mode and the small-amplitude oblique modes are found to be phase-locked. Actually, the SI modes include both the 3D travelling waves  and the stationary streak mode, and the amplitude of the latter was found to be much greater than those of the former \citep{Brad_2009,Chou_2011,Chynoweth_2019,sivasubramanian_fasel_2015,hader_fasel_2019}. However, the latter phenomenon so far is not well explained  from the dynamic viewpoint.

For convenience of illustration,  each Fourier component with a frequency $m\omega_0$ and a spanwise wavenumber $n\beta_0$ is denoted by $(m,n)$, where $\omega_0$ and $\beta_0$ are the fundamental frequency and spanwise wavenumber, respectively. It is seen that in both the SR and FR regimes, the dominant perturbation in the early nonlinear phase is 2D. If we choose the frequency of the 2D mode to be the fundamental frequency $\omega_0$, then the 2D fundamental mode is denoted by (1,0). For the SR, the most unstable 3D travelling waves are components $(1/2,n)$, where $n$ is an integer to represent the spanwise wavenumber. The components (1,0), $(1/2,n)$  and $(1/2,-n)$ with a non-zero $n$ form a triad resonance system, for which the mutual interaction of any two components seeds for the growth of the third one. When the dominant 2D mode (1,0) reaches a nonlinear saturation phase, the oblique modes $(1/2,n)$  and $(1/2,-n)$ could amplify with a greater rate,  interpreted as an SI regime. As the oblique pair $(1/2,\pm n)$ reach finite amplitudes, their interaction could also drive a streak mode $(0,2n)$, similar to the oblique-mode breakdown regime.
However, for the FR, the most unstable 3D travelling waves are $(1,n)$ with $n$ being an integer, and the Fourier components $(1,0)$, $(1, n)$ and $(1,-n)$ with a non-zero $n$ do not form a triad resonance system. Although  SI analyses have confirmed the rapid amplification of $(1,\pm n)$ in the nonlinear phase \citep{sivasubramanian_fasel_2015,chen_zhu_lee_2017,hader_fasel_2019}, the energy transfer among different Fourier components due to their nonlinear interaction is far beyond obvious. Actually, the SI modes include a set of oblique modes with the same frequency $(1,n)$ and  stationary streak modes $(0,n)$, and the Fourier components (1,0), $(1,n)$ and $(0,n)$ could form a triad resonance system. Inclusion of the streak mode in the triad resonance determines the key role of the streak mode in the SI process of the FR, which is in contrast to the SR. Unfortunately, such a dynamic mechanism has not been formulated theoretically, especially in the hypersonic boundary layers, which is the main task of the present work.

The rest part of this paper is structured as follows. In \S  \ref{sec:2}, we introduce the physical model  and the numerical treatment (NPSE approach) for the FR, and the NPSE calculations showing the cruial role of the streak mode are demonstrated in  $\S$ \ref{sec:FR_NPSE}. In $\S$ \ref{sec:asymptotic}, we develop an asymptotic theory to describe the dynamic mechanism of the FR, whose accuracy is confirmed by the NPSE calculations for moderate Reynolds numbers in  $\S$ \ref{sec:compare} and by SI analysis for sufficiently high Reynolds numbers in  $\S$ \ref{sec:compare_SIA}. The concluding remarks are present in \S  \ref{sec:Conclusion}.

\section{Physical model and governing equations\label{sec:2}}
\subsection{Physical model}

\begin{figure}
\centering
\includegraphics[width=0.95\textwidth]{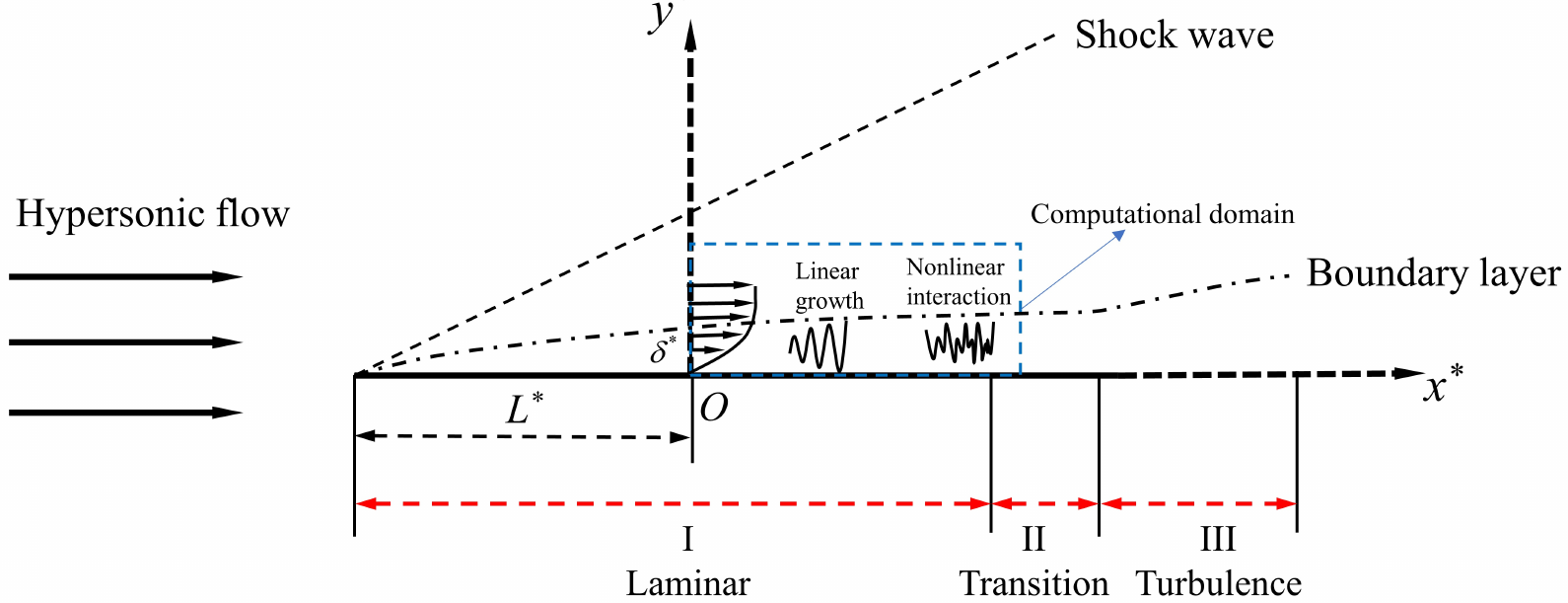}
\caption{Sketch of the physical model.}
\label{fig:sketch}
\end{figure}

The physical model to be studied is a flat plate inserted into a perfect-gas hypersonic stream with zero angle of attack, as sketched in figure~\ref{fig:sketch}. The plate is assumed to be infinitely thin such that a rather weak oblique shock forms from its leading edge, and the flow quantities behind the shock are rather close to those of the oncoming stream. 
A viscous boundary layer forms in the close neighbourhood of the wall.

We study the evolution of a set of Mack instability modes in a selected computational domain, whose inlet boundary is located at a distance $L^*$ downstream of the leading edge, where the  Mack  modes are already unstable. The inflow perturbations consist of a dominant 2D Mack mode and a pair of small 3D Mack modes with the same frequency.
The problem is described by the Cartesian coordinate $\textbf x^*=(x^*,y^*,z^*)$ with its origin $O$ at the inlet of the computational domain. The reference length is selected to be the boundary-layer characteristic thickness  at the origin $\delta^*=\sqrt{\nu^{*}_\infty  L^{*}/U^{*}_\infty}$, where $U_\infty^*$ and $\nu^{*}_\infty$ denote the velocity and the kinematic viscosity of the oncoming flow. The dimensionless coordinate and time are expressed as
\begin{equation}
\textbf x=(x,y,z)=\textbf x^*/\delta^*,\quad t=t^*U^{*}_\infty/\delta^*.
\end{equation}
In what follows, the superscript $^*$ denotes the dimensional quantities.
The velocity field $\textbf u=(u,v,w)$, density $\rho$, temperature $T$, pressure $p$ are normalised by $U^{*}_\infty$, $\rho^{*}_\infty$, $T^{*}_\infty$, $\rho^{*}_\infty U^{{*}2}_\infty$, respectively, where the subscript $\infty$ denotes the quantities at the oncoming stream. 
 For  unsteady perturbations, the frequency $\omega$, streamwise wavenumber $\alpha$ and spanwise wavenumber $\beta$ are normalised as
\begin{equation}
\omega=\omega^* \delta^*/U^{*}_\infty,\quad \alpha=\alpha^* \delta^*,\quad\beta=\beta^*\delta^*,
\end{equation}

The flow field is governed by two dimensionless parameters,  the Reynolds number $R=U^{*}_\infty\delta^*/\nu^{*}_\infty$ and the Mach number $M=U_\infty^*/a_\infty^*$, with $a_\infty^*$ denoting the sound speed of the oncoming stream.

\subsection{Governing equations}

The dimensionless governing equations are
 \begin{subequations}
\begin{align}
&\frac{{\partial \rho }}{{\partial t}} + \nabla  \cdot \left( {\rho {\bf{u}}} \right) = 0,\\
&\rho  {\frac{{D{\bf{u}}}}{{Dt}} }  =  - \nabla {p}  + \frac{1}{{{\mathop{ R}\nolimits} }}\left[ {2\nabla  \cdot \left( {\mu {\bf{S}}} \right) - \frac{2}{3}\nabla \left( {\mu \nabla  \cdot {\bf{u}}} \right)} \right],\\
&\frac{1}{\gamma }\rho \frac{{DT}}{{Dt}} - \frac{{\gamma  - 1}}{\gamma }T\frac{{D\rho }}{{Dt}} = \frac{1}{{PrR}}\nabla  \cdot \left( {\mu \nabla T} \right) + \frac{{\left( {\gamma  - 1} \right){M^2}}}{{R}}\left[ {2\mu {\bf{S}}:{\bf{S}} - \frac{2}{3}\mu {{\left( {\nabla  \cdot {\bf{u}}} \right)}^2}} \right], \\
& {p} = \frac{{\rho }{T}}{\gamma {M}^2}, \label{eq:state}
  \end{align}
\label{eq:NS_eq_dimless}
 \end{subequations}
where ${\bf{S}} = {{\left[ {\nabla {\bf{u}} + {{\left( {\nabla {\bf{u}}} \right)}^T}} \right]} \mathord{\left/
 {\vphantom {{\left[ {\nabla u + {{\left( {\nabla u} \right)}^T}} \right]} 2}} \right.
 \kern-\nulldelimiterspace} 2}$ is the strain-rate tensor, $\mu$ is the dynamic viscosity satisfying the Sutherland law ($\mu=(1+ { C})T^{\frac{3}{2}}/(T+ { C}) $ with ${ C}=110.4K/{T_\infty^{*}}$),  $Pr$ is the Prandtl number, $\gamma$ is the ratio of the specific heats, and $\frac{D}{Dt} = \frac{\partial}{\partial t} + \bf{u} \cdot \nabla$ denotes the material derivative. 
In this paper, we take $Pr=0.72$ and $\gamma=1.4$.
 
The no-slip, non-penetration and isothermal boundary conditions are applied at the wall,
\begin{equation}
    (u,v,w,T)=(0,0,0,T_w) \quad \text{at} \quad y=0,
     \label{eq:lower_BL}
\end{equation}
where $T_w$ is the dimensionless wall temperature. In the far field, all the perturbations damp exponentially (the radiating mode as in \cite{chuvakhov_fedorov_2016} is not considered here), and the upper boundary conditions read
 \begin{equation}
    (\rho, u,v,w,T)\rightarrow(1,1,0,0,1) \quad \text{as} \quad y \to \infty.
    \label{eq:upper_BL}
\end{equation}

 \subsection{ \label{sec:perturbation} Perturbation field}

 The instantaneous flow field $\phi \equiv (\rho ,u,v,w,  T)$ can be decomposed into a steady base flow $\Phi_B$ and an unsteady perturbation $\tilde \phi$,
\begin{equation}
 \phi  = \Phi_B (x,y)  +  \tilde \phi (x,y,z,t).
\label{eq:decomposite}
\end{equation}
where $\Phi_B=(1/T_B,U_B,V_B,0,T_B)$ is the compressible Blasius solution. Substituting (\ref{eq:decomposite}) into (\ref{eq:NS_eq_dimless}), and subtracting the base flow out, we obtain the nonlinear equations governing the perturbations,

\begin{equation}
 \begin{split}
  & {\bf { G }}\frac{{\partial \tilde \phi }}{{\partial t}} + {\bf{A}}\frac{{\partial \tilde \phi }}{{\partial x}} + {\bf{B}}\frac{{\partial \tilde \phi }}{{\partial y}} + {\bf{C}}\frac{{\partial \tilde \phi }}{{\partial z }} + {\bf{D}}\tilde \phi  + {{\bf{V}}_{xx}}\frac{{{\partial ^2}\tilde \phi }}{{\partial {x^2}}} + {{\bf{V}}_{yy}}\frac{{{\partial ^2}\tilde \phi }}{{\partial {y^2}}} +\\
& {{\bf{V}}_{z z }}\frac{{{\partial ^2}\tilde \phi }}{{\partial {z ^2}}} + {{\bf{V}}_{xy}}\frac{{{\partial ^2}\tilde \phi }}{{\partial x\partial y}} + {{\bf{V}}_{xz }}\frac{{{\partial ^2}\tilde \phi }}{{\partial x\partial z }} + {{\bf{V}}_{yz }}\frac{{{\partial ^2}\tilde \phi }}{{\partial y\partial z }} = \bf{ F},
\label{eq:dis_e}
 \end{split}
\end{equation}
where the coefficient matrices $\bf {G}$, $\bf{A}$, $\bf{B}$, $\bf{C}$, $\bf{D}$, ${\bf{V}}_{xx} $, ${\bf{V}}_{yy} $, ${\bf{V}}_{z z} $, ${\bf{V}}_{xy} $, ${\bf{V}}_{yz} $ and ${\bf{V}}_{x z} $ and the nonlinear forcing $\bf{F}$ can be found in Appendix \ref{sec:Appb}. The pressure perturbation $\tilde p$ has been eliminated by the equation of the state (\ref{eq:state}).

In this paper, we particularly focus on the FR regime, therefore, the introduced perturbations $ \tilde \phi$ at the inlet of the computational domain should include a 2D Mack second mode with a frequency $\omega_0$ and a pair of 3D Mack second modes with the same frequency $\omega_0$ but opposite spanwise wavenumbers $\pm \beta_0$, where $\omega_0$ and $\beta_0$ are referred to as the fundamental frequency and fundamental spanwise wavenumber, respectively. For convenience, we use $(m,n)$ to denote a perturbation with a frequency $m\omega_0$ and a spanwise wave number $n\beta_0$. Thus, the three introduced perturbations are denoted by (1,0), (1,1) and (1,-1), respectively. The inflow perturbation can be expressed as
\begin{equation}
\begin{split}
 \tilde \phi(0,y,z,t) &=\epsilon_{10}\hat \phi_{10}(y)\exp(-\ri\omega_0 t)+\epsilon_{11}\hat \phi_{11}(y)\exp[\ri(\beta_0z-\omega_0 t)] \\
 &+\epsilon_{1-1}\hat \phi_{1-1}(y)\exp[\ri(-\beta_0z-\omega_0t)]+ \text{c.c.},   
\end{split}
\label{eq:initial_perturbations}
\end{equation}
where $\epsilon_{mn}$ measures the initial amplitude of the introduced perturbation $(m,n)$, $\hat{\phi}_{mn}$ denotes the perturbation profile for the $(m,n)$ component, $\text{c.c.}$ denotes the complex conjugation and $\text{i} \equiv \sqrt{-1}$. For a hypersonic boundary layer, the 2D second mode is usually more unstable, and the amplitude of  mode (1,0) should be much greater than modes $(1,\pm 1)$ due to the historical accumulative effect. Thus, the  amplitude of the introduced 2D mode $\epsilon_{10}$ is taken to be much greater than those of the 3D modes $\epsilon_{11}$ and $\epsilon_{1-1}$. Also, the linear growth rates of the two oblique modes are the same, and so we let $\epsilon_{11}=\epsilon_{1-1}$. Note that in reality the spanwise wavenumbers of the oblique modes may be not exactly opposite, and their amplitudes may be different, but our selection is still a good demonstration to reveal their resonance mechanism. 

 \subsubsection{ \label{sec:LST} Linear stability theory (LST)}
 
The perturbation profile $\hat\phi$ for each Fourier mode of the inflow perturbation is obtained by the linear stability theory (LST). Introducing the parallel-flow assumption, the perturbation with a frequency $\omega$, a streamwise wavenumber $\alpha$ and a spanwise wavenumber $\beta$ is expressed in terms of a travelling-wave form,
\begin{equation}
 \tilde \phi = \epsilon_L  \hat{\phi}(y) \exp [\ri (\alpha x + \beta z -\omega t) ] + \text{c.c.} , 
   \label{eq:LST}
\end{equation}
where $\epsilon_L \ll 1$ measures its amplitude. Substituting (\ref{eq:LST}) into the system (\ref{eq:dis_e}) with $O(\epsilon_L^{2})$ terms neglected, we obtain the compressible Orr-Sommerfeld (O-S) equations,
\begin{equation}
    {\bf{\tilde B}}\frac{{\partial \hat \phi }}{{\partial y}} + {{\bf{V}}_{yy}}\frac{{{\partial ^2}\hat \phi }}{{\partial {y^2}}} + {\bf{\tilde D}}\hat \phi  = 0,
    \label{eq:OS}
\end{equation}
where 
\begin{equation}
\begin{split}
&{\bf{\tilde B}} = {\bf{B}} + {\rm{i}}\alpha {{\bf{V}}_{xy}} + {\rm{i}}\beta{{\bf{V}}_{yz }},\\
&{\bf{\tilde D}} =  - {\rm{i}}\omega {\bf{G}}  + {\rm{i}}\alpha {\bf{A}} + {\rm{i}}\beta{\bf{C}} + {\bf{D}} - {\alpha ^2}{{\bf{V}}_{xx}} - {\beta^2}{{\bf{V}}_{zz }} - \alpha \beta{{\bf{V}}_{xz }}  .  
 \label{eq:matrix_LST}
\end{split}
\end{equation}
Introducing the homogeneous boundary conditions, 
\begin{equation}
\begin{split}
[\hat{u},\hat{v},\hat{w},\hat{T}]=0 \quad \text{at} \quad y=0; \quad \hat{\phi} \to 0 \quad \text{as} \quad y\to \infty,
\end{split}
\label{eq:BC_OS}
\end{equation}
 we arrive at an eigenvalue problem.
For the spatial mode, $\omega$ and $\beta$ are given to be real, and the eigenvalue $\alpha= \alpha_r + \text{i} \alpha_i$ is complex with the opposite of its imaginary part representing the growth rate. Usually, the imaginary part is much smaller than the real part in the boundary-layer flow, i.e., $|\alpha_i| \ll |\alpha_r|$. 
The numerical details to solve the eigenvalue system (\ref{eq:OS}) with (\ref{eq:BC_OS}) can be found in our previous papers \citep{dong_liu_wu_2020,Song_2020,dong_zhao_2021,li2021}.

 \subsubsection{ \label{sec:FR_cal_NPSE} Nonlinear parabolised stability equations (NPSE)}

The NPSE approach \citep{bertolotti_herbert_spalart_1992, chang_malik_1994}  is considered as a more accurate means because it allows the slow streamwise variation of the perturbation profiles and takes into account the non-parallelism of the base flow. The only approximation is that the $\partial_{xx}$ terms are neglected to reduce the elliptic system to a parabolised system, which is quite reasonable for a boundary layer with a smooth wall. Expressing $\tilde \phi$ and $\bf{F}$ in terms of the Fourier series with respect to $z$ and $t$, we obtain 
\begin{equation}
\begin{split}
  &  \tilde \phi (x,y,z,t)  = \sum\limits_{m = -M_e}^{{M_e}} {\sum\limits_{n =  - {N_e}}^{{N_e}}  {{\mathord{\buildrel{\lower3pt\hbox{$\scriptscriptstyle\smile$}} 
\over \phi } }_{mn}}(x,y) \text{exp} [\text{i} \left(n {\beta_0} z - m {\omega_0} t\right) ] }  , \\
 & {\bf{F}} (x,y,z,t) = \sum\limits_{m = -M_e}^{{M_e}} {\sum\limits_{n =  - {N_e}}^{{N_e}} {{{{\bf{\tilde F}}}_{mn} (x,y) }{ \text{exp} [ {\text{i}\left( {n{\beta_0}z  - m{\omega _0}t} \right)} ] }} }  ,
    \label{eq:tidle_phi_F}
    \end{split}
  \end{equation}
   where $M_e$ and $N_e$ denote the orders of the Fourier-series truncation. In this paper, we choose $M_e =5$ and $N_e =5$, which has been confirmed to be sufficient via resolution tests.
  Considering that the perturbations are propagating with two length scales, a fast one with an oscillatory manner and a slow one related to the non-parallelism, we express the perturbation profile ${\mathord{\buildrel{\lower3pt\hbox{$\scriptscriptstyle\smile$}} 
\over \phi } }$ in terms of a Wentzel-Kramers-Brillouin (WKB) form,
\begin{equation}
\begin{split}
  &     {{\mathord{\buildrel{\lower3pt\hbox{$\scriptscriptstyle\smile$}} 
\over \phi } }_{mn}} (x,y) = {{\check \phi }_{mn}}\left( {x,y} \right){\text{exp}({i \int\limits_{{x_0}}^x {{\alpha _{mn}} (\bar x) d\bar x}  })} , 
    \label{eq:tidle_phi_1}
    \end{split}
  \end{equation}
where each Fourier component is denoted by $(m,n)$, $\omega_{0}$ and $n_{0}$ are the fundamental frequency and spanwise wavenumber, respectively, and $\alpha _{mn}$ represents the complex streamwise wavenumber of $(m,n)$. The shape function $\check{\phi}_{mn}$ varies slowly with $x$. The integral in (\ref{eq:tidle_phi_1})  starts from a reference streamwise position $x_0$, which is selected as the inlet of the computational domain for the numerical calculations in this paper, namely, $x_0 \equiv 0$.


Neglecting the $\partial_{xx} \check{\phi}_{mn}$ terms,  system (\ref{eq:dis_e}) is reduced to 
\begin{equation}
   {{{\bf{\tilde A}}}_{mn}}\frac{{\partial {{\check \phi }_{mn}}}}{{\partial x}} + {{{\bf{\tilde B}}}_{mn}}\frac{{\partial {{\check \phi }_{mn}}}}{{\partial y}} + {{\bf{V}}_{yy}}\frac{{{\partial ^2}{{\check \phi }_{mn}}}}{{\partial {y^2}}} + {{{\bf{\tilde D}}}_{mn}}{{\check \phi }_{mn}} =  {{{{{\bf{\check F}}}_{mn}}}},
   \label{eq:PSE}
\end{equation}
where the matrices ${{{\bf{\tilde A}}}_{mn}}$, ${{{\bf{\tilde B}}}_{mn}}$ and ${{{\bf{\tilde D}}}_{mn}}$ are given by
\begin{equation}
\begin{split}
&{{{\bf{\tilde A}}}_{mn}} = {\bf{A}} + 2{\rm{i}}{\alpha _{mn}}{{\bf{V}}_{xx}} + {\rm{i}}n{\beta_0}{{\bf{V}}_{xz }},\\
&{{{\bf{\tilde B}}}_{mn}} = {\bf{B}} + {\rm{i}}{\alpha _{mn}}{{\bf{V}}_{xy}} + {\rm{i}}n{\beta_0}{{\bf{V}}_{yz }},\\
&{{{\bf{\tilde D}}}_{mn}} =  - {\rm{i}}m{\omega _0} {\bf {G}}  + {\rm{i}}{\alpha _{mn}}{\bf{A}} + {\rm{i}}n{\beta_0}{\bf{C}} + {\bf{D}}- {n^2}\beta_0^2{{\bf{V}}_{z z }}\\
& \quad \quad \quad - \left( {\alpha _{mn}^2 - {\rm{i}}\frac{{d{\alpha _{mn}}}}{{dx}}} \right){{\bf{V}}_{xx}}  - n{\alpha _{mn}}{\beta_0}{{\bf{V}}_{xz }}, \\
 & {{{{{\bf{\check F}}}_{mn}}}}= {{{{{\bf{\tilde F}}}_{mn}}}}{{{\text {exp}({-\text{i}\int\limits_{{x_0}}^x {{\alpha _{mn}} (\bar x)d\bar x} })}}}.
    \end{split}
    \label{eq:tidle_phi}
\end{equation}
The inflow perturbations are given by (\ref{eq:initial_perturbations}) and the lower and upper boundary conditions are 
\begin{equation}
    \begin{split}
&({{\check u}_{mn}},{{\check v}_{mn}},{{\check w}_{mn}},{{\check T}_{mn}}) = (0,0,0,0) \quad \text{at} \quad y = 0 ,\\
&({{\check \rho}_{mn}},{{\check u}_{mn}},{{\check v}_{mn}},{{\check w}_{mn}},{{\check T}_{mn}}) \to (0,0,0,0,\textcolor{red}{0}) \quad \text{as} \quad y \to \infty.
    \end{split}
\end{equation}
To solve for the complex streamwise wavenumber $\alpha_{mn}$ and the profiles $\check{\phi}$, an iterative procedure is employed, which can be found in \cite{Zhao_2016}, and our code validation is provided in the appendix of \cite{Song_2022}.

Additionally, if $\check {\bf{F}}_{mn}$ is set to be zero, then equation (\ref{eq:PSE}) is recast to the linear PSE (LPSE), which can be used to track the evolution of each linear mode individually. To be distinguished, the PSE approach with $\check {\bf{F}}_{mn}$ being retained is referred to as the nonlinear PSE (NPSE) in this paper.

\subsubsection{\label{sec:SIA}Secondary instability analysis (SIA) for a wavy base flow}
 When the amplitude of the fundamental perturbation (1,0) has reached a finite level, the rapid growth of the infinitesimal perturbations can be explained by  
 the SI based on a wavy profile driven by a 2D quasi-saturated travelling mode.

Since the growth rate of the fundamental mode is usually much smaller than that of the SI mode, as confirmed by many numerical studies such as \cite{sivasubramanian_fasel_2015}, \cite{chen_zhu_lee_2017} and \cite{hader_fasel_2019}, we take $\alpha_{10}$ to leading order to be real and introduce $ \breve \alpha \equiv \Re (\alpha_{10})  $. Thus, the base flow for the SIA is a superposition of the steady base flow and a series of quasi-saturated travelling waves. In a moving frame, the base flow $\breve{\Phi}_{B} \equiv [\breve \rho_B,\breve U_B,0,0,\breve T_B] $ is expressed as
\begin{equation}
    {\breve {\Phi}} _{B} (\tilde x, y) =  (1/T_{B},  {U_B}-c_r, 0,  0,  T_{B})(y)  + \sum\limits_{m =  - M_{W}}^{M_{W}} {{{\mathord{\buildrel{\lower3pt\hbox{$\scriptscriptstyle\smile$}} 
\over \phi } }_{m0}}\left( y \right){\text{exp}({\text{i}  m \breve \alpha \tilde x}}}) +\cdots  , \label{eq:floquet_base_flow}
\end{equation}
where $c_r= \omega_{0}/{\breve \alpha}$, $\tilde x = x  - c_rt$ is the Galilean transformed coordinate and $M_W$ denotes the order of the Fourier-series truncation. The laminar base flow $\Phi_{B}$ develops with a length scale much greater than the wavelength of the fundamental mode $2 \pi/ {\breve \alpha}$, and so the non-parallelism of $\Phi_B$ in the local region is neglected in the present analysis, rendering a periodic feature of $\breve \Phi_B$ in the streamwise direction.

According to the Floquet theory, the periodic base flow supports the instability modes $\tilde \phi_W$ which can be expressed as
\begin{equation}
\begin{split}
  {{\tilde \phi_W }(\tilde x,y,z,t)} &=  \epsilon_W {\breve \phi _W}\left( {\tilde x,y} \right){\exp{[\tilde \sigma( \tilde x+c_rt)+\text{i} \beta z+{\text{i}{\tilde \sigma _d} \breve \alpha \tilde x}]}} + \text{c.c.}, \\
    \quad {\breve \phi _W}\left( {\tilde x,y} \right) &= \sum\limits_{n =  - {N_W}}^{{N_W}} {{{\hat \phi }_{W,n}}\left( y \right)\exp \left( { \text{i} n \breve \alpha \tilde x} \right)},  
\end{split}
\label{eq:floquet_dis}
    \end{equation}
where $\tilde \sigma$ represents the growth rate, $ \beta$ is the spanwise wavenumber, $\tilde \sigma_{d}$ is the detuning parameter, and $N_W$ is the order of the Fourier-series truncation and $\epsilon_W \ll 1$ measures the amplitude. For the fundamental resonance, we take $\tilde \sigma _d=0$. The component $\hat{\phi}_{W,0}$ denotes the streak component, and $\hat{\phi}_{W,n}$ with $n \ne 0$ represents the travelling mode. Substituting (\ref{eq:floquet_base_flow}) and (\ref{eq:floquet_dis}) into (\ref{eq:dis_e}) with $O(\epsilon_W^{2})$ terms neglected, we arrive at a linear system,
\begin{equation}
    \left( {{{{\bf M}_0}} + \tilde \sigma  { {{\bf M}_1}} + {\tilde \sigma^2} { {{\bf M}_2}} }\right){ \breve  \phi _W}\left( {\tilde x,y} \right) = 0,
    \label{eq:floquet_eigen}
\end{equation}
where
\begin{equation}
\begin{split}
{ {{\bf M}_0}} &= \left( { {\bf{A}} + {\rm{i}}  \beta { {{\bf{V}}_{ x z}}}} \right)\frac{\partial }{{\partial \tilde x}} + \left( {{\bf {B}} + {\rm{i}}  \beta { {{\bf V}_{y z}}}} \right)\frac{\partial }{{\partial y}} + \left[ { {\bf {D}} + {\rm{i}}  \beta {\bf { C}} + {{( {{\rm{i}}  \beta } )}^2}{ { {{\bf V}_{ z z}}}}} \right]  \\
&+{{{\bf V}_{xx}}}\frac{{{\partial ^2}}}{{\partial {{\tilde x}^2}}} +  { {{\bf V}_{yy}}}\frac{{{\partial ^2}}}{{\partial {y^2}}} +  { {{\bf V}_{xy}}}\frac{{{\partial ^2}}}{{\partial \tilde x \partial y}},    \\
 { {{\bf M}_1}} &= \left( { {\bf {A}} + \text{i}  \beta { {{\bf V}_{ x z}}}} \right) + 2 { {{\bf V}_{xx}}}\frac{\partial }{{\partial \tilde x}} + {{{\bf V}_{xy}}}\frac{\partial }{{\partial y}} + c  {\bf {G}} ,  \\
{ {{\bf M}_2}} & = { {{\bf V}_{xx}}}.
\end{split}
\end{equation}
The wall-normal boundary conditions read
\begin{equation}
\begin{split}
    & {\breve u}_{W}= {\breve v}_{W}={\breve w}_{W}={\breve T}_{W}=0  \quad \text{at} \quad y=0, \\
&({\breve \rho}_{W},{\breve u}_{W}, {\breve v}_{W},{\breve w}_{W},{\breve T}_{W}) \to 0 \quad \text{as} \quad y \to \infty.  
\end{split}
\label{eq:SIA_BC}
\end{equation}
The linear system (\ref{eq:floquet_eigen}) with the homogeneous boundary conditions forms an eigenvalue problem with the growth rate $\tilde \sigma$ being the eigenvalue. In our paper, we choose $(M_W,N_W)=(3,6)$, which has been confirmed to be of sufficient accuracy. Such an analysis has also been used in the study of the SI of 2D Mack second modes in  hypersonic boundary layers \citep{chen_zhu_lee_2017,Xu_2020}, and our code validation and discretisation method are provided in \cite{Song_Zhao_Dong_2023}.

\section{\label{sec:FR_NPSE}Demonstration of the fundamental resonance regime by NPSE calculations}
\subsection{Case studies}
\begin{table}
  \begin{center}
\def~{\hphantom{0}}
  \begin{tabular}{cccccccccccc}
     Case& &$M$ & & $T^{*}_{\infty}$ &  &$T_w$&& $T_w/T_{ad}$& &   $\delta_{99}$  \\ [10pt]
      A & &5.92   & &48.69K& &6.95&&1& &19.95  \\
      B &&5.92   & &48.69K& &3.47&&0.5& &13.94   \\
  \end{tabular}
  \caption{Parameters characterising the flow condition}
  \label{tab:bf_condition}
  \end{center}
\end{table}
For demonstration of the FR, we select a wind-tunnel condition in \cite{maslov_shiplyuk_sidorenko_arnal_2001}, for which the Mach number and temperature of the oncoming stream are 5.92 and 48.69K, respectively. Such an oncoming condition was also used in \cite{dong_liu_wu_2020} and \cite{dong_zhao_2021}. Two wall temperatures as listed in Table \ref{tab:bf_condition} are selected, which are equal to and a half of the adiabatic wall temperature $T_{ad}$, respectively, where $T_{ad}$ is estimated by an empirical formula in White (\citeyear{white2006}, p.512),
\begin{equation}
    T_{ad}=1+\sqrt{Pr}(\gamma-1) M^{2}/2.
\end{equation}
The nominal boundary-layer thickness for each case is also listed in the table.
\subsection{Base flow and its linear instability}
\begin{figure}
\centering
\begin{overpic}[width=0.96 \textwidth]{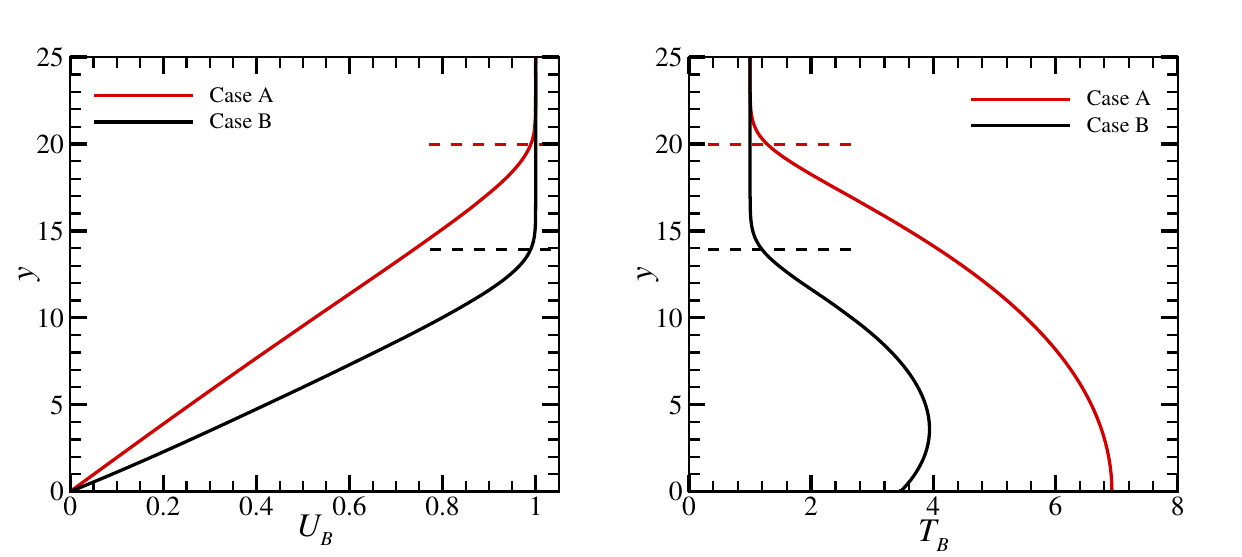}
 \put(-1.5,39.5){(a)}
\put(48,39.5){(b)}
\end{overpic}
\caption{ The streamwise velocity (a) and temperature (b) of the compressible Blasius solution at $x=0$ for cases A and B. The red and black horizontal lines denote the nominal boundary-layer thicknesses for the two cases, respectively.      }
\label{fig:bf_1}
\end{figure}
\begin{figure}
\centering
\begin{overpic}[width=0.96 \textwidth]{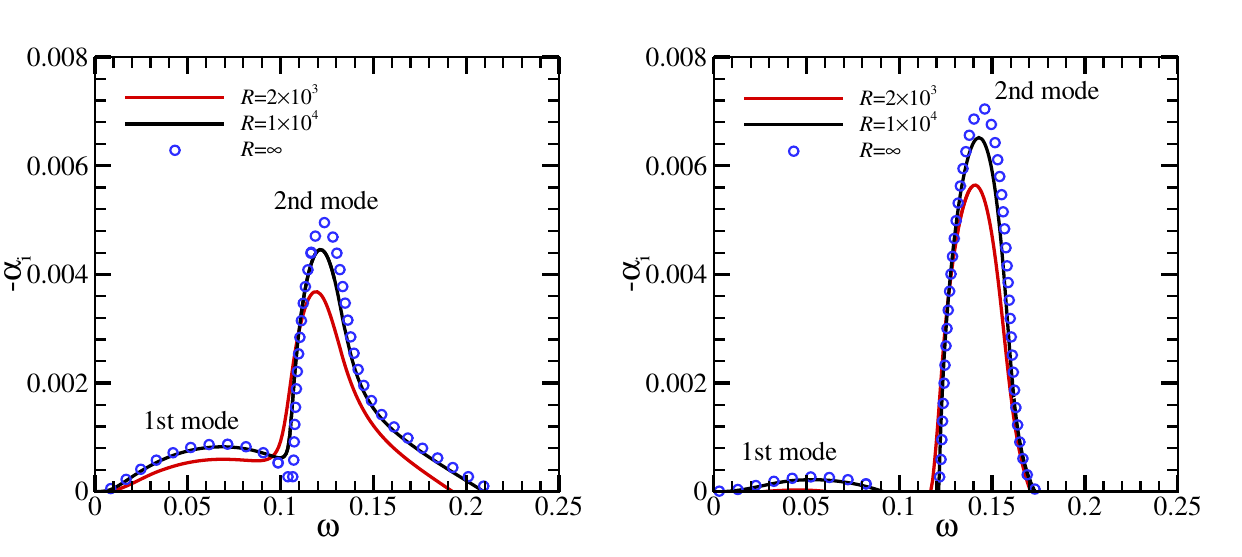}
 \put(-2,39.5){(a)}
\put(47.5,39.5){(b)}
\end{overpic}
\caption{ Dependence on the frequency $\omega$ of the growth rate $-\alpha_i$ of 2-D modes for case A (a) and case B (b).      }
\label{fig:gr_w_1st_2nd-new}
\end{figure}
 The base-flow profiles of $U_B$ and $T_B$ at $x=0$ for the two cases are shown in figure~\ref{fig:bf_1}. As the wall temperature decreases, the boundary-layer thickness is reduced and the shear rates of $U_B$ and $T_B$ at the wall increase. Solving the O-S equations numerically based on these base-flow profiles, we obtain the dependence of the growth rates $-\alpha_i$ of 2D Mack modes on the frequency $\omega$ for the two cases, as shown in figures~\ref{fig:gr_w_1st_2nd-new}-(a) and (b), respectively. Two distinguished unstable zones appear for each case, which are marked by the Mack first and second modes, respectively \citep{Mack_1987}. The second mode is more unstable than the first mode, and decrease of the wall temperature leads to an enhancement of the second mode and suppression of the first mode overall. In each panel, we show the results for three Reynolds numbers, namely, $R=2 \times 10^{3}$, $1\times 10^{4}$ and $\infty$ (for this case the O-S equations reduce to the Rayleigh equations, which will be shown in $\S$ \ref{sec:asy_2D}). Overall, increase of the Reynolds number leads to a greater growth rate, indicating the inviscid nature of the 2D Mack modes.

 \begin{figure}
\centering
\begin{overpic}[width=0.96 \textwidth]{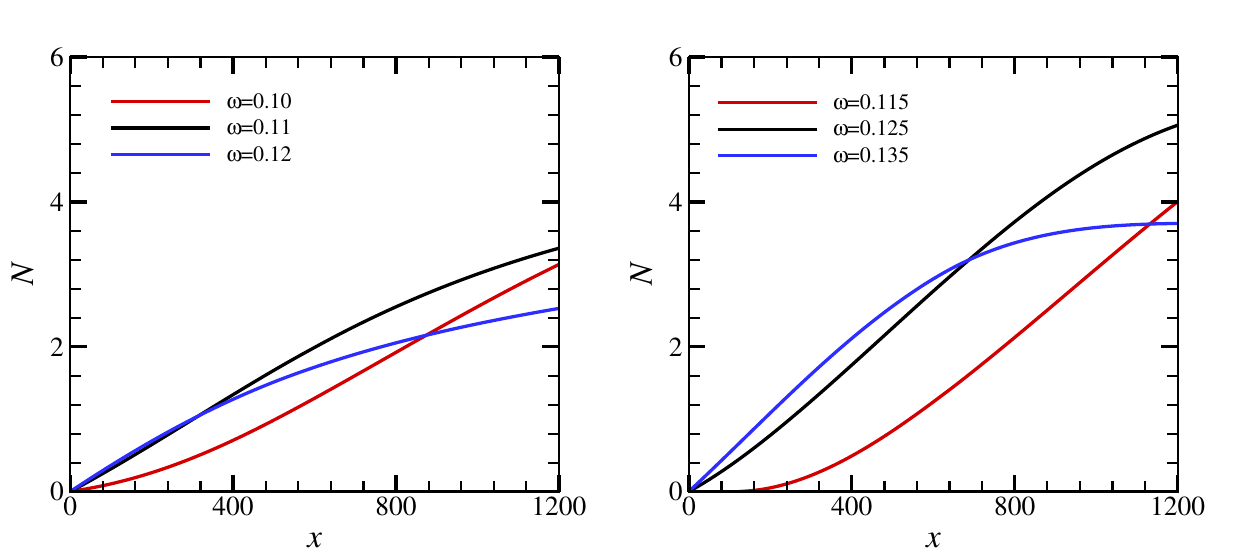}
 \put(-2,39.5){(a)}
\put(47.5,39.5){(b)}
\end{overpic}
\caption{ Streamwise evolution of the $N$ factors of 2-D second modes for case A (a) and case B (b) at $R=2 \times 10^{3}$.      }
\label{fig:N_2nd}
\end{figure}

 The accumulated amplitude of each Fourier mode can be quantified by an $N$ factor according to the LST, defined by 
\begin{equation}
    N(x) =\exp\Big[ \int\limits_{{0}}^x { - {\alpha _i} (\bar x) d\bar x}\Big].
\end{equation}
 In  figure~\ref{fig:N_2nd}, we plot the streamwise evolution of the $N$ factors of 2D second modes with representative frequencies in the second-mode frequency band for $R=2 \times 10^{3}$. It is observed that for cases A and B, the frequencies of the most amplified second modes from $x=0$ to $1200$ are  $\omega=0.11$ and $\omega=0.125$, respectively, and they are selected as the fundamental frequencies $\omega_0$ in the following NPSE calculations.

\subsection{\label{sec:cal_demon}Calculations of the fundamental resonance}

\begin{table}
  \begin{center}
\def~{\hphantom{0}}
  \begin{tabular}{ccccccccccccccccc}
     Case&& $M$&&$T_w/T_{ad}$ & & $R$ &  &$\omega_0$&& $\beta_{0}$& &   $\epsilon_{10}$ &&$\epsilon_{1\pm1}$ \\ [10pt]
   Case A & &5.92&&1   & &$2\times 10^{3}$& &0.11&&0.1 & &$2.5 \times 10^{-3}$&&$2.5 \times 10^{-5}$  \\
   Case B & &5.92&&0.5   & &$2\times 10^{3}$& &0.125&&0.1 & &$2.5 \times 10^{-3}$&&$2.5 \times 10^{-5}$  \\
  \end{tabular}
  \caption{Parameters for case studies in $\S$\ref{sec:cal_demon}. }
  \label{tab:case_demon}
  \end{center}
\end{table}

\begin{figure}
\centering
\begin{overpic}[width=0.93 \textwidth]{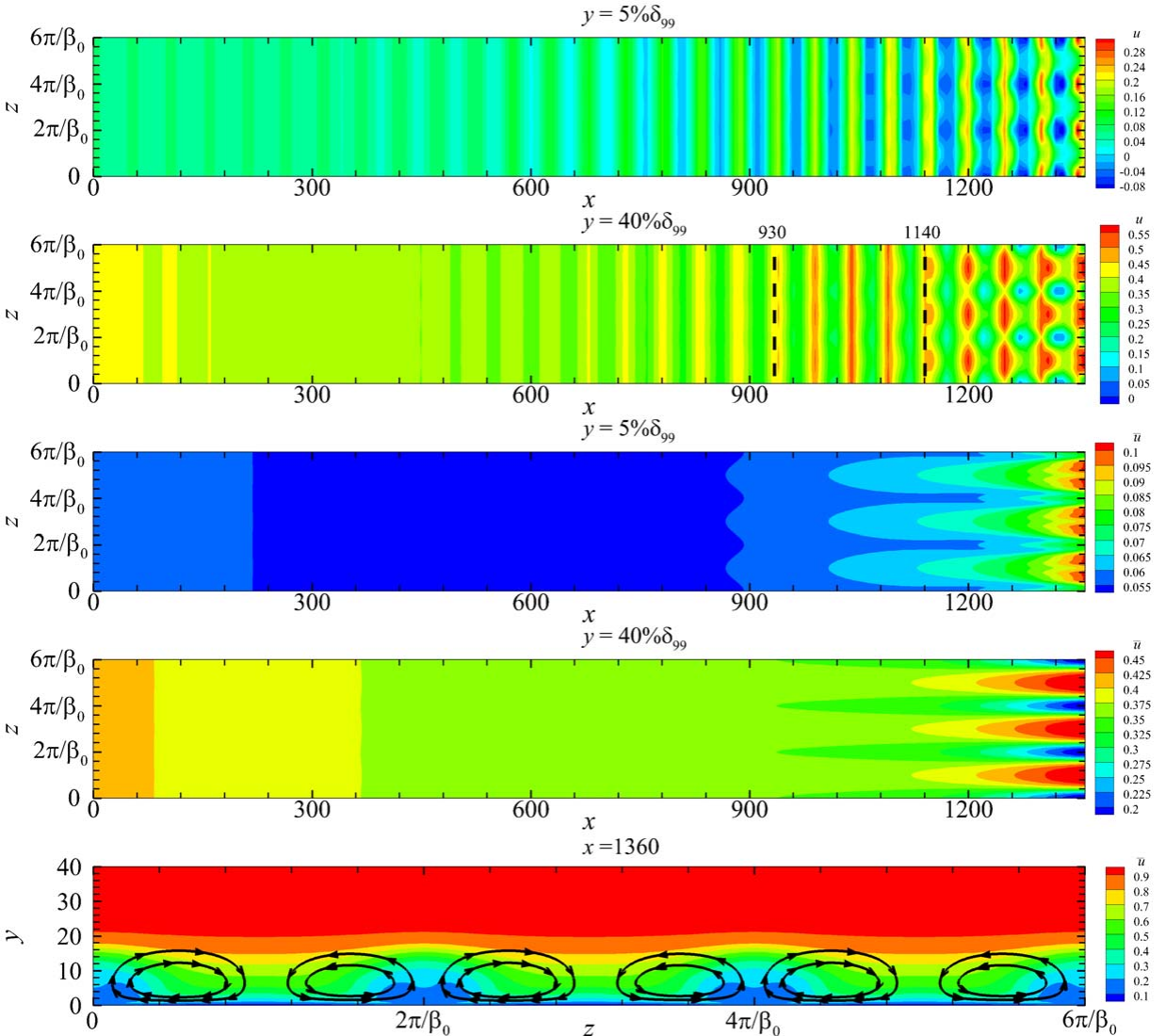}
\put(-4,85){(a)}
\put(-4,67){(b)}
\put(-4,49){(c)}
\put(-4,31){(d)}
\put(-4,13){(e)}
\end{overpic}
\caption{ Contours of the velocity field obtained by the NPSE for case A. (a) and (b) are for the instantaneous velocity $u$ in the $x$-$z$ planes at $y=5\% \delta_{99}$ and $y=40\% \delta_{99}$, respectively. (c) and (d) are for the time-averaged velocity in the $x$-$z$ planes at $y=5\% \delta_{99}$ and $y=40\% \delta_{99}$, respectively. \textcolor{magenta}{(e) is for the time-averaged velocity in the $y-z$ plane at $x=1360$, where the arrayed curves show the streamlines.} }
\label{fig:bf_dis}
\end{figure}
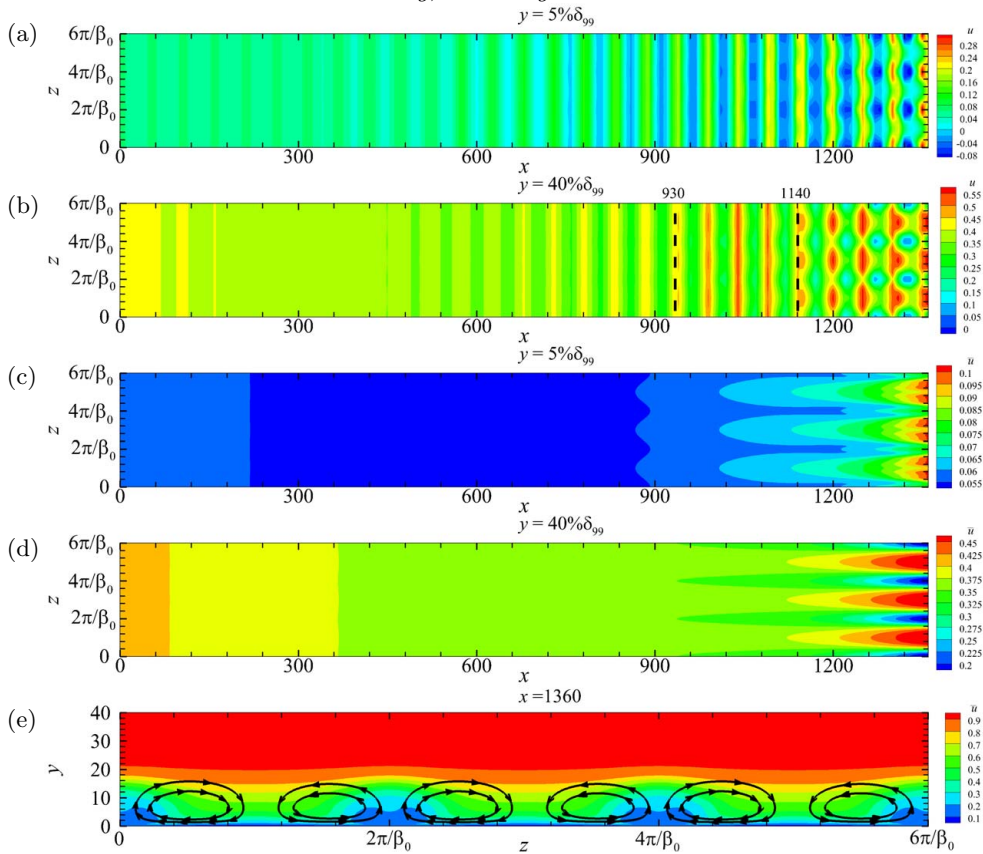

For each case, we calculate the nonlinear evolution of the initial perturbations (\ref{eq:initial_perturbations}) using the NPSE approach, until the calculation blows up, indicating the emergence of the transition onset in a short distance downstream \citep{AMM2008_Dong}. The parameters for cases A and B are summarised in table~\ref{tab:case_demon}. For case A, the contours of the instantaneous velocity $u$ in the $x-z$ plane at two wall-normal positions are shown in figures~\ref{fig:bf_dis}-(a) and (b), respectively. For $x<930$, the perturbation field is dominated by planar waves; however, 3-D structures appear in further downstream locations $(x>930)$ and grow with a high rate. Panels (c) and (d) show the time-averaged streamwise velocity at the same wall-normal positions for comparison, and the low- and high-speed streaks are observed evidently in the late nonlinear phase. \textcolor{magenta}{In panel (e), we plot the contours of the time-averaged streamwise velocity in the $y-z$ plane at $x=1360$, where the spanwise localised blue structure indicates the low-speed streaks. The streamlines show the counter-rotating roll structures of the streamwise vorticies, which push the near-wall fluids upward, showing a lift-up mechanism for the formation of  the low-speed streaks.}

\begin{figure}
\centering
\begin{overpic}[width=0.96 \textwidth]{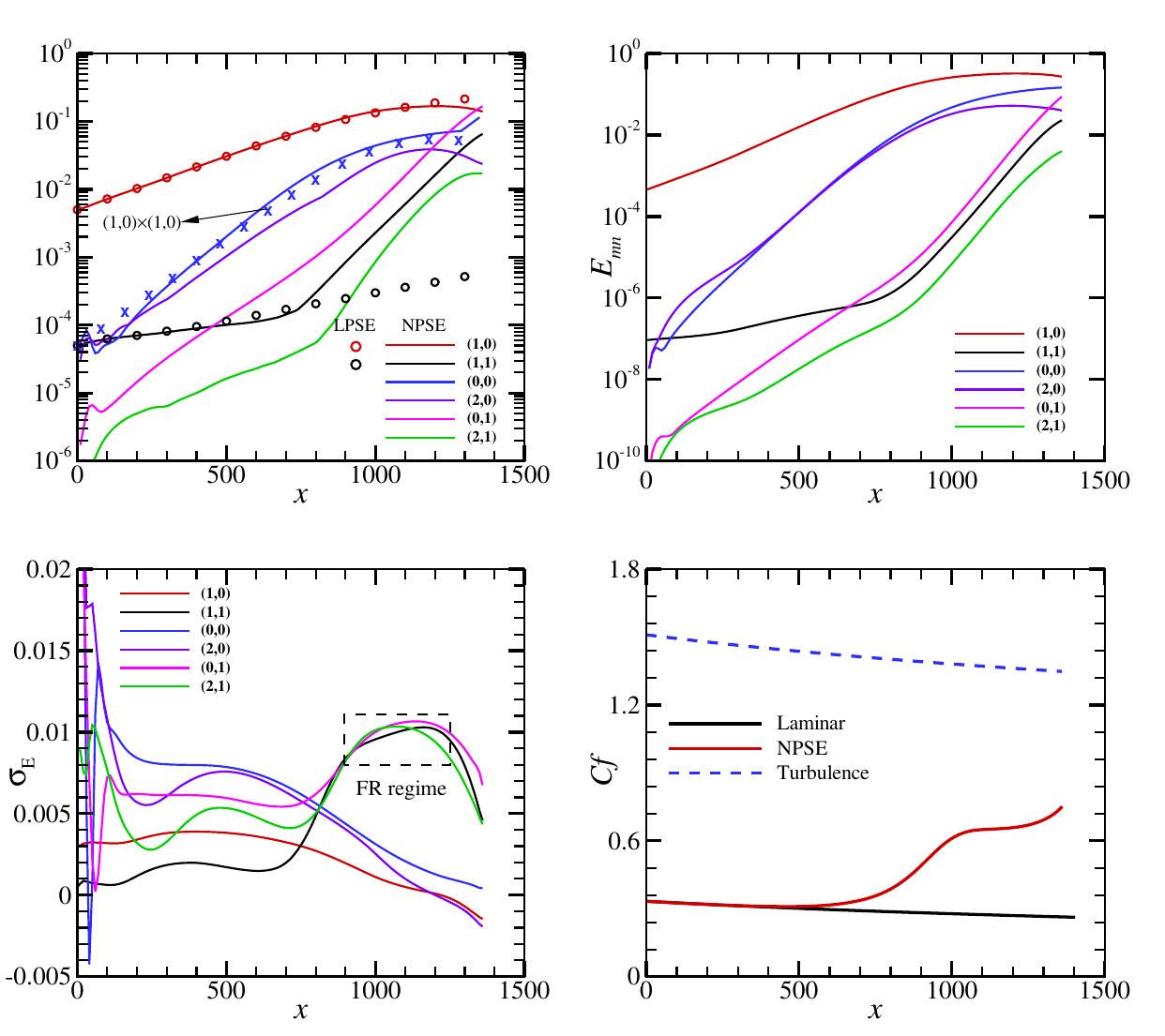}
 \put(-2,83.5){(a)}
\put(47.5,83.5){(b)}
\put(-1,64){\begin{turn}{90}${\tilde u}_{max}$\end{turn}}
\put(-2,38.5){(c)}
\put(47.5,38.5){(d)}
\end{overpic}
\caption{\textcolor{red}{Streamwise evolution of the NPSE results for case A. (a): Amplitude  of each Fourier component $\hat u_{max}^{(m,n)}$; (b): perturbation energy of each Fourier component $E_{mn}$ ;} (c): energy growth rate of each Fourier component $\sigma_E$; \textcolor{magenta}{(d): Coefficient of surface friction, where the Cf curve for  turbulence is given by the empirical formula in White (\citeyear{white2006}, p.553) .} }
\label{fig:Au_evolution_large}
\end{figure}
The amplitude of each Fourier component in the physical space can be expressed as 
\begin{equation}
{\tilde u_{\max }^{(m,n)}(x)} = \left\{ \begin{array}{l}
{\max _y}\left| {{{\mathord{\buildrel{\lower3pt\hbox{$\scriptscriptstyle\smile$}} 
\over u} }_{mn}}\left( {x,y} \right) + c.c} \right|, \quad {{   m \ne 0 \lor n \ne 0}},\\
{\max _y}\left| {{{\mathord{\buildrel{\lower3pt\hbox{$\scriptscriptstyle\smile$}} 
\over u} }_{mn}}\left( {x,y} \right) } \right|,\quad {{   m=0 \land n=0}}.
\end{array} \right. 
\label{eq:A}
\end{equation}
In figure~\ref{fig:Au_evolution_large}-(a), we plot the streamwise evolution of ${\tilde u}_{max}^{(m,n)}$, shown by the solid lines. The amplitude of the fundamental mode (1,0) agrees well with the linear result predicted by LPSE (shown by the red circles) until $x \approx 1100$, after which it saturates due to the nonlinearity. The oblique waves $(1,\pm 1)$ grow with exactly the same rate, and so only the curve for ${\tilde u}_{max}^{(1,1)}$ is plotted. It agrees with the linear prediction shown by the black circles until $x \approx 700$, after which a
drastic amplification is observed. Although the other Fourier components are not introduced as initial perturbations, they are excited due to the mutual interaction of the introduced modes. The mean-flow distortion (MFD) (0,0) and the harmonic mode (2,0) are driven by the self-interaction of mode (1,0), and therefore, their growth rates are almost twice that of (1,0)
in most of the computational domain, as confirmed by comparison with the blue crosses. For $x > 700$, the streak component (0,1) and the high-order harmonics (2,1) grow at almost the same rate as (1,1), but their amplitudes differ by a remarkable amount; the amplitude of the streak mode (0,1) is the greatest among the three. When the streak mode (0,1) overwhelms the fundamental mode (1,0) and becomes the dominant perturbation, the calculation blows up, indicating that the transition to turbulence is not far.  

\textcolor{red}{Alternatively, one can trace the evolution of the perturbation energy of each Fourier component, which is defined as \citep{Chu_1965}
\begin{equation}
\begin{split}
   \quad E_{mn}=\int_{0}^{\infty}{\mathord{\buildrel{\lower3pt\hbox{$\scriptscriptstyle\smile$}} 
\over \phi } }_{mn}^{\dag} M{{\mathord{\buildrel{\lower3pt\hbox{$\scriptscriptstyle\smile$}} 
\over \phi } }_{mn}} dy, \quad  M=diag(\frac{T_B}{ \gamma M^2 \rho_B}, \rho_B,\rho_B,\rho_B,\frac{\rho_B}{ \gamma (\gamma - 1) M^{2} T_B } )   ,
\end{split}
\end{equation}
where the superscript $\dag$ denotes the complex conjugate with respect to its argument. The streamwise evolution of $E_{mn}$ for each Fourier component is shown in figure~\ref{fig:Au_evolution_large}-(b), and overall the same feature as in panel (a) is observed. This is quite predictable, because $\tilde u$ and $\tilde T$ are the dominant components in $\tilde \phi$ with similar evolution trend, and the evolution of the perturbation energy should agree overall with that of each dominant component. Figure~\ref{fig:Au_evolution_large}-(c) further plots the evolution of the growth rate of the perturbation energy, defined as
\begin{equation}
\begin{split}
    \sigma_E^{(m,n)} =\frac{1}{2E_{mn}} \frac{ d E_{mn}}{dx}. 
\end{split}
\label{eq:sigma}
\end{equation}
Remarkably, in the interval of $x \in [900,1200]$ as highlighted by the dashed box, the growth rates of (0,1), (1,1) and (2,1) are almost identical, and much greater than that of the fundamental mode in the linear phase. 
}
 This is a representative feature of the FR, as also observed in  \cite{sivasubramanian_fasel_2015}, \cite{chen_zhu_lee_2017} and \cite{hader_fasel_2019}. 
Such a high growth rate was explained by the SIA as introduced in $\S$ \ref{sec:SIA}. {In the SIA, the base flow is regarded as a superposition of the time- and spanwise-averaged mean flow and the quasi-saturated 2D fundamental mode, together with its high-order harmonics,} and the perturbation fields, including the streak mode, the 3D travelling mode and higher-order harmonics with the same spanwise wavenumber, are governed by a linear eigenvalue system (\ref{eq:floquet_eigen}), with the growth rate $\tilde \sigma$ appearing as the eigenvalue.

\textcolor{magenta}{In figure~\ref{fig:Au_evolution_large}-(d), the streamwise evolution of the coefficients of the skin friction 
\begin{equation}
    C_f = ( \frac{{2\underline \mu }}{R}\frac{{\partial \underline {u}}}{{\partial y}})_{y=0},
\end{equation}
are plotted, where $ \underline {u}$ and $\underline \mu$ represents the temporal and spanwise average of the streamwise velocity and the dynamic viscosity. The $C_f$ curve obtained by the NPSE calculation decreases with $x$ gradually at the beginning, agreeing with the unperturbed laminar-flow state, but it starts to deviate from the laminar state at $x \approx 500$, indicating a moderate MFD appearing there. The $C_f$ curve reaches its first peak at $x \approx 1000$, followed by a plateau until $x \approx 1300$, after which it shows another increase until the blowup position. The double-increase phenomenon is typical for the
fundamental resonance regime, as also reported by precious
works \citep{chen_zhu_lee_2017,hader_fasel_2019}. The first increase is associated with the strong MFD induced by the finite-amplitude fundamental mode, and the following plateau agrees with the region of FR. Due to the FR, the streak mode becomes the dominant perturbation in the late phase, which, together with the travelling modes,  may drive another type of secondary instability to support the growth of the high-frequency perturbations. Since these secondary instability modes amplify with high growth rates, which produce sufficient Reynolds stress to cause the rapid distortion of the mean flow, the parabolised assumption in the NPSE approach ceases to be valid, leading to the blowup of the NPSE calculation eventually.}

\begin{figure}
\centering
\begin{overpic}[width=0.6 \textwidth]{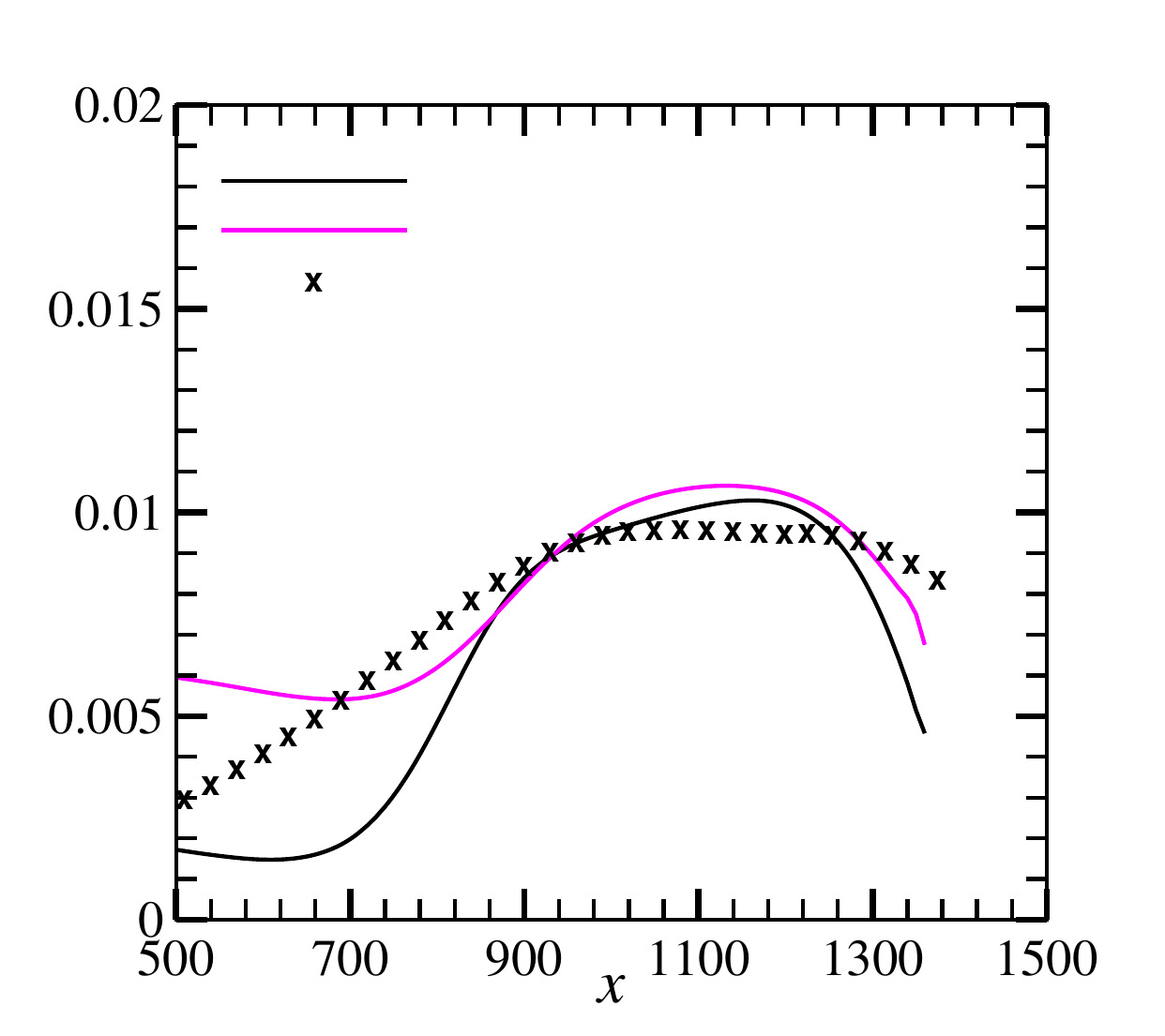}
\put(-1,40){\begin{turn}{90}$\sigma_E$, $\tilde \sigma$ \end{turn}}
\put(39,73){$\sigma_E^{(1,1)}$}
\put(39,67.5){$\sigma_E^{(0,1)}$}
\put(39,63){$\tilde \sigma$}
\end{overpic}
\caption{Comparison of the growth rate $\sigma_E$ obtained by the NPSE calculation with the SIA prediction $\tilde \sigma$.}
\label{fig:gr_compare_SIA_NPSE}
\end{figure}

\begin{figure}
\centering
\begin{overpic}[width=0.96 \textwidth]{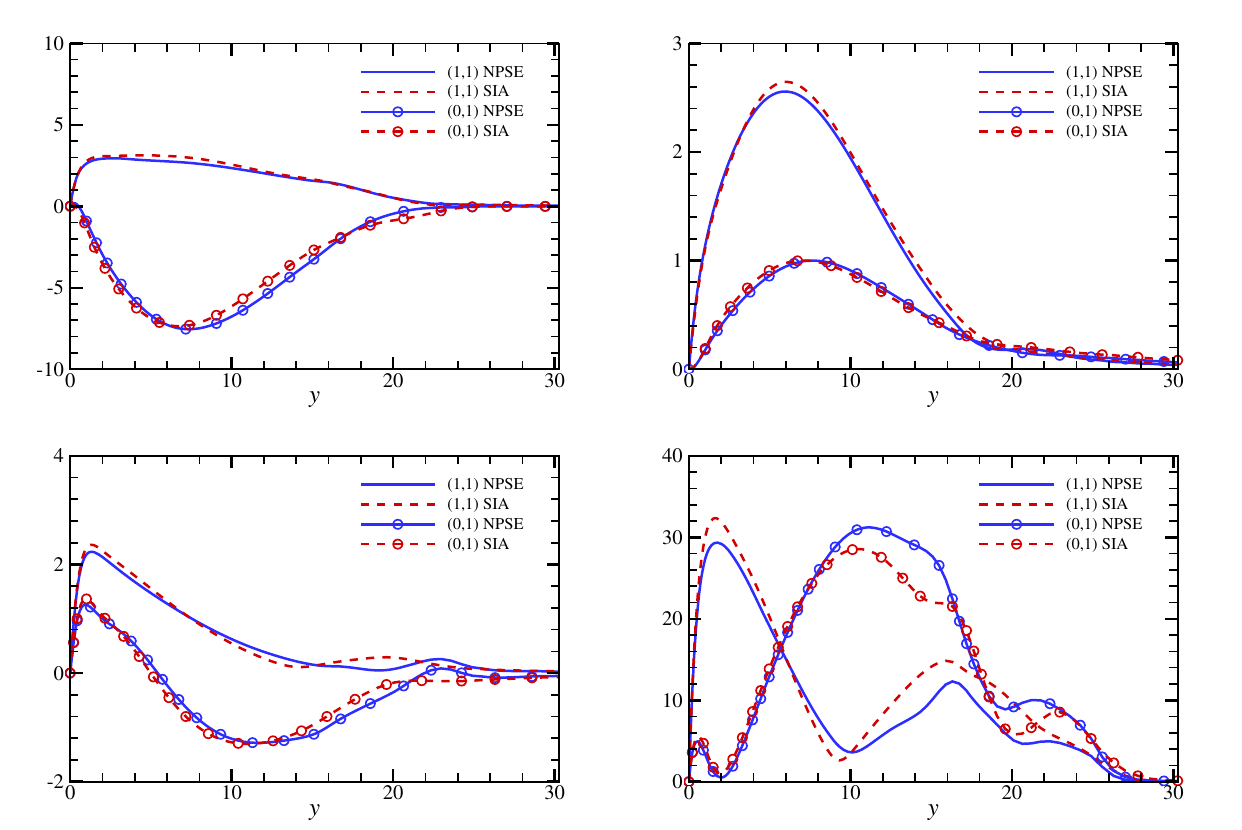}
\put(-1,62){(a)}
\put(48,62){(b)}
\put(-1,29){(c)}
\put(48,29){(d)}
\put(-1,45){\begin{turn}{90} $\hat{u}_{01}$, $|\hat{u}_{11}|$ \end{turn}}
\put(8,40){$\hat{u}_{01}$}
\put(8,51){$|\hat{u}_{11}|$}
\put(48,45){\begin{turn}{90} $\hat{v}_{01}$, $|\hat{v}_{11}|$ \end{turn}}
\put(62,42){$\hat{v}_{01}$}
\put(61,56){$|\hat{v}_{11}|$}
\put(-1,10){\begin{turn}{90} $-\ri \hat{w}_{01}$, $|\hat{w}_{11}|$ \end{turn}}
\put(8,7){$- \ri \hat{w}_{01}$}
\put(8,23){$|\hat{w}_{11}|$}
\put(48,10){\begin{turn}{90} $\hat{T}_{01}$, $|\hat{T}_{11}|$ \end{turn}}
\put(67,26){$\hat{T}_{01}$}
\put(57,26){$|\hat{T}_{11}|$}
\end{overpic}
\caption{ Comparison of the perturbation profiles for (0,1) and (1,1) obtained by the SIA and NPSE approaches for case A at $x=1000$. The profiles are normalised by the maximum of ${\hat v}_{01}$. } 
\label{fig:01_11_SIA_NPSE}
\end{figure}
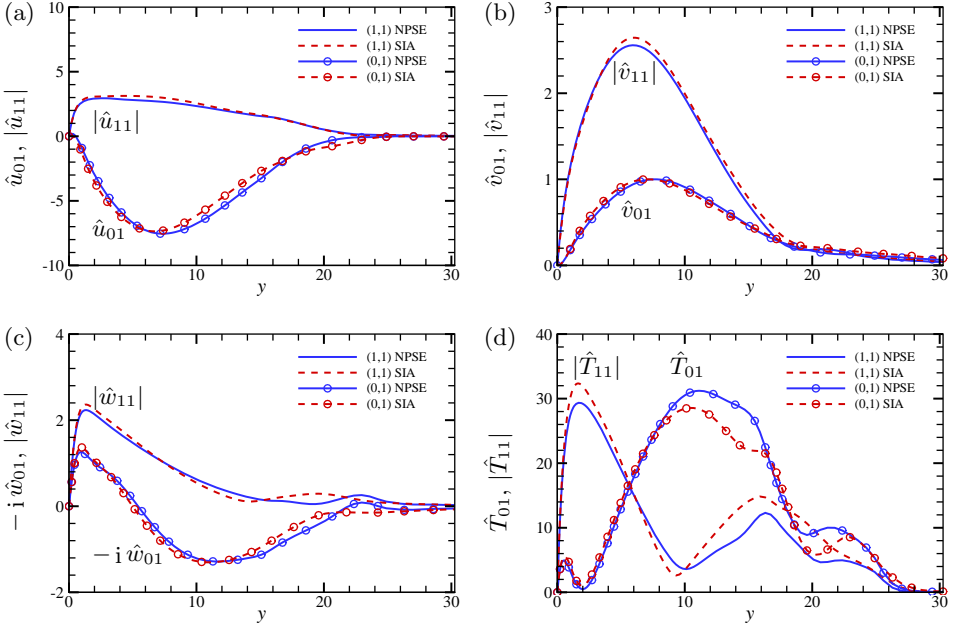

In figure~\ref{fig:gr_compare_SIA_NPSE}, we compare the growth rates of modes (1,1) and (0,1), $\sigma_E^{(1,1)}$ and $\sigma_E^{(0,1)}$ with that predicted by SIA, $\tilde \sigma$. In the interval $x \in [900,1200]$, the three curves agree with each other. The perturbation profiles of (0,1) and (1,1) obtained by the two approaches also agree well, as shown in figure~\ref{fig:01_11_SIA_NPSE}. In fact, $\hat{u}_{01}$, $\hat{v}_{01}$ and $\hat{T}_{01}$ are real and $\hat{w}_{01}$ is pure imaginary, which will be discussed in $\S$ \ref{sec:leading_order}. Observations in figure~\ref{fig:01_11_SIA_NPSE} also indicate that the amplitude of the streamwise velocity of the streak mode $\hat{u}_{01}$ is much greater than that of the 3D traveling mode $\hat{u}_{11}$, agreeing with the amplitude evolution in figure~\ref{fig:Au_evolution_large}-(a). For the profiles of the temperature perturbation, the streak mode in the bulk of the boundary layer $\hat{T}_{01}$ is also much greater than the 3D traveling mode $\hat{T}_{11}$, but it is much weaker in the near-wall region. Although the SIA can predict  quantitatively the growth rates and profiles of both the streak mode and 3D traveling modes in the interval [900,1200], the underlining mechanism determining the dominant role of the streak mode in the bulk region is not obvious. To answer these questions, a more in-depth analysis is required as will be introduced in the following section. Numerical results for case B show the same feature, which will be illustrated in detail in $\S$\ref{sec:compare}.




\section{Asymptotic analysis for the principle of the fundamental resonance}
\label{sec:asymptotic}
\subsection{\label{sec:asy_2D}Flow decomposition and the 2D fundamental mode}
To reveal the principle mechanism of the FR, we perform a weakly nonlinear analysis based on the high-Reynolds-number asymptotic technique.
In the weakly nonlinear phase, the perturbation field $\tilde \phi=(\tilde \rho,\tilde u, \tilde v, \tilde w,\tilde T)$ defined in (\ref{eq:decomposite}) includes a set of harmonic perturbations,
\begin{equation}
\tilde \phi(x,y,z,t)=\bar \epsilon_{00}\tilde\phi_{00}+ \bar \epsilon_{10}\tilde\phi_{10}
+\bar \epsilon_{11}\tilde\phi_{11}+\bar \epsilon_{1-1}\tilde\phi_{1-1}+\bar \epsilon_{01}\tilde\phi_{01}+\text{c.c.} +h.o.t.,
\end{equation}
where $\bar\epsilon_{mn} \ll1$ denotes the amplitude of each component in the nonlinear phase, $h.o.t$ denotes the high-order terms, and the subscripts $00$, 10, 11 ( and 1-1) and 01 denote the MFD, the fundamental mode, the 3D travelling modes and the streak mode, respectively. 
According to the numerical results in $\S$\ref{sec:cal_demon}, we know that the amplitude of the fundamental mode $\bar \epsilon_{10}$ is much greater than those of the 3D travelling modes and the streak mode, namely, 
\begin{equation}
\bar \epsilon_{10}\gg \bar \epsilon_{11},\quad \bar \epsilon_{10}\gg \bar \epsilon_{01}.
\end{equation}

 In the following analysis, we will focus on the region $x>900$, for which the fundamental mode $\tilde \phi_{10}$ evolves either in the linear phase or in the nonlinear phase, but the streak mode and the 3D travelling modes undergo drastic amplification, as shown in figure~\ref{fig:Au_evolution_large}-(a). The base flow for the nonlinear analysis is chosen as the time- and spanwise-averaged mean flow, which includes the Blasius solution and the MFD,
\begin{equation}
(\bar U,\bar T)(x,y)=( U_B, T_B)(x,y)+\bar \epsilon_{00}(\tilde u_{00},\tilde T_{00})(x,y).
\label{eq:baseflow}
\end{equation}
In the early nonlinear phase, the MFD is mainly driven by the fundamental mode, which also acts back on the fundamental mode to leads to its saturation eventually. Since the streamwise length scale of the mean flow is much greater than the Mack-mode wavelength, the non-parallelism of the base flow is negligible in the following analysis.

From the linear stability analysis based on the parallel mean flow (\ref{eq:baseflow}) at a chosen streamwise location $x$, we find that the growth rate of the fundamental Mack mode is much smaller than its wavenumber. We can express the perturbation profiles $\tilde\phi_{10}$ in terms of
\begin{equation}
\tilde\phi_{10}(x,y,t)=\hat \phi_{10}(y;x)\re^{\ri(\int^x_{0}\alpha d  x-\omega_0 t)},
\end{equation}
where the streamwise wavenumber $\alpha$ is almost real and $O(1)$, the frequency $\omega_0$ is also taken to be $O(1)$ and $\hat \phi_{10}$ is the eigenfunction of the fundamental mode.
Asymptotic analyses as in \cite{dong_liu_wu_2020} and \cite{dong_zhao_2021} showed that the Mack mode shows a double-deck structure in the high-$R$ approximation, namely, a main layer where $y=O(1)$ and a viscous Stokes layer where $y=O(R^{-1/2})$.

The eigenfunction of the fundamental mode $\hat{\varphi}_{10} = (\hat{v}_{10},\hat{p}_{10})$ satisfies the Rayleigh equation in the main layer based on the mean flow $(\bar U, \bar T)$,
\begin{equation}
    {\cal L}_{R} \hat{\phi}_{10} \equiv (d_y - {\cal H}_{0} ) \hat{\varphi}_{10}=0 ,
    \label{eq:Rayleigh}
\end{equation}
where
\begin{equation}
  {\cal H}_{0}= \left( \begin{array}{l}
 {{\bar U_{y}}}/({{\bar U - c}})\quad {{ - {M^2}S_0 - {\alpha ^2}\bar T}}/{{{S_0}}}\\
 - {{{S_0}}}/{{\bar T}}\quad \quad \quad \quad 0
\end{array} \right)
\end{equation}
with $S_0 = \ri \alpha (\bar U -c)$ and $c \equiv \omega /\alpha$. The boundary conditions read
\refstepcounter{equation}
$$
\hat v_{10}(0)=0;\quad \hat p_{10}(\infty)\rightarrow 0.
\eqno{(\theequation{a,b})}
$$
Such a linear, homogeneous system leads to an eigenvalue problem. For a spatial mode, the frequency $\omega$ is given to be real, and following the numerical method as in \cite{dong_liu_wu_2020}, \cite{dong_zhao_2021} and \cite{Zhao_Dong_2022}, we can calculate the eigenvalue $\alpha=\alpha_r+\ri \alpha_i$ $(|\alpha_i| \ll |\alpha_r|)$ and the eigenfunctions. If, for a particular $\omega$, the Mack mode is neutral, i.e., $\alpha_i=0$, then the system (\ref{eq:Rayleigh}) becomes singular, and a critical layer around the location where $\bar U=\omega/\alpha$ appears. For a linear critical layer, the leading-order balance in this layer is between the inertial and viscous terms, and in the numerical process, the solution can be obtained by detouring the integrating path around the critical point, as illustrated in Schmid \& Henningson (\citeyear{Schmid2001}, pp.44-45); an example can be found in \cite{dong_liu_wu_2020}. Other perturbation quantities such as $\hat{u}_{10}$ and $\hat{T}_{10}$ can be obtained by
\begin{equation}
    \hat{u}_{10}=-\frac{\bar U_{y} \hat{v}_{10}+ \ri \alpha \hat{p}_{10}}{S_{0}},\quad \hat{T}_{10}=-\frac{\bar T_{y} \hat{v}_{10}}{S_{0}}+(\gamma-1)M^{2}\bar T \hat{p}_{10}.
\end{equation}
Additionally, we take $\text{max}_{y} |\hat{u}_{10}(y)|=1$ for normalisation.


Because the Rayleigh solution does not satisfy the no-slip condition at the wall, a viscous Stokes layer needs to be taken into account, whose solutions can be found in \cite{dong_liu_wu_2020} and so are not presented here.

\subsection{The 3D travelling modes and the streak mode}
It is seen from figure~\ref{fig:Au_evolution_large}-(a) that after the fundamental 2D mode reaches a finite amplitude, both the 3D travelling modes $(1,\pm 1)$ and the streak mode $(0,1)$ amplify with the same rate, which is much greater than that of the fundamental mode, showing a FR phenomenon. To reveal this mechanism, we probe the evolution of the perturbations $\tilde \phi_{1\pm 1}$ and $\tilde \phi_{01}$, which are expressed as
\begin{equation}
\tilde\phi_{11}=\hat\phi_{11}(y)\re^{\ri(\alpha_r x+\beta_0 z-\omega_0 t)+\sigma x},\quad \tilde\phi_{1-1}=\hat\phi_{1-1}(y)\re^{\ri(\alpha_r x-\beta_0 z-\omega_0 t)+\sigma x}
\end{equation}
\begin{equation}
\tilde\phi_{01}=\hat\phi_{01}(y)\re^{\ri\beta_0 z+\sigma x},
\end{equation}
where $\sigma$ denotes their common growth rate. Due to the fundamental resonance, the streamwise wavenumber $\alpha_r$ of modes $(1,\pm 1)$ is the same as that of the fundamental 2D mode, and all these modes grow with the same rate $\sigma$.
Although the growth rate $\sigma$ could be much greater than that of the fundamental mode, its magnitude is still much smaller than unity, i.e., 
\begin{equation}
    |\alpha_i|\ll \sigma\ll 1.
    \end{equation}

\subsubsection{\label{sec:leading_order}Leading-order asymptotic solutions in the main layer}
In the main layer where $y\sim 1$, balancing the governing equations and taking into account the scalings of $\alpha$, $\beta$ and $\omega$, we obtain that all the quantities of the 3D travelling mode, $\hat {\textbf{u}}_{1 \pm 1}$, $\hat \rho_{1 \pm 1}$, $\hat T_{1 \pm 1}$, $\hat p_{1 \pm 1}$ are of the same order, but those of the streak mode are not.
The streamwise wavenumber of the streak mode is zero, and its streamwise growth rate $\sigma$ is small. Thus, the streamwise derivative of $\hat \phi_{01}$ with respect to $x$ is only $O(\sigma\hat \phi_{01})$, much smaller than its derivatives with respect to $y$ and $z$. Then, from the balance of the continuity equation, we obtain that  the streamwise perturbation $\hat u_{01}$ must be much greater than $\hat v_{01}$ and $\hat w_{01}$. For convenience, we let $\hat u_{01}\sim 1$ (because its magnitude is measured by $\bar\epsilon_{01}$), then balance of the continuity equation leads to
\begin{equation}
\hat v_{01}\sim\hat w_{01}\sim \sigma.
\label{eq:v01_w01}
\end{equation}
$\tilde \phi_{11}$ is driven by the nonlinear interaction of $\tilde \phi_{01}$ and $\tilde \phi_{10}$, therefore, we have
\begin{equation}
\bar \epsilon_{11}=\bar \epsilon_{10} \bar \epsilon_{01}.
\label{eq:epsilon_11}
\end{equation}
In the spanwise momentum equation of the streak mode, the inertial term $uw_x\sim \bar \epsilon_{01}\sigma\bar U \hat w_{01}$, the pressure-gradient term $p_z\sim \bar \epsilon_{01}\ri\beta\hat p_{01}$, and the nonlinear terms  $O(\bar \epsilon_{10} \bar \epsilon_{11})$ should balance, {\color{red}namely,
\begin{equation}
 \bar\epsilon_{01}\sigma^2\sim\bar\epsilon_{01}\hat p_{01}\sim \bar\epsilon_{10}\bar\epsilon_{11}, 
\end{equation}}
leading to
\refstepcounter{equation}
$$
\sigma\sim \bar \epsilon_{10},\quad \hat p_{01}\sim \sigma^{2}\sim \bar \epsilon_{10}^2.
\label{eq:relation_sigma_delta}
\eqno{(\theequation{a,b})}$$
{\color{red}In the balance of the streamwise momentum equation, the nonlinear terms $O(\bar\epsilon_{10}\bar\epsilon_{11})$ are much smaller than the inertia term $O(\bar\epsilon_{01}\sigma)$, and so they do not appear in the leading-order balance.}
For convenience, we introduce
\begin{equation}
\bar \sigma=\bar \epsilon_{10}^{ -1}{\sigma}=O(1).
\end{equation}
\textcolor{red}{It should be noted that one may be dissatisfied with (\ref{eq:relation_sigma_delta}$a$) if the NPSE results are re-examined, because figure~\ref{fig:Au_evolution_large}-(a) shows that in the FR region, $\bar\epsilon_{10} \sim 0.1$,  but figure \ref{fig:gr_compare_SIA_NPSE} indicates that $\sigma \sim 0.01$. Actually, the scaling relation (\ref{eq:relation_sigma_delta} $a$) is from \emph{a priori} analysis from the properties of the governing equations, instead of an observation from the numerical results at a finite Reynolds number. The purpose to prescribe this scaling relation is to construct a consistent mathematical description to explain the physics of FR, which should agree with the reality as long as the Reynolds number is sufficiently high. Therefore, the ratio of $\sigma$ and $\bar\epsilon_{01}$ is expected to approach $O(1)$ as $R$ increases, which will be displayed in figure \ref{fig:large_R_validate_juere_gr}.}

Likewise, balance of the energy equation and the equation of the state for $\hat \phi_{01}$, we obtain that $\hat T_{01}\sim\hat \rho_{01}\sim 1$.
Thus, we introduce
\begin{equation}
(\hat u_{01},\hat v_{01},\hat w_{01},\hat p_{01},\hat \rho_{01},\hat T_{01})=(\breve u_{01},{\bar \epsilon_{10}}\breve v_{01},\bar \epsilon_{10}\breve w_{01},\bar \epsilon_{10}^{2}\breve p_{01},\breve \rho_{01},\breve T_{01})+\cdots.
\label{eq:main_layer_expansion}
\end{equation}
The physical quantities for the 3D travelling waves $\hat\phi_{1\pm 1}$ are all $O(1)$. For convenience, both $\hat{\phi}_{1+1}$ and $\hat{\phi}_{11}$ are used to denote the 3D travelling mode (1,1) in this paper.

Collecting the leading-order terms from the governing equations of the 3D travelling modes $\tilde \phi_{1 \pm 1}$, we obtain 
 \begin{subequations}
\begin{align}
&\hat{S}_0 M^2\hat p_{1 \pm 1}+\ri\alpha_r \hat u_{1 \pm 1}+\hat v_{1 \pm 1}' \pm \ri\beta \hat w_{1 \pm 1}=F_{1 \pm 1,1}\equiv \nonumber \\ 
&- {M^2}\ri\alpha_r {\hat p_{10}} {\breve u}_{01} + \left( {\hat{S}_0{{\hat \rho }_{10}} - \frac{{\bar T_{y}}}{{{{\bar T}^2}}}{{\hat v}_{10}}} \right) {\breve T}_{01} - \bar T\left( {\ri\alpha_r {{\hat u}_{10}} + {{\hat v'}_{10}}} \right) {\breve \rho}_{01} ,\\
&\hat{S}_0 \hat u_{1 \pm 1}+\bar U_{y}\hat v_{1 \pm 1}+\ri\alpha_r \bar T\hat p_{1 \pm 1}=F_{1 \pm 1,2} \equiv -\hat v_{10}\breve u_{01}'-\ri\alpha_r \hat u_{10}\breve u_{01}+\ri\alpha_r\bar T^2\hat p_{10}\breve \rho_{01},\\
&\hat{S}_0 \hat v_{1 \pm 1}+\bar T\hat p_{1 \pm 1}'=F_{1 \pm 1,3}\equiv-\ri\alpha_r \hat v_{10}\hat u_{01}-\hat{S}_0 \bar T\hat v_{10}\hat \rho_{01},\\
&\hat{S}_0 \hat w_{1 \pm 1} \pm \ri\beta\bar T\hat p_{1 \pm 1}=F_{1 \pm 1,4}\equiv 0,\\
&\hat{S}_0 {{\hat T }_{1 \pm 1}} + \bar T_{y}{{\hat v}_{1 \pm 1}} - \left( {\gamma  - 1} \right){M^2}\bar T \hat{S}_0 {{\hat p}_{1 \pm 1}} =F_{1 \pm 1,5}\equiv  \nonumber \\
 &- {{\hat v}_{10}} {\breve T '}_{01}  - \ri\alpha_r \left[ {{{\hat T }_{10}} - \left( {\gamma  - 1} \right){M^2}\bar T{{\hat p}_{10}}} \right] {\breve u}_{01} - \bar T\left( {\bar T_{y}{{\hat v}_{10}}-\hat{S}_0 {{\hat T }_{10}} } \right) \breve \rho_{01},\\
 & - \frac{{{{\hat T }_{1 \pm 1}}}}{{\bar T}} + \gamma {M^2}{\hat p_{1 \pm 1}} - \bar T{\hat \rho _{1 \pm 1}} = F_{1 \pm 1,6}\equiv  {\hat \rho _{10}} {\breve T}_{01} + {\hat T _{10}} {\breve \rho}_{01}.
\end{align}
\label{eq:11}
 \end{subequations}
where $\hat{S}_0 = -\ri\omega+\ri\alpha_r \bar U $ and in what follows, the prime denotes the derivative with respect to its argument. The equation (\ref{eq:11} $a$) is obtained by eliminating $\hat \rho_{1\pm 1}$ and $\hat T_{1\pm 1}$ from the continuity equation, the energy equation and the equation of the state. It is seen that a critical layer appears at a position where $\bar U=\omega/\alpha_r$, and we use the standard treatment for the linear critical layer  as in Schmid \& Henningson (\citeyear{Schmid2001}, pp.44-45) to avoid the singularity. 

If the profiles of the streak mode $\breve\phi_{01}$ are known, then equations (\ref{eq:11}) form an inhomogeneous linear system. However, they are coupled with the unknown vector of the streak mode $\breve\phi_{01}$, rendering a triad resonance system.

The leading-order governing equations for the streak mode are
  \refstepcounter{equation}
  \label{eq:01}
$$
\bar\sigma\breve u_{01}+\breve v_{01}'+\ri\beta\breve w_{01}=0,
\eqno{(\theequation{a})}
$$
$$
\bar\sigma\bar U\breve u_{01}+\bar U_{y}\breve v_{01}=0,
\eqno{(\theequation{b})}$$
$$
\bar\sigma\bar U\breve v_{01}+\bar T\breve p_{01}'= F_{01,3}\equiv -\hat v_{10}^\dag (\hat v_{11}'-\ri\alpha_r\hat u_{11}-\bar T \hat{S}_{0} \hat{ \rho}_{11})-\Big(2(\hat v_{10}^\dag)'-\frac{\bar T_{y}}{\bar T}\hat v_{10}^\dag\Big)\hat v_{11}+c.c.,
\label{eq:v01}
\eqno{(\theequation{c})}$$
$$
\bar\sigma\bar U\breve w_{01}+\ri\beta\bar T\breve p_{01}= F_{01,4}\equiv-(\hat v_{10}^\dag\hat w_{11})'+\frac{\bar T_{y}}{\bar T}\hat v_{10}^\dag\hat w_{11}-c.c.,\label{eq:w01}
\eqno{(\theequation{d})}$$
$$
\bar\sigma\bar U\breve T_{01}+\bar T_{y}\breve v_{01}=0,
\quad
\breve T_{01}+\bar T^2\breve \rho_{01}=0.
\eqno{(\theequation {e,f})}$$
 From the symmetric feature of these equations, we know that $\breve u_{01}$, $\breve v_{01}$, $\breve T_{01}$ and $\breve p_{01}$ are real and $\breve w_{01}$ is pure imaginary. However, the detouring method near the critical layer is used and thus breaks the symmetric feature slightly. But this effect is very weak, and therefore, to the leading-order approximation, we also consider $\breve u_{01}$, $\breve v_{01}$, $\breve T_{01}$ and $\breve p_{01}$ to be real and $\breve w_{01}$ to be pure imaginary.

{\color{red}Remarkably,} it is seen from (\ref{eq:v01}) that the transverse and lateral velocities of the streak mode are forced by the nonlinear interaction of the 2D fundamental and 3D travelling modes, while its streamwise velocity with a greater amplitude is driven by the linear lift-up mechanism. Such a mechanism is a reminiscence of the stronger amplification of the streak mode in the oblique breakdown regime \cite{Song_2022}.

The attenuation condition is imposed in the far field, and the analysis of the wall layer as will be shown in $\S$\ref{sec:wall_layer} implies that
 the non-penetration condition is imposed at the lower boundary,
\refstepcounter{equation}
$$
( \hat p_{11},\hat p_{1-1},\breve p_{01})\rightarrow 0\quad\mbox{as } y\rightarrow \infty,
\label{eq:eigen_BC}
\eqno{(\theequation{a,b,c})}$$
$$
\hat v_{11}(0)=\hat v_{1-1}(0)=\breve v_{01}(0)=0.
\label{eq:zero_condition}
\eqno{(\theequation{d,e,f})}$$
Combining (\ref{eq:11}) and (\ref{eq:01}), we obtain a six-order linear differential system,
\begin{equation}
\frac{d \hat\varphi}{dy }=\mathbf A\hat\varphi,
\label{eq:eigenvalue}
\end{equation}
where $\hat\varphi=(\hat v_{11},\hat p_{11},\hat v_{1-1},\hat p_{1-1},\breve v_{01},\breve p_{01})$ and $\mathbf A$ can be deduced readily from the equations (\ref{eq:11}) and (\ref{eq:01}). Such a homogeneous system with the homogeneous boundary conditions forms an eigenvalue system, with $\bar \sigma$ being the eigenvalue. The numerical approaches as in \cite{Malik1990} can be employed to solve this system.

Now we are interested in  the near-wall behaviours of the perturbation fields for components (0,1) and (1,1). Although $\hat v_{11}$ and $\breve v_{01}$ satisfy the non-penetration condition at the wall, the perturbations of the streamwise velocity, spanwise velocity and temperature may be finite or even blow up, \textcolor{red}{disagreeing} with the no-slip and isothermal boundary conditions. From a scaling estimate, we find that as $y\to 0$,
\begin{equation}
\breve v_{01}\sim y\ln y,\quad \Big(\breve u_{01},\breve T_{01}\Big)\sim \ln y,\quad \breve w_{01}\sim 1,\quad \breve p_{01}\sim 1,
\end{equation}
\begin{equation}
\hat v_{11}\sim y\ln y,\quad\Big(\hat u_{11},\hat T_{11}\Big)\sim \ln y, \quad\hat w_{11}\sim 1,\quad \hat p_{11}\sim 1.
\end{equation}

In fact, the inhomogeneous forcing terms in (\ref{eq:v01} $d$) in the vicinity of the wall are expanded as
\begin{equation}
 F_{01,4}=\hat A_2+\hat A_3y+\cdots,
\end{equation}
where
\begin{equation}
\hat A_2=-2\ri\Im[\lambda_v^\dag\hat w_{11}(0)],\quad \hat A_3=-2\ri\Im\Big[2\lambda_v^\dag\hat w_{11}'(0)+\Big(\bar\lambda_v^\dag-\frac{\lambda_T\lambda_v^\dag}{T_w}\Big)\hat w_{11}(0)\Big],
\end{equation}
\begin{equation}
\hat w_{11}(0)=\frac{\beta T_w}{\omega}\hat p_{11}(0),\quad \hat w_{11}'(0)=\Big[\frac{\alpha \beta\lambda T_w}{\omega^2}+\frac{\beta \lambda_T}{\omega}\Big]\hat p_{11}(0),
\end{equation}
with $\lambda_v=\hat v_{10,y}(0)$, $\bar\lambda_v=\hat v_{10,yy}(0)$, $\lambda=\bar U_{y}(0)$ and $\lambda_{T}=\bar T_{y}(0)$. Thus, applying $y\rightarrow 0$ to (\ref{eq:v01}), we have
\refstepcounter{equation}
$$
\breve v_{01}'=\frac{\breve v_{01}}{y}+\frac{\ri\beta}{\bar\sigma \lambda y}\Big[\ri\beta (T_w+\lambda_Ty)\breve p_{01}-\hat A_2\Big]-\frac{\ri\beta\hat A_3}{\bar\sigma \lambda}+\cdots,
\eqno{(\theequation{a})}$$
$$
\breve p'_{01}=-\frac{\bar\sigma\lambda y}{T_w}\breve v_{01}+\cdots.\eqno{(\theequation{b})}
$$
The solutions of the transverse velocity and pressure of the streak mode are
\begin{equation}
\breve v_{01}\to\frac{\ri\beta }{\bar\sigma\lambda}\Big(-\hat A_3+\frac{\lambda_T\hat A_2}{T_w}\Big)y\ln y+\cdots,
\quad
\breve p_{01}\rightarrow \frac{\hat A_2}{\ri\beta T_w}+\cdots,
\label{eq:main_v01}
\end{equation}
and the other perturbations behave like
\begin{equation}
(\breve u_{01},\breve T_{01})\to\Big(1,\frac{\lambda_T}{\lambda}\Big)\frac{\ri\beta }{\bar\sigma^2\lambda}\Big(\hat A_3-\frac{\lambda_T\hat A_2}{T_w}\Big)\ln y+\cdots,\quad
\breve w_{01}\to \frac{1}{\lambda\bar\sigma}\Big(\hat A_3-\frac{\lambda_T\hat A_2}{T_w}\Big) +\cdots.
\label{eq:streak_main_asymp}
\end{equation}
Obviously, $ \breve u_{01}$ and $\breve T_{01}$ are unbounded at the wall, which requires consideration of the viscosity in a thin layer, referred to as the viscous wall layer.

\subsection{Viscous wall layer}
\label{sec:wall_layer}
Since the streak mode is singular at the wall, a viscous wall layer has to be taken into account. Balance of the inertial and viscous terms, we obtain the thickness of the wall layer, $y\sim (\bar \epsilon_{10}R)^{-1/3}$.
For convenience, we introduce a local coordinate
\begin{equation}
Y=\epsilon^{-1} y=O(1),\quad \epsilon\equiv (\bar \epsilon_{10}R)^{-1/3}.
\label{eq:wall_Y}
\end{equation}

From the main-layer estimate (\ref{eq:streak_main_asymp}), we obtain the magnitude of the streak mode in the wall layer,
\begin{equation}
(\breve u_{01},\breve v_{01},\hat p_{01},\breve T_{01},\breve \rho_{01})\sim (\ln \epsilon,\epsilon \ln \epsilon, 1,\ln\epsilon,\ln\epsilon).
\end{equation}
However, to satisfy the wall-layer governing equations, the perturbation spanwise velocity must come to the leading-order continuity equation. Thus, we let
\begin{equation}
\breve w_{01}\sim\ln \epsilon,
\end{equation}
which decays algebraically as $Y\to \infty$. In the spanwise momentum equation, the leading-order balance is between the pressure gradient and the inhomogeneous forcing, and the inertial term $\sigma\bar U\hat w_{01}$ appears only in the second order, which \textcolor{red}{must} balance with the second-order pressure gradient $\ri \beta\hat p_{01}^{(2)}$, where $\hat p_{01}^{(2)}$ denotes the second-order pressure perturbation. This balance leads to $\hat p_{01}^{(2)}\sim \bar\epsilon_{10}^2\epsilon\ln\epsilon$, and as $Y\to \infty$, the spanwise velocity perturbation $\hat w_{01}\to \bar\epsilon_{10}\ln \epsilon Y^{-1}$, or $\breve w_{01}\to \ln \epsilon Y^{-1}$.

Therefore, the streak-mode perturbation field is expanded as
\refstepcounter{equation}
$$
\Big(\breve u_{01},\breve v_{01},\breve w_{01},\breve T_{01}\Big)=\ln\epsilon\Big(\breve U_0,\epsilon\breve V_0,\breve W_0,\breve T_0\Big){\color{red}+\Big(\breve U_1,\epsilon\breve V_1,\breve W_1,\breve T_1\Big)}+\cdots,
\eqno{(\theequation{a})}\label{eq:wall_u01}$$
$$
\breve p_{01}=\breve P_0+\epsilon\ln\epsilon \breve P_1+\cdots,\eqno{(\theequation{b})}
$$
where $\breve P_0=\hat A_2/(\ri\beta T_w)$. The presence of $\epsilon\ln\epsilon \breve P_1$ and $\epsilon\breve V_1$ induces an $O(\epsilon\ln\epsilon)$ correction to the main-layer solution.

The perturbation quantities of the corresponding
  3D travelling mode in the wall layer are expanded as
\begin{equation}
\Big(\breve u_{11},\breve v_{11},\breve w_{11},\breve p_{11},\breve T_{11},\breve \rho_{11}\Big)=\Big(\ln \epsilon\hat U_{0},\epsilon\ln \epsilon\hat V_{0},\hat W_{0},\hat P_{0},\ln \epsilon\hat T_{0},\ln \epsilon\hat R_{0}\Big)+\cdots.
\end{equation}

The leading-order governing equations for the streak mode read
  \refstepcounter{equation}
$$
\bar\sigma\breve U_0+\breve V_0'+\ri\beta \breve W_0=0,
\label{eq:Wall_EQ1}
\eqno{(\theequation{a})}
$$
$$
\lambda\bar\sigma Y\breve U_0+\lambda \breve V_0-C_w\breve U_0''=0,\quad \breve P_{1}'=0,
\label{eq:Wall_EQ2}
\eqno{(\theequation{b,c})}
$$
$$\ri\beta T_w\breve P_0=\hat A_2,\quad
\lambda\bar\sigma Y\breve W_0-C_w\breve W_0''+\ri\beta T_w \breve P_1=0,
\label{eq:Wall_EQ4}\eqno{(\theequation{d,e})}
$$
$$
\lambda \bar \sigma Y \breve T_0+\lambda_T\breve V_0-\frac{C_w}{Pr}\breve T_0''=0,
\eqno{(\theequation{f})}
$$
where  $C_w=\mu_wT_w$.
This is a linear homogeneous system, in which the nonlinear interaction of the fundamental mode and the 3D travelling mode does not appear in the leading-order balance. Because the viscosity appears in the leading order of the streak-mode equations, the no-slip conditions are satisfied at the lower boundary,
\begin{equation}
\breve U_0(0)=\breve V_0(0)=\breve W_0(0)=\breve T_0(0)=0.
\end{equation}
In the upper limit, the perturbation field must match the main-layer solutions, namely,
\begin{equation}
(\breve U_0,\breve V_0,\breve W_0,\breve T_0)\rightarrow \Big(1,-\bar\sigma Y,O(1/Y),\frac{\lambda_T}{\lambda} \Big)\frac{\ri\beta}{\lambda\bar\sigma^2}\Big(\hat A_3-\frac{\lambda_T\hat A_2}{T_w}\Big)\quad\mbox{as }Y\rightarrow \infty.
\label{eq:V0_matching}
\end{equation}

From (\ref{eq:Wall_EQ2} $d,e$), the matching condition  and the no-slip condition ($\breve W_0(0)=0$),  we find that
\begin{equation}
\breve P_0=\frac{\hat A_2}{\ri\beta T_w},\quad \breve W_0=-\frac{\ri \pi\beta T_w\breve P_1}{C_w^{1/3}(\lambda\bar\sigma)^{2/3}}\Big(\mbox{Gi}(\eta)-\frac{\mbox{Gi}(0)}{\mbox{Ai}(0)}\mbox{Ai}(\eta)\Big),
\end{equation}
where $\eta=(\lambda\bar\sigma/C_w)^{1/3}Y$, $\mbox{Ai}$ and $\mbox{Gi}$ are the Airy's functions of the first kind and the Scorer’s function respectively; see Abramowitz \& Stegun (\citeyear{Handbook1964}, p.448).

Equating (\ref{eq:Wall_EQ1} $a,b,e$) and eliminating $\breve U_0$, $\breve W_0$ and $\breve P_0$,  we obtain
\begin{equation}
C_w\breve V_0^{(4)}-\lambda\bar\sigma Y\breve V_0''=0,
\end{equation}
whose solution reads
\begin{equation}
\breve V_0=\bar C\int_0^{\eta}\int_0^{\bar\eta}\mbox{Ai}(\hat\eta)d\hat\eta d\bar\eta,
\label{eq:V0_solution}
\end{equation}
where $\bar C$ is a constant to be determined later.
The upper limit of $\breve V_0$ is
\begin{equation}
\breve V_0\to \Big[\frac{1}{3}(\lambda\bar\sigma/C_w)^{1/3}Y-0.2588\Big]\bar C\quad\mbox{as }Y\rightarrow \infty.
\label{eq:wall_01_matching}
\end{equation}
Comparing with the matching condition (\ref{eq:V0_matching}), we obtain
\begin{equation}
{\bar C}=-\frac{3\ri\beta C_w^{1/3}}{(\lambda\bar\sigma)^{4/3}}\Big(\hat A_3-\frac{\lambda_T\hat A_2}{T_w}\Big).
\end{equation}
The implication is that in the main layer, the second-order perturbation is driven by an outflux $\epsilon\ln \epsilon V_\infty$ with
\begin{equation}
V_\infty=-0.2588\bar C=\frac{0.7764\ri\beta C_w^{1/3}}{(\lambda\bar\sigma)^{4/3}}\Big(\hat A_3-\frac{\lambda_T\hat A_2}{T_w}\Big).
\label{eq:V_infty}
\end{equation}
 Thus, to predict the growth rate $\bar\sigma$ more accurately, an improved boundary condition will be introduced in $\S$ \ref{sec:improved}.

Substituting  (\ref{eq:V0_solution}) into the continuity equation (\ref{eq:Wall_EQ1} $a$), we obtain
\begin{equation}
\breve U_0=-\frac{1}{\bar\sigma}\Big((\lambda\bar\sigma/C_w)^{1/3}\bar C\int_0^\eta\mbox{Ai}(\bar\eta)d\bar\eta+\ri\beta\breve W_0\Big).
\end{equation}
The energy equation (\ref{eq:Wall_EQ1} $e$) leads to
\begin{equation}
\breve T_0=\frac{Pr^{1/3}\lambda_T\pi}{C_w^{1/3}\lambda\bar\sigma}
\Big[-\mbox{Ai}\int_0^{\xi}\breve V_0\mbox{Bi}( \bar \xi)d \bar \xi+\mbox{Bi}\int_0^\xi \breve V_0\mbox{Ai}(\xi)d\xi\Big],
\end{equation}
where $\xi=(\lambda\bar\sigma Pr/C_w)^{1/3} Y$.
Applying (\ref{eq:Wall_EQ2} $b$) at $Y=0$, we obtain $\breve U_0''(0)=0$, which leads to
\begin{equation}
\breve P_1=-\frac{3\ri \mbox{Ai}'(0)C_w^{1/3}}{T_w\beta(\lambda\bar\sigma)^{1/3}}\Big(\hat A_3-\frac{\lambda_T\hat A_2}{T_w}\Big).
\end{equation}

The governing equations for the 3D travelling mode are

 \begin{subequations}
\begin{align}
\ri\alpha\hat U_{0}+\hat V_{0}'=-\ri\alpha M^2\hat p_{w0}\breve U_{0}-\ri\omega \hat\rho_{w0}\breve T_0&+(\ri\alpha \hat u_{w0}+\lambda_v)T_w^{-1}\breve T_0
\label{eq:EOC_1},\\
-\ri\omega\hat U_{0}=-\ri\alpha\hat u_{w0}\breve U_{0}-\lambda_v Y\breve U_0'&-\ri\omega T_w^{-1}\hat u_{w0}\breve T_0,
\label{eq:eq:EOC_2}\\
\ri\omega \hat W_0+\ri\beta T_w\hat P_0&=0, \label{eq:eq:EOC_3}\\
 -\ri\omega \hat T_{0}=-\ri\alpha [\hat \theta_{w0}-(\gamma-1)M^2T_w\hat p_{w0}]\breve U_0&-\lambda_vY\breve T_0'-\ri\omega T_w^{-1}\hat \theta_{w0}\breve T_0,
\label{eq:EOE_1}
  \end{align}
\label{eq:EO}
 \end{subequations} 
where $(\hat p_{w0},\hat u_{w0},\hat \theta_{w0},\hat \rho_{w0})=\Big( \hat{p}_{10}(0),\hat u_{10}(0),\hat T_{10}(0),\hat \rho_{10}(0)\Big)$. It is seen that $\breve P_0$, $\breve V_0$, $\breve W_0$ and $\breve P_1$ do not appear in the leading-order balance. From the Rayleigh equation (\ref{eq:Rayleigh}) we know that
\begin{equation}
\Big(\hat u_{w0},\lambda_v,\hat \rho_{w0},\hat \theta_{w0}\Big)=\Big(\frac{\alpha T_w}{\omega},(\ri\omega M^2+\frac{\alpha^2T_w}{\ri\omega}),\frac{M^2}{T_w},(\gamma-1)M^2T_w\Big) \hat{p}_{w0}.
\end{equation}

The viscous effect of the 3D-travelling-mode is secondary in the wall layer, and so only the non-penetration condition is satisfied at the lower boundary,
\begin{equation}
\hat V_0(0)=0.
\end{equation}
Solving the above system, we obtain the solutions for the streamwise velocity and temperature of the 3D travelling mode,
\begin{equation}
\hat U_{0}=\Big[ \frac{\alpha^2 T_w}{\omega^2}\breve U_{0}+\Big( M^2-\frac{\alpha^2T_w}{\omega^2}\Big) Y\breve U_0'+\frac{\alpha}{\omega}\breve T_0 \Big] \hat{p}_{w0},
\end{equation}
\begin{equation}
\hat T_{0}=\Big[ \Big(M^2-\frac{\alpha^2T_w}{\omega^2}\Big)Y\breve T_0'+ (\gamma-1)M^2\breve T_0 \Big] \hat{p}_{w0}.
\label{eq:T0}
\end{equation}
It is seen that the no-slip and isothermal conditions are satisfied automatically, i.e., $\hat U_0(0)=\hat T_0(0)=0$. Integrating (\ref{eq:EOC_1}), we obtain the solution for the transverse velocity,
\begin{equation}
\hat V_0=\int_0^Y -\ri \alpha \hat{p}_{w0}  \Big[ (M^2 +\frac{\alpha^2}{\omega^2} )\breve U_0 +(M^2 -\frac{\alpha^2}{\omega^2} ) \bar Y \breve U_0' +\frac{\alpha}{\omega} \breve T_0      \Big] d \bar Y
\end{equation}
In the upper limit, we can estimate, by dropping the $O(Y)$ unbounded part, the outflux to the main layer,
\begin{equation}
V_{1\infty}\equiv \frac{\alpha \beta \hat{p}_{w0} C_{w}^{1/3} }{{\bar \sigma}^{7/3} \lambda^{4/3} } \Big[ C_1 (M^2+\frac{\alpha^2}{\omega^2} )+ C_2 (M^2-\frac{\alpha^2}{\omega^2} ) +C_3 \frac{\lambda_T \alpha }{\lambda \omega}     \Big]\Big(\hat A_3-\frac{\lambda_T\hat A_2}{T_w}\Big).
\label{eq:V1_infty}\end{equation}
We obtain from numerical calculations that $C_1=-0.8869$, $C_2=0.1139$ and $C_3=-1.0221$.

Interestingly, from (\ref{eq:T0}) we find that the second term of $\hat T_{0}$ appears as $(\gamma-1)M^{2}\breve T_{0}$. Although we take $M=O(1)$ in this paper, the factor $M^2$ may be numerically large for a hypersonic case. Thus, for a finite-$R$ case, $\bar \epsilon_{11} \hat T_{0}$ could be greater than $\bar \epsilon_{01} \breve T_{0}$ in the near-wall region, as observed in figure~\ref{fig:01_11_SIA_NPSE}.

If we go to the second order, the perturbation of the streamwise velocity and temperature of the 3D travelling waves would also approach constants, which requires a Stokes layer to satisfy the no-slip condition. 
Balance of the unsteady and viscous terms, we obtain that the thickness of the Stokes layer is $y\sim R^{-1/2}$ for $\omega=O(1)$. Since they do not affect the leading-order balance, the Stokes-layer analysis is omitted in this paper; the detailed Stokes-layer solution can be found  in \cite{dong_liu_wu_2020}.

{\color{red}There is another issue that we would like to emphasize. From (\ref{eq:main_v01}) we know that \begin{equation}
 \breve v_{01}\sim \hat C y\ln y \quad\mbox{as }y\to 0,  
\end{equation}
where $\hat C$ is a constant. Under the wall-layer coordinate, this asymptotic behaviour is translated to \begin{equation}
\breve v_{01}\sim \epsilon \ln \epsilon Y\hat C+\epsilon \hat CY \ln Y.   
\label{eq:matching_01}
\end{equation} 
In the wall layer where $Y=O(1)$, the two terms on the right-hand side separate into two different scales. The wall-layer expansion (\ref{eq:wall_u01}) indicates that the above scale separation leads to two orders of solutions $(\breve U_1,\breve V_1,\breve W_1,\breve T_1)$ in the wall layer. The leading-order solution (\ref{eq:wall_01_matching}) matches  the leading-order term in (\ref{eq:matching_01}), where the $O(1)$ part of (\ref{eq:wall_01_matching}) induces an outflux to the main layer. In order to avoid lengthy mathematical argument, we do not show the second-order solution in the wall layer. Although ignoring the second-order solution of the wall layer will prevent us from constructing a consistent composite solution for the whole boundary layer, the leading-order solution is sufficient to derive a viscous correction to the main-layer solution, as will be demonstrated in the next subsection. Therefore, the numerical justification of the perturbation profiles, as will be displayed in figures~\ref{fig:egf_compare_2000-10000} and  \ref{fig:egf_compare_2000-10000_dengwen}, will be only focus on the comparison of the asymptotic predictions and the NPSE calculations in the main layer.}

\subsection{Improved asymptotic theory}
\label{sec:improved}
Indeed, the second-order terms in the main-layer expansion (\ref{eq:main_layer_expansion}) should be of $O(\epsilon\ln \epsilon)$, which is driven by the wall-layer outflux $\epsilon\ln \epsilon V_\infty$ and $\epsilon \ln \epsilon V_{1\infty}$. Since the value of $\epsilon \ln \epsilon$ is usually not quite small, neglecting it may lead to a quantitatively large error. As demonstrated by \cite{dong_liu_wu_2020}, if the corrections from the lower boundary is taken into account, the accuracy of the main-layer solutions could be improved significantly. 
Thus, the improved boundary conditions for (\ref{eq:zero_condition} $d, e, f$) are derived,
\begin{equation}
\hat v_{01}(0)=\epsilon\ln\epsilon V_{\infty},\quad \hat v_{11}(0)=\hat v_{1-1}(0)=\epsilon \ln \epsilon V_{1\infty},
\label{eq:improved_BC}
\end{equation}
where $V_\infty$ and $V_{1\infty}$ were defined in (\ref{eq:V_infty}) and (\ref{eq:V1_infty}), respectively.
Since $\hat A_3$ and $\hat A_2$ in $V_\infty$ and $V_{1\infty}$ are functions of $\hat p_{11}(0)$, the boundary condition (\ref{eq:improved_BC}) is homogeneous. Now the improved strategy is to solve the eigenvalue system (\ref{eq:eigenvalue}) with boundary conditions (\ref{eq:eigen_BC} $a, b, c$) and (\ref{eq:improved_BC}). The improved approach includes the impact of Reynolds number explicitly.

\subsection{Discussion}
{\color{red}From the above asymptotic analysis, we have described the skeleton of the fundamental resonance in hypersonic boundary layers by a triad resonance system appearing among the 2D fundamental mode (1,0), the 3D travelling mode (1,1) and the streak mode (0,1). Here, the most distinguished feature is that the amplitude of the streamwise velocity component  of the streak mode  (streak component) is much greater than those of the transverse and lateral velocity components (roll components), therefore, the magnitude of the terms in the momentum equation governing the streak component is much greater than that governing the roll components. The mutual interaction of modes (0,1) and (1,1) could drive the formation of the roll components (see (\ref{eq:v01} $c,d$)), but is too small to affect the leading-order streamwise momentum equation of the streak mode. }

\textcolor{magenta}{The conditions for the triad resonance  include: (1)  the dimensionless growth rates of the streak mode (0,1) and the 3D travelling modes ($1,\pm 1$) are of the same order as the dimensionless amplitude of the fundamental mode $\bar \epsilon_{10}$ (see (\ref{eq:relation_sigma_delta} $a$)); (2) the magnitude of the roll structure is smaller by a factor of $O(\sigma)$ than that of the streak structure (see (\ref{eq:v01_w01})).
Such scaling relations could not be prescribed by the SIA, which can   also be regarded as a distinguished feature of the FR regime.}

\begin{figure}
\centering
\includegraphics[width=0.98\textwidth]{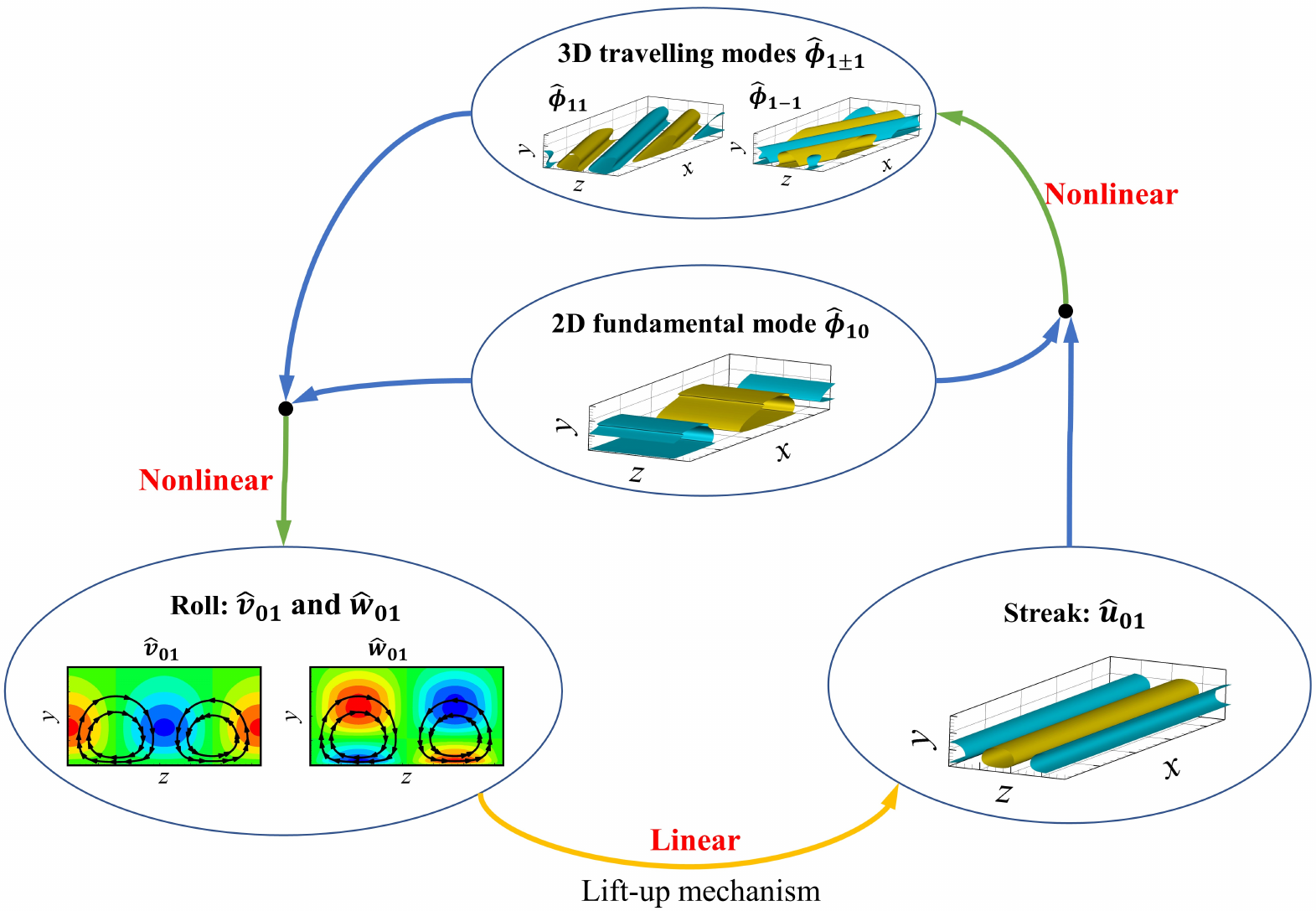}
\caption{Sketch of the principle of the FR. }
\label{fig:sketch_mechanism}
\end{figure}

\begin{figure}
\centering
\begin{overpic}[width=0.98 \textwidth]{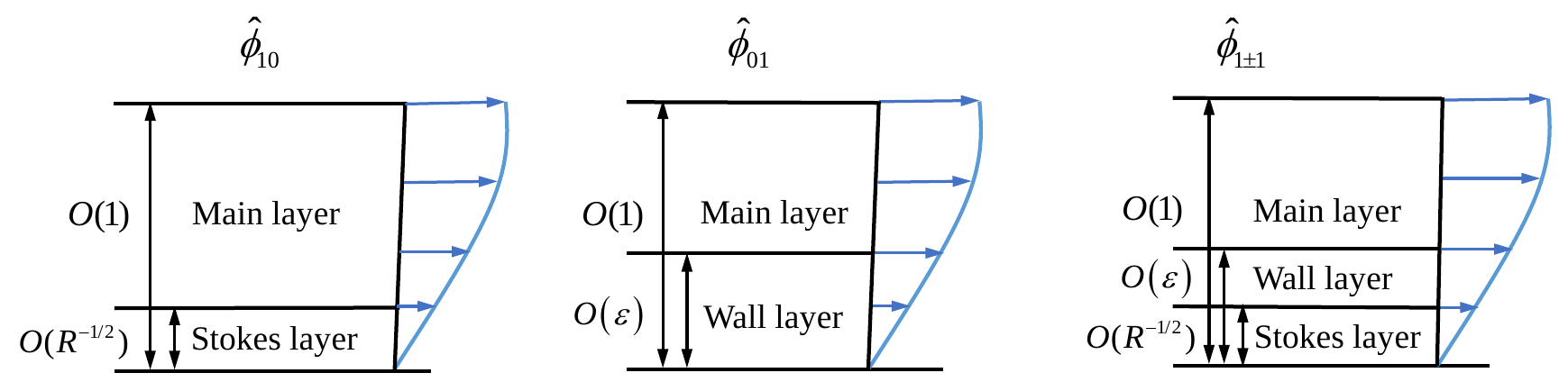}
 \put(2,20){(a)}
 \put(35,20){(b)}
  \put(69,20){(c)}
\end{overpic}
\caption{Asymptotic structures of $\hat{\phi}_{10}$ (a), $\hat{\phi}_{01}$ (b) and $\hat{\phi}_{1\pm1}$ (c), where $\epsilon \equiv (\bar \epsilon_{10} R)^{-1/3}$.}
\label{fig:sketch_structure}
\end{figure}
The value of the present asymptotic analysis is to reveal the intrinsic mechanism of the fundamental resonance from the dynamic point of view, as summarised in the sketchmatic in figure~\ref{fig:sketch_mechanism}. The nonlinear interaction of the fundamental mode and the streak mode seeds for the growth of the 3D travelling mode; the nonlinear interaction of the fundamental Mack mode $\hat \phi_{10}$ and the 3D travelling modes $\hat \phi_{1\pm 1}$ drives the roll components of the streak mode, $\hat v_{01}$ and $\hat w_{01}$; the stronger amplification of the streak component of the streak mode is due to the linear lift-up mechanism. 

The aforementioned analyses also indicate that the transverse asymptotic structures for different Fourier components are different. As shown in figure~\ref{fig:sketch_structure}, the 2D fundamental mode $\hat \phi_{10}$ shows a double-deck structure, namely, a main layer of $O(1)$ and a thinner stokes layer of $O(R^{-1/2})$; the streak mode $\hat \phi_{01}$ shows an asymptotic structure with a main layer of $O(1)$ and a wall layer of $O(\epsilon)$; for the 3D traveling mode $\hat{\phi}_{11}$, a main layer of $O(1)$, a wall layer of $O(\epsilon)$ and a more thinner stokes layer of $O(R^{-1/2})$ are observed. The wall-layer solutions of $\hat{\phi}_{01}$ and $\hat{\phi}_{1\pm1}$ communicate with the main-layer solutions via outflux velocities, which are both $O(\epsilon \ln \epsilon)$. Inclusion of these effects leads to construction of the improved boundary conditions of the main-layer equations, which could increase the accuracy of the asymptotic predictions of both the growth rates and the perturbation profiles in the main layer.

\section{\label{sec:compare} Verification of the asymptotic theory by NPSE calculations for moderate $R$ values}

 %
\subsection{Parameters of the case studies}

\begin{table}
  \begin{center}
\def~{\hphantom{0}}
  \begin{tabular}{ccccccccccccccccc}
     Case&& $M$&&$T_w/T_{ad}$ & & $R$ &  &$\omega_0$&& $\beta_{0}$& &   $\epsilon_{10}$ &&$\epsilon_{1\pm1}$ \\ [10pt]
     Case A1 & &5.92&&1   & &$2\times 10^{3}$& &0.11&&0.1 & &$2.5 \times 10^{-3}$&&$2.5 \times 10^{-9}$  \\
      Case A2 & &5.92&&1   & &$4\times 10^{3}$& &0.11&&0.1& &$2.5 \times 10^{-3}$&&$2.5 \times 10^{-9}$    \\
      Case A3 & &5.92&&1   & &$6\times 10^{3}$& &0.11&&0.1& &$2.5 \times 10^{-3}$ &&$2.5 \times 10^{-9}$   \\
      Case A4 & &5.92&&1   & &$8\times 10^{3}$& &0.11&&0.1& &$2.5 \times 10^{-3}$ &&$2.5 \times 10^{-9}$   \\
      Case A5 & &5.92&&1   & &$1\times 10^{4}$& &0.11&&0.1& &$2.5 \times 10^{-3}$&&$2.5 \times 10^{-9}$    \\
      Case B1 & &5.92&&0.5   & &$2\times 10^{3}$& &0.125&&0.1& &$2.5 \times 10^{-3}$&&$2.5 \times 10^{-9}$   \\
      Case B2 & &5.92&&0.5   & &$4\times 10^{3}$& &0.125&&0.1& &$2.5 \times 10^{-3}$ &&$2.5 \times 10^{-9}$   \\
      Case B3 & &5.92&&0.5   & &$6\times 10^{3}$& &0.125&&0.1& &$2.5 \times 10^{-3}$ &&$2.5 \times 10^{-9}$   \\
      Case B4 & &5.92&&0.5   & &$8\times 10^{3}$& &0.125&&0.1& &$2.5 \times 10^{-3}$&&$2.5 \times 10^{-9}$    \\
      Case B5 & &5.92&&0.5   & &$1\times 10^{4}$& &0.125&&0.1& &$2.5 \times 10^{-3}$ &&$2.5 \times 10^{-9}$   \\
  \end{tabular}
  \caption{Parameters for case studies in $\S$\ref{sec:compare}. }
  \label{tab:case_R}
  \end{center}
\end{table}

\begin{figure}
\centering
\begin{overpic}[width=0.96 \textwidth]{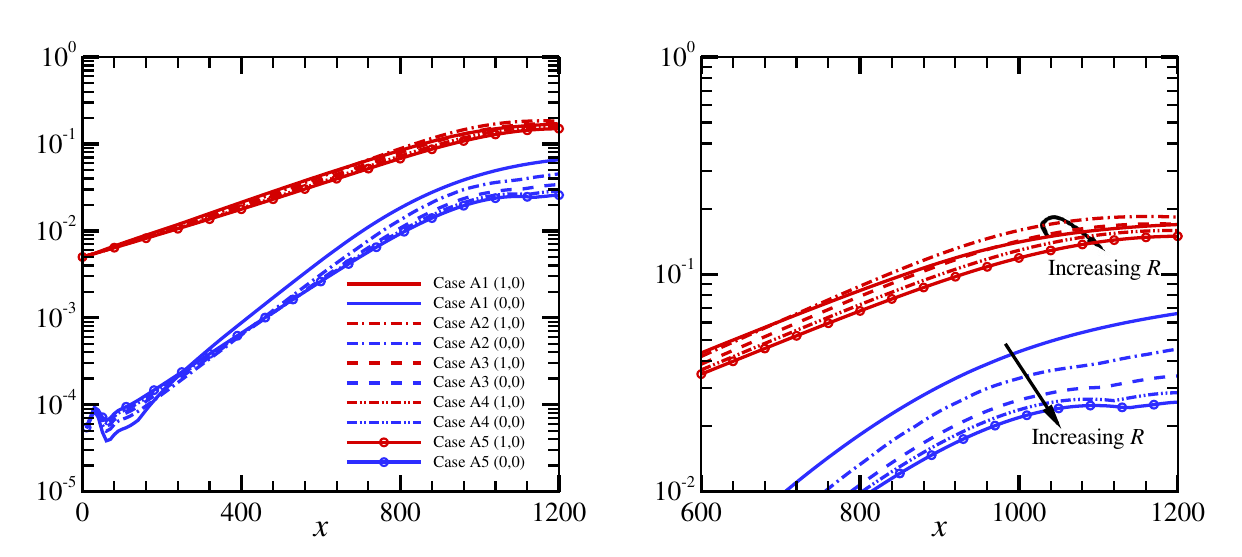}
 \put(-2,39.5){(a)}
\put(47.5,39.5){(b)}
\put(-1,20){\begin{turn}{90}${\tilde u}_{max}$\end{turn}}
\end{overpic}
\begin{overpic}[width=0.96 \textwidth]{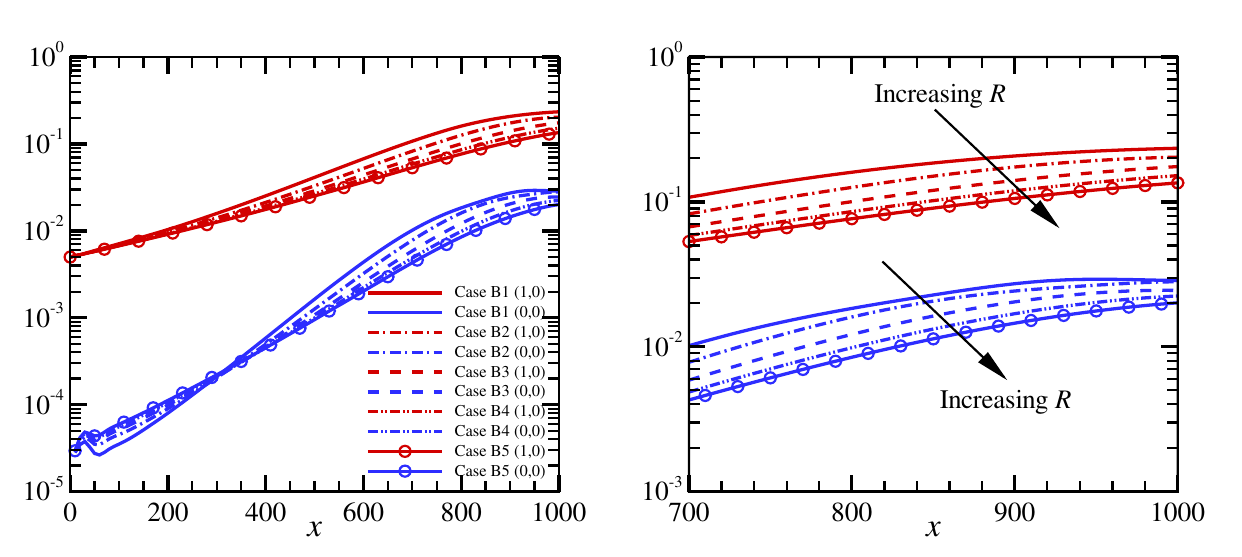}
 \put(-2,39.5){(c)}
\put(47.5,39.5){(d)}
\put(-1,20){\begin{turn}{90}${\tilde u}_{max}$\end{turn}}
\end{overpic}
\caption{ The evolution of the amplitude of modes (1,0) and (0,0) obtained by NPSE. (a,b): Results for cases A1 to A5; (c,d): results for cases B1 to B5. (b) and (d) are the zoom-in plots of (a) and (c), respectively.      }
\label{fig:10_00_different_R}
\end{figure}
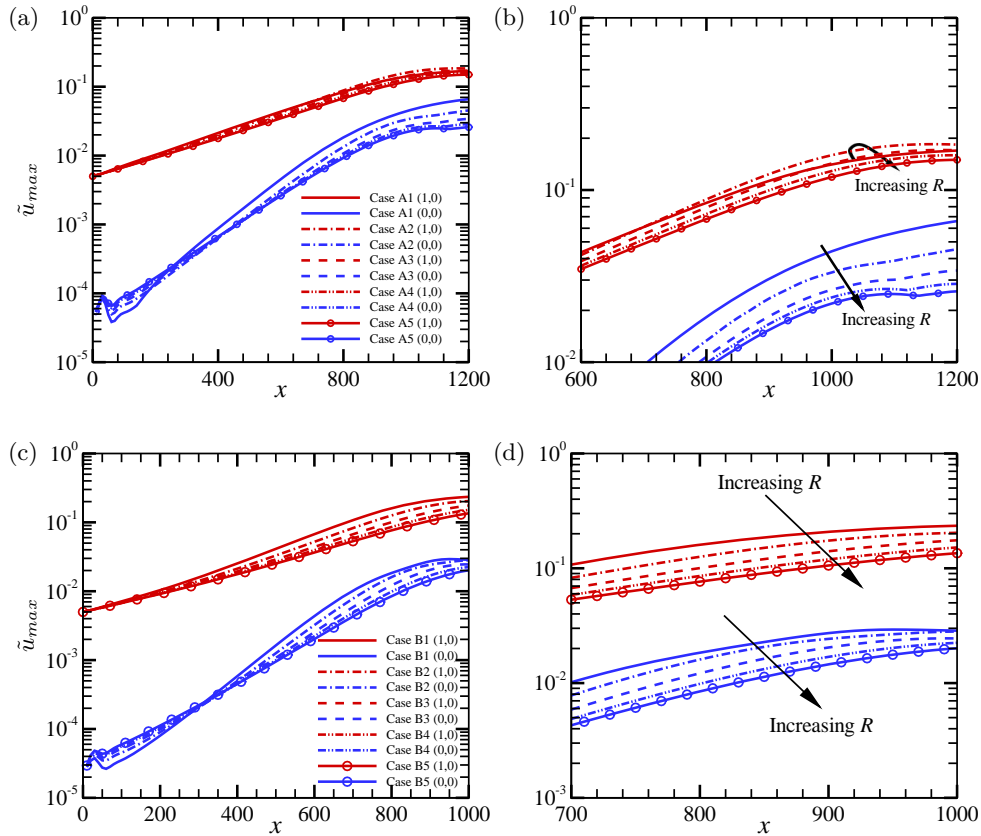

To verify the asymptotic theory in $\S$\ref{sec:asymptotic}, we choose a set of case studies with the same Mach number but different wall temperatures and Reynolds numbers as listed in table~\ref{tab:case_R}. Each case is labeled by a two-digit character, the first and second of which distinguish the wall temperature and the Reynolds number, respectively. The  frequency $\omega_0$, spanwise wavenumber $\beta_0$ and initial amplitude $\epsilon_{10}$ of the fundamental 2D mode  are the same as those in $\S$\ref{sec:FR_NPSE}, but the initial amplitudes of the oblique modes $\epsilon_{1\pm 1}$ are reduced to be $2.5\times 10^{-9}$, which enlarges the streamwise region of the fundamental resonance.

\subsection{Evolution of the fundamental modes}
Figures~\ref{fig:10_00_different_R}-(a) and (c) show the amplitude evolution of the fundamental mode (1,0) and the MFD (0,0) obtained by NPSE calculations for cases A1 to A5 and cases B1 to B5, respectively, and figure~\ref{fig:10_00_different_R}-(b) and (d) display their zoom-in plots in the nonlinear phase.
For all the Reynolds numbers, the mode (1,0) become \textcolor{red}{saturated} at $x\approx  1000$ for cases A and  $x\approx  800$ for cases B. Overall, the saturated amplitudes of both  (1,0) and (0,0) components decrease with increase of $R$, except mode (1,0) for case A1. Since the MFD is mainly driven by the self interaction of the fundamental mode, its order of magnitude is square of that of the latter.
\begin{figure}
\centering
\begin{overpic}[width=0.96 \textwidth]{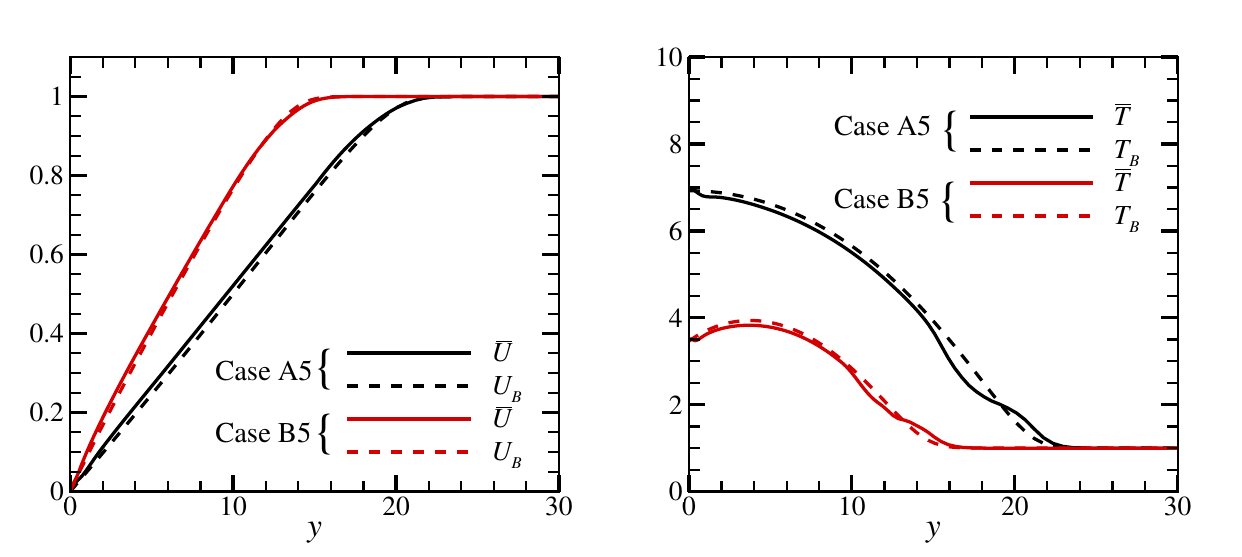}
\put(-2,20){\begin{turn}{90} $U_B , \bar U$ \end{turn}}
\put(50,20){\begin{turn}{90} $T_B , \bar T$ \end{turn}}
 \put(-1.5,39.5){(a)}
\put(48,39.5){(b)}
\end{overpic}
\caption{ Comparison of the base flow $\Phi_{B}$ and the time- and spanwise-averaged mean flow $\bar \Phi$ for case A5 and case B5 at $x=1000$. (a): Streamwise velocity; (b) temperature.    }
\label{fig:U00_T00}
\end{figure}
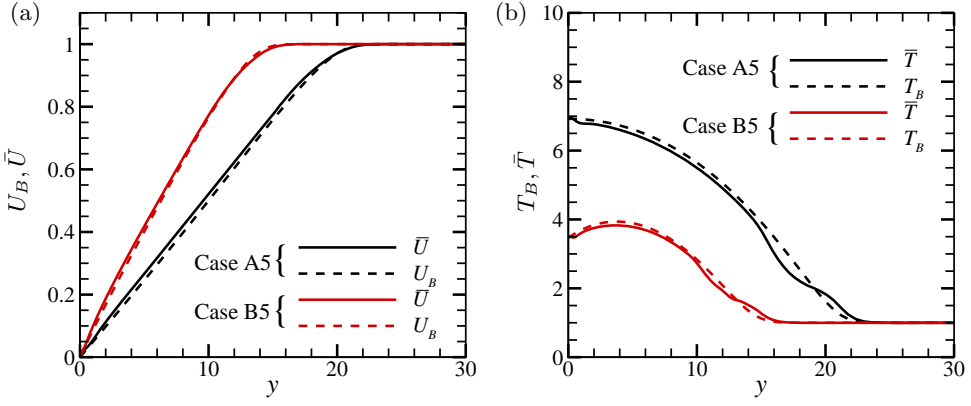
Figure~\ref{fig:U00_T00} compares the profiles of the streamwise velocity and temperature of the base flow $(U_B,T_B)$ with those of the mean flow $(\bar U,\bar T)$. The difference between the two families of curves is quite limited.


\begin{figure}
\centering
\begin{overpic}[width=0.96 \textwidth]{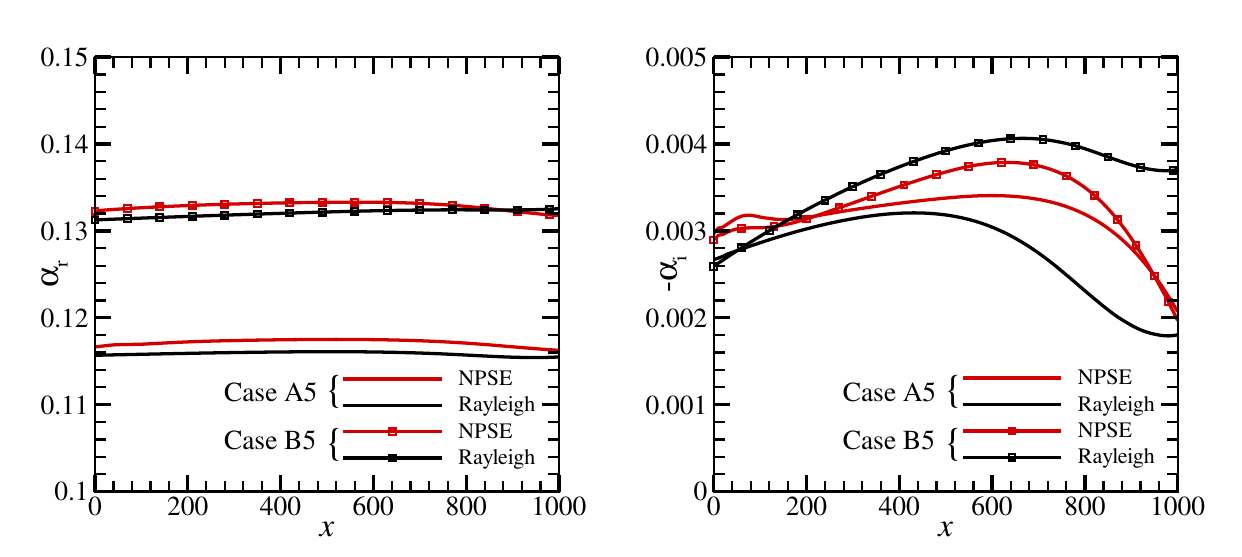}
 \put(-1.5,39.5){(a)}
\put(48,39.5){(b)}
\end{overpic}
\caption{ Comparison of the wavenumber $\alpha_r$ (a) and the growth rate $-\alpha_i$ (b) of mode (1,0) obtained by the NPSE calculations and the Rayleigh solutions for case A5 and case B5. }
\label{fig:gr_compare_Rayleigh_NPSE_10}
\end{figure}

\begin{figure}
\centering
\begin{overpic}[width=0.96 \textwidth]{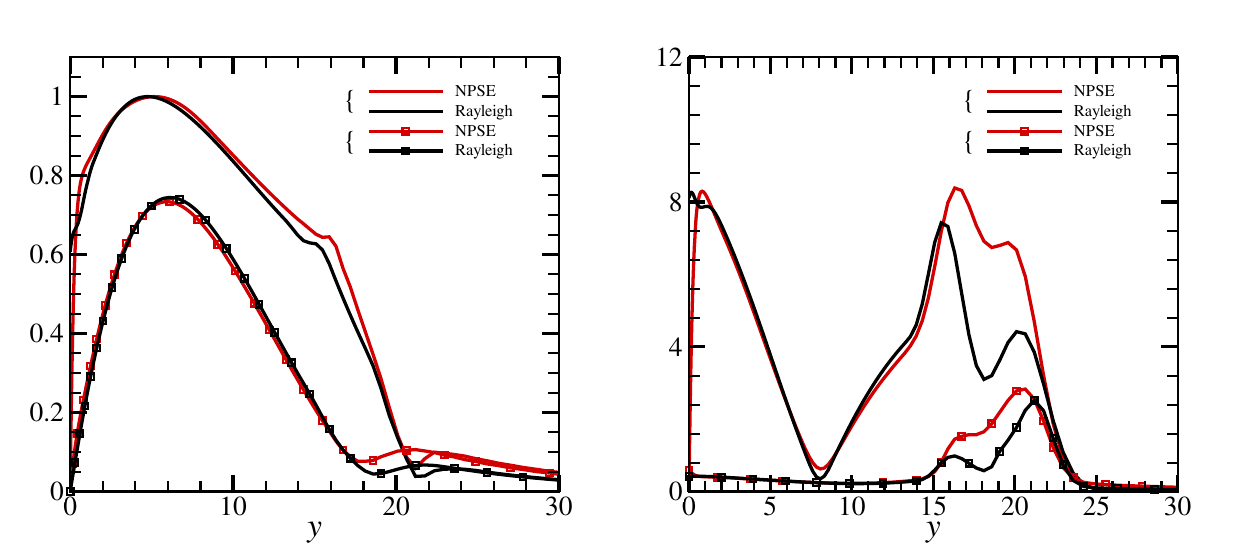}
\put(-2,20){\begin{turn}{90} $|\hat{u}_{10}| , |\hat{v}_{10}|$ \end{turn}}
\put(50,20){\begin{turn}{90} $|\hat{\rho}_{10}| , |\hat{T}_{10}|$ \end{turn}}
\put(22,36){$|\hat{u}_{10}|$}
\put(22,32.5){$|\hat{v}_{10}|$}
\put(71,36){$|\hat{T}_{10}|$}
\put(71,32.5){$|\hat{\rho}_{10}|$}
 \put(-1.5,39.5){(a)}
\put(48,39.5){(b)}
\end{overpic}
\begin{overpic}[width=0.96 \textwidth]{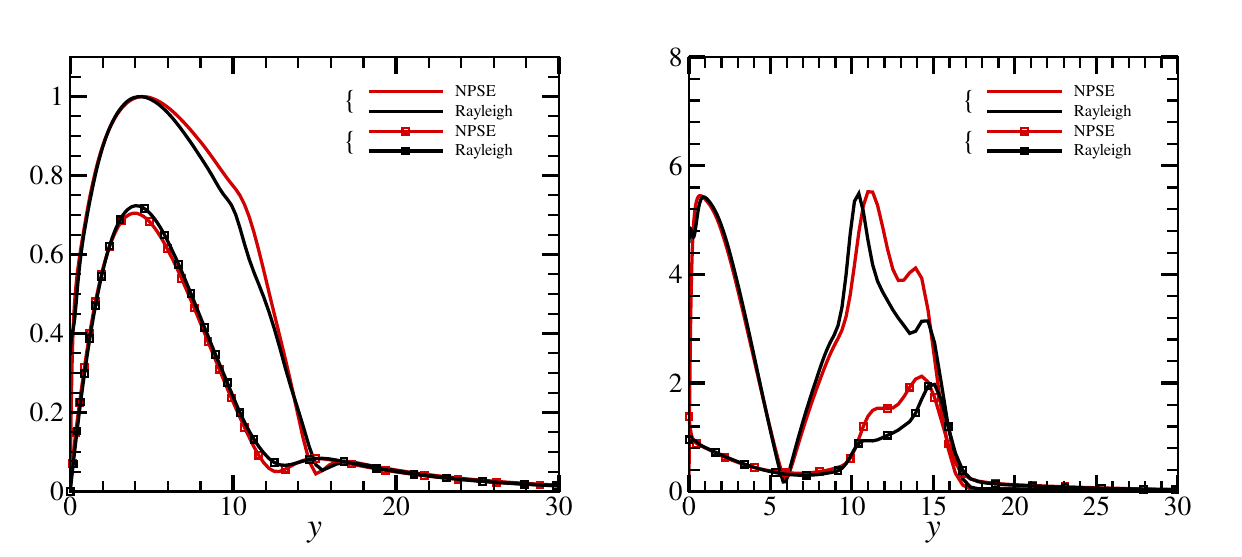}
\put(-2,20){\begin{turn}{90} $|\hat{u}_{10}| , |\hat{v}_{10}|$ \end{turn}}
\put(50,20){\begin{turn}{90} $|\hat{\rho}_{10}| , |\hat{T}_{10}|$ \end{turn}}
\put(22,36){$|\hat{u}_{10}|$}
\put(22,32.5){$|\hat{v}_{10}|$}
\put(71,36){$|\hat{T}_{10}|$}
\put(71,32.5){$|\hat{\rho}_{10}|$}
 \put(-1.5,39.5){(c)}
\put(48,39.5){(d)}
\end{overpic}
\caption{ Comparison of the perturbation profiles of the fundamental mode (1,0) obtained by the NPSE calculation and the Rayleigh solutions for case A5 (a,b) and case B5 (c,d) at $x=1000$. }
\label{fig:10_R=1e4}
\end{figure}

In figure~\ref{fig:gr_compare_Rayleigh_NPSE_10}, we compare the wavenumber and growth rate of the Rayleigh solutions and NPSE calculations, where only cases A5 and B5 are chosen for demonstration. Although there exist small discrepancies between the two families of curves, the overall trends agree. The error is attributed to the nonlinear, non-parallel and viscous effects.

Figure~\ref{fig:10_R=1e4} shows the comparison of the perturbation profiles obtained by Rayleigh and NPSE calculations. \textcolor{magenta}{The largest error appears in the near-wall region} and the critical layer, because the viscosity, neglected in the Rayleigh equation (\ref{eq:Rayleigh}), is more important there. However, the overall agreement is quite satisfactory.



\subsection{\label{sec:compare_3D_streak} The 3D travelling mode and the streak mode}

\begin{figure}
\centering
\begin{overpic}[width=0.96 \textwidth]{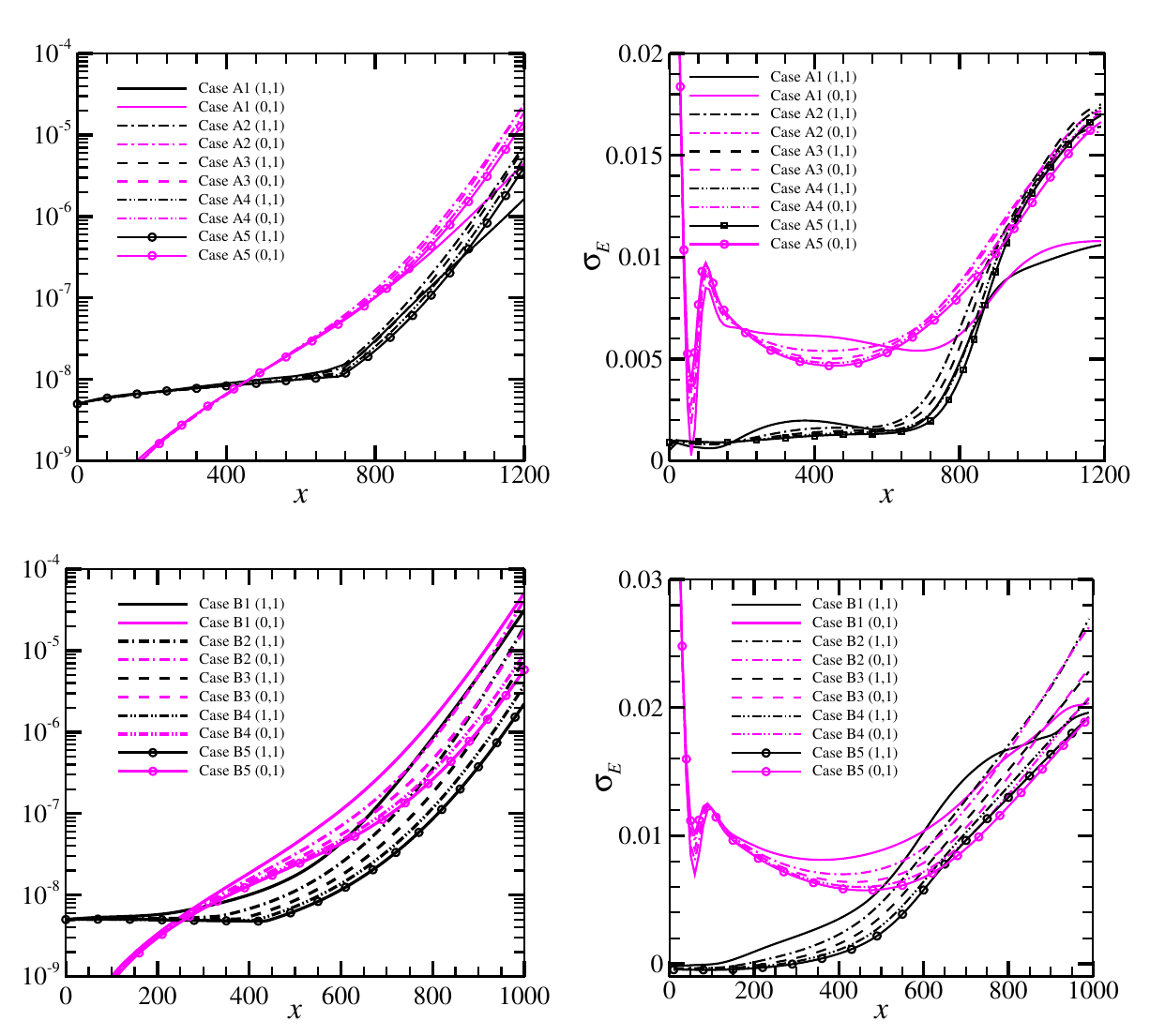}
 \put(-2,83.5){(a)}
\put(47.5,83.5){(b)}
 \put(-2,39){(c)}
\put(47.5,39){(d)}
\put(-2,64){\begin{turn}{90}${\tilde u}_{max}$\end{turn}}
\put(-2,19.5){\begin{turn}{90}${\tilde u}_{max}$\end{turn}}
\end{overpic}
\caption{ The evolution of the amplitude (a,c) and growth rate (b,d) of (1,1) and (0,1) components. Top row: cases A1 to A5; bottom row: cases B1 to B5.       }
\label{fig:11_01_different_R}
\end{figure}

Figure~\ref{fig:11_01_different_R}-(a) shows the evolution of the amplitudes  of (1,1) and (0,1) obtained by NPSE calculations for cases A1 to A5. Being similar to figure~\ref{fig:Au_evolution_large}, a mild increase for the 3D travelling mode (1,1) before $x \approx 720$ is observed, which is determined by its linear instability.  After that, a drastic amplification occurs for both (1,1) and (0,1). In panel (b), we also plot their growth rates, which are nearly identical and increases monotonically in the interval of $x \in [900,1200]$, indicating the FR phenomenon. The increase of the growth rates is attributed to the increase of the amplitude of the fundamental mode (1,0), as predicted by the asymptotic analysis (\ref{eq:relation_sigma_delta} $a$). Similar observations can be found in panels (c) and (d) for cases B1 to B5. However, the position where FR appears is promoted to $x \approx 650$, because the fundamental mode reaches a finite amplitude at an earlier position due to its higher growth rate.

\textcolor{red}{Comparing with figure~\ref{fig:gr_compare_Rayleigh_NPSE_10}-(b), we find that the FR appears before the mode (1,0) reaches the nonlinear saturation phase. At the FR regions ($x\in [900,1200]]$ for case A and $x\in [650,1000]$ for case B), both the amplitude of the fundamental mode $\bar\epsilon_{01}$ and the growth rate of the secondary instability mode $\sigma$ increase with $x$, agreeing qualitatively with the scaling relation
(\ref{eq:relation_sigma_delta} $a$).  }


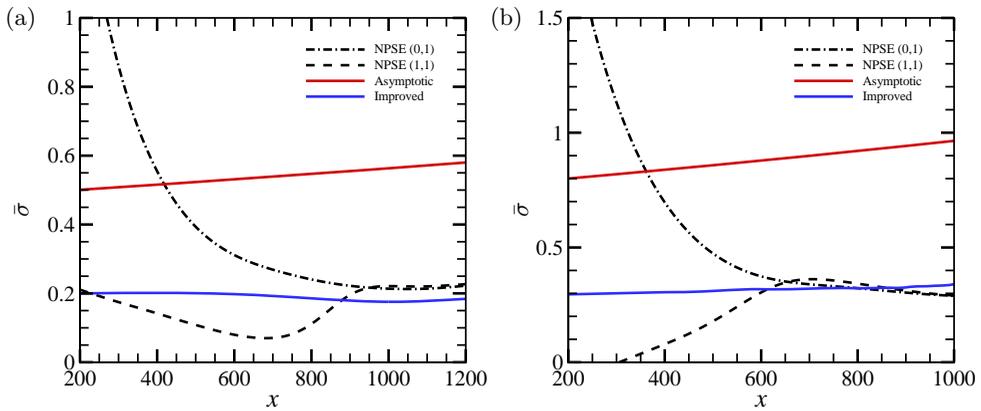
\begin{figure}
\centering
\begin{overpic}[width=0.96 \textwidth]{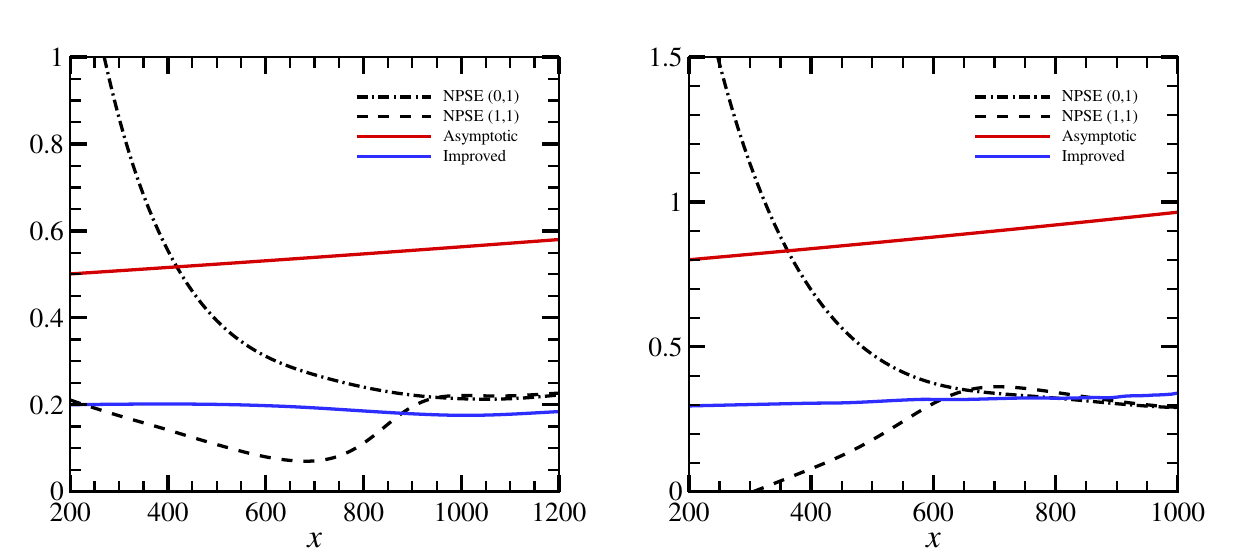}
 \put(-2,39.5){(a)}
\put(47.5,39.5){(b)}
\put(-1,20){\begin{turn}{90}${\bar \sigma}$\end{turn}}
\put(50,20){\begin{turn}{90}${\bar \sigma}$\end{turn}}
\end{overpic}
\caption{ Dependence on $x$ of the normalised growth rate obtained by the NPSE calculation, the asymptotic prediction and the improved asymptotic prediction for case A5 (a) and case B5 (b). }
\label{fig:gr_x}
\end{figure}
{\color{red}To confirm quantitatively the asymptotic prediction of the growth rate of the secondary instability mode, figure~\ref{fig:gr_x} compares the NPSE results with the asymptotic predictions of both the original and improved versions.}  The growth rate of the NPSE calculation is normalised by the amplitude of the fundamental mode at each $x$, 
\begin{equation}
     \bar \sigma^{(m,n)}(x)=  \frac{\sigma_{E}^{(m,n)}(x)}{ \frac{1}{2} \tilde {u}_{max}^{(1,0)}(x) },
\end{equation}
 where $\sigma_{E}^{(m,n)}$ and ${u}_{max}^{(1,0)}$ are defined in (\ref{eq:sigma}) and (\ref{eq:A}), respectively. Here,  the growth rates of  (0,1) and (1,1) obtained by NPSE are both shown. In the FR regions, $x\in[900,1200]$ for case A5 and $x\in[650,1000]$ for case B5, the normalised growth rates $\bar \sigma^{(0,1)}$ and $\bar\sigma^{(1,1)}$ remain almost constant, confirming the scaling relation (\ref{eq:relation_sigma_delta} $a$). The growth rate of the original asymptotic theory $\bar \sigma^{\text{Asmp}}$ is obtained by solving (\ref{eq:eigenvalue}) with (\ref{eq:zero_condition}), shown by the red lines in the figure. Overall, $\bar \sigma^{\text{Asmp}}$ does not change much with $x$ in the FR region, showing the same feature as the NPSE calculations, but its value is almost 3 times greater than $\bar \sigma^{(0,1)}$ or $\bar \sigma^{(1,1)}$. If the impact of the viscous wall layer is taken into account, as predicted by the improved asymptotic theory by solving (\ref{eq:eigenvalue})  with (\ref{eq:zero_condition} $a,b,c$) and (\ref{eq:improved_BC}), then the agreement to the NPSE results is much better, as indicated by the blue lines in the figure.
 

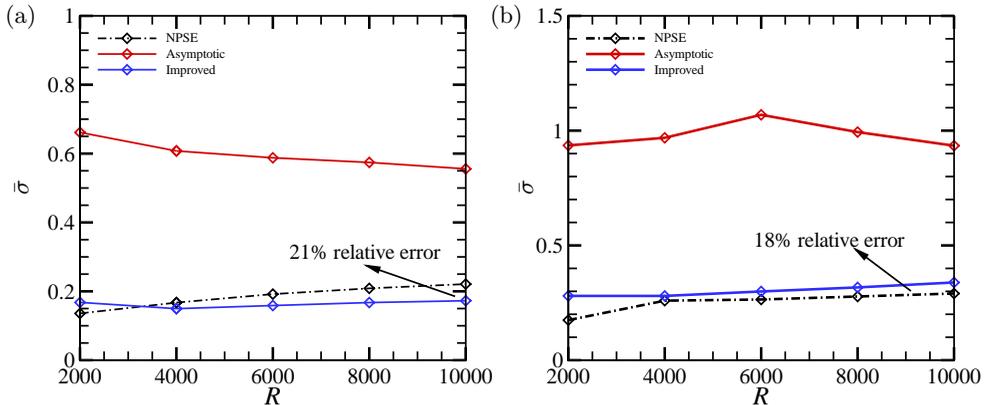
\begin{figure}
\centering
\begin{overpic}[width=0.96 \textwidth]{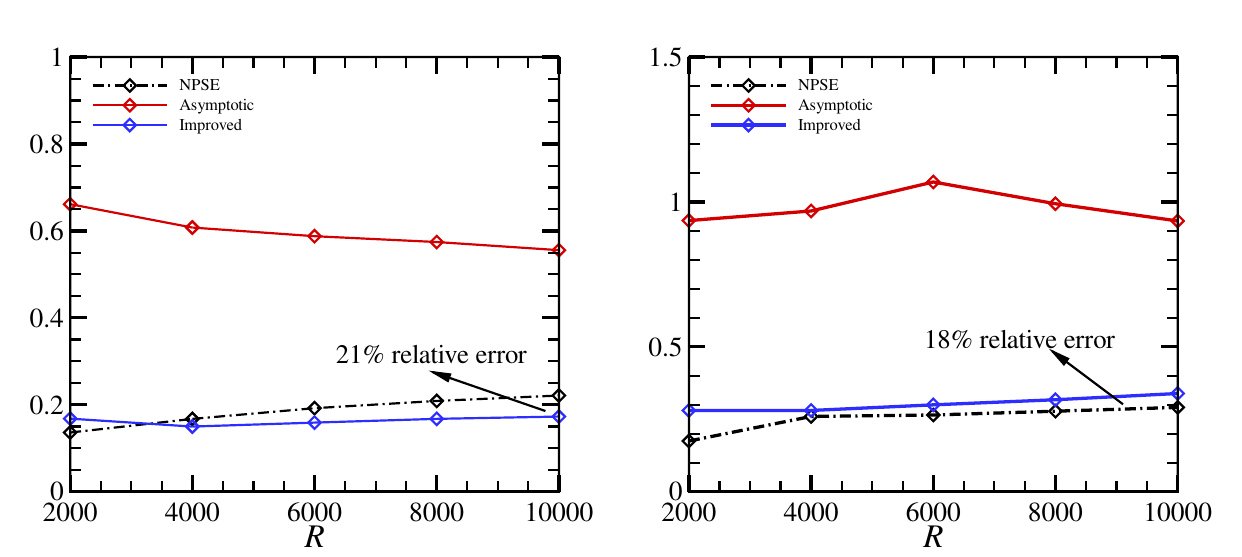}
 \put(-2,39.5){(a)}
\put(47.5,39.5){(b)}
\put(-1,22){\begin{turn}{90}${\bar \sigma}$\end{turn}}
\put(50.5,22){\begin{turn}{90}${\bar \sigma}$\end{turn}}
\end{overpic}
\caption{ Dependence on $R$ of the normalised growth rate $\bar \sigma$ obtained by the NPSE calculation, the asymptotic prediction and the improved asymptotic prediction at $x=1000$. (a): cases A1 to A5; (b): cases B1 to B5. }
\label{fig:gr_R}
\end{figure}

In figure~\ref{fig:gr_R}, we show the impact of $R$ on the predictions of the growth rate $\bar \sigma$, where only the curves for $\bar\sigma^{(0,1)}$ are plotted for the NPSE calculations. Again, the improved asymptotic predictions show a better agreement with the NPSE calculations than the original asymptotic prediction, and its relative error is only about $20\%$. The discrepancy is acceptable because in the asymptotic theory, the impact of the high-order viscosity, the non-parallelism and the higher-order Fourier components are excluded.

\begin{figure}
\centering
\begin{overpic}[width=0.96 \textwidth]{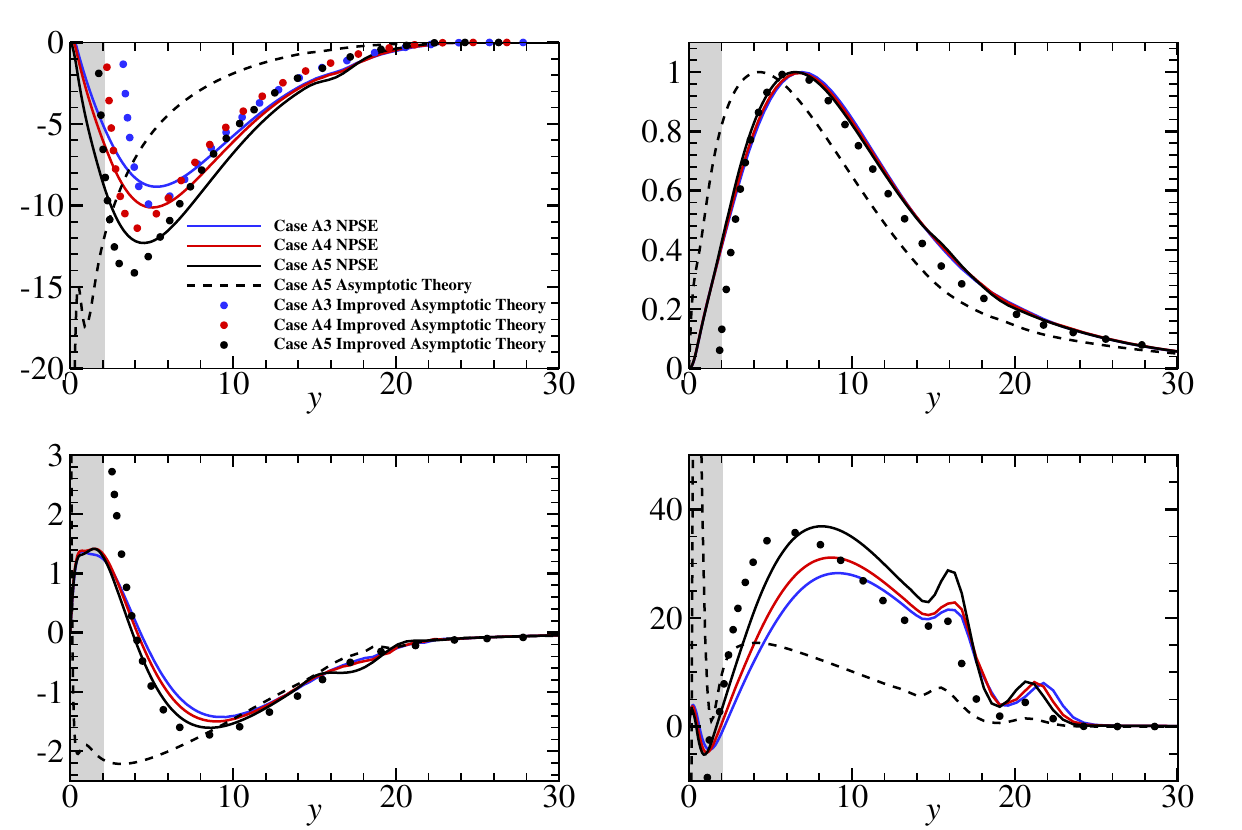}
\put(-1,62){(a)}
\put(48,62){(b)}
\put(-1,29){(c)}
\put(48,29){(d)}
\put(-1,48){\begin{turn}{90} $\Re(\hat{u}_{01})$ \end{turn}}
\put(48,48){\begin{turn}{90} $\Re(\hat{v}_{01})$ \end{turn}}
\put(-1,15){\begin{turn}{90} $\Im (\hat{w}_{01})$ \end{turn}}
\put(48,15){\begin{turn}{90} $\Re(\hat{T}_{01})$ \end{turn}}
\end{overpic}
\begin{overpic}[width=0.96 \textwidth]{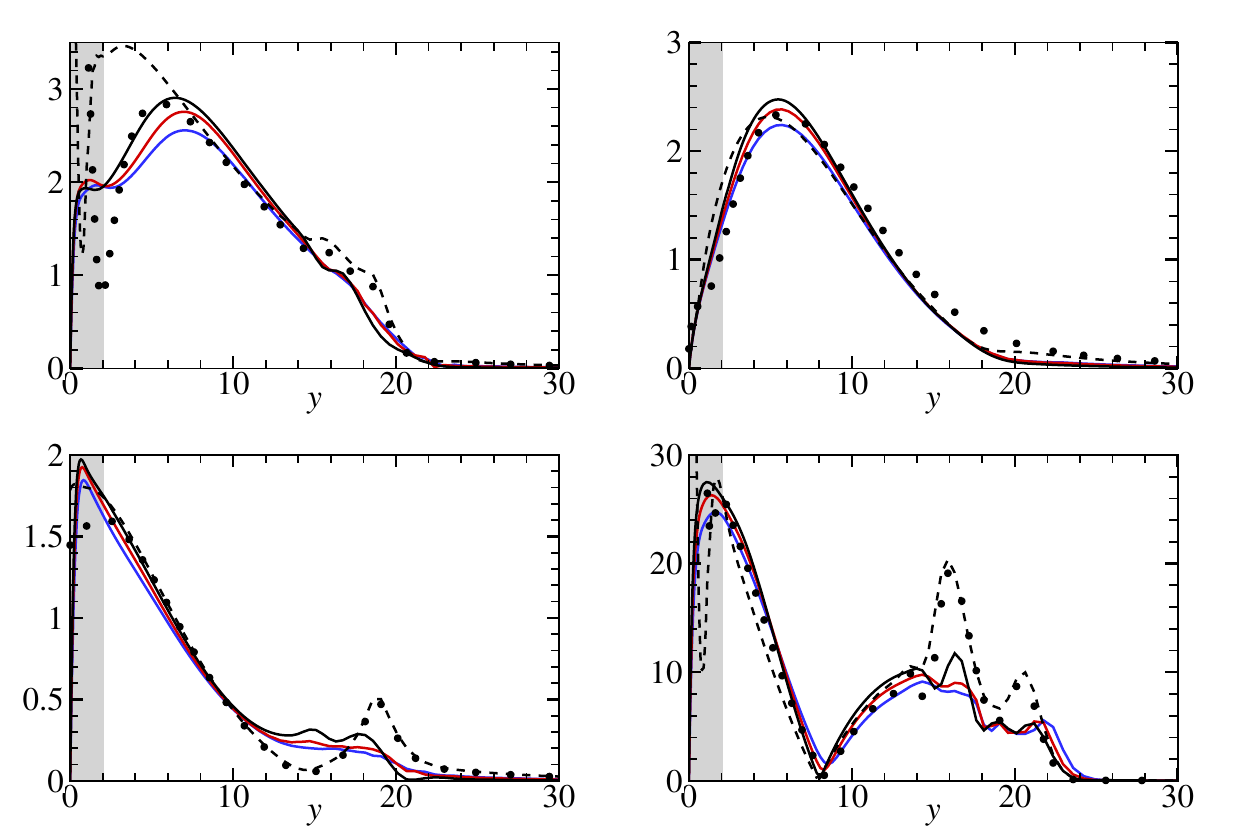}
\put(-1,62){(e)}
\put(48,62){(f)}
\put(-1,29){(g)}
\put(48,29){(h)}
\put(-1,48){\begin{turn}{90} $|\hat{u}_{11}|$ \end{turn}}
\put(48,48){\begin{turn}{90} $ |\hat{v}_{11}|$ \end{turn}}
\put(-1,15){\begin{turn}{90} $|\hat{w}_{11}|$ \end{turn}}
\put(48,15){\begin{turn}{90} $ |\hat{T}_{11}|$ \end{turn}}
\end{overpic}
\caption{ Perturbation profiles of (0,1) and (1,1) at $x=1000$ for case A. Solid lines: NPSE results; dashed lines: asymptotic predictions; symbols: improved asymptotic predictions. The profiles are normalized by the maximum of $\Re(\hat{v}_{01})$. } 
\label{fig:egf_compare_2000-10000}
\end{figure}

\begin{figure}
\centering
\begin{overpic}[width=0.96 \textwidth]{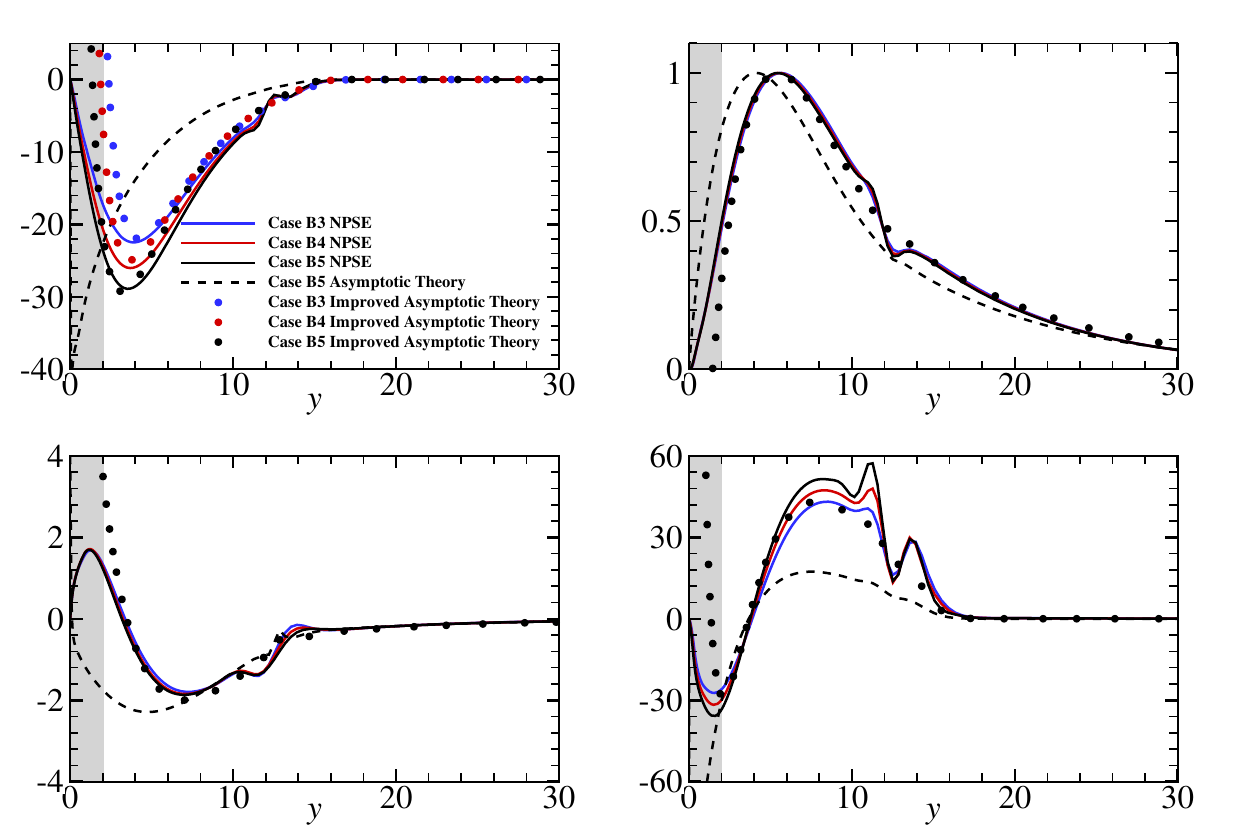}
\put(-1,62){(a)}
\put(48,62){(b)}
\put(-1,29){(c)}
\put(48,29){(d)}
\put(-1,48){\begin{turn}{90} $\Re(\hat{u}_{01})$ \end{turn}}
\put(48,48){\begin{turn}{90} $\Re(\hat{v}_{01})$ \end{turn}}
\put(-1,15){\begin{turn}{90} $\Im (\hat{w}_{01})$ \end{turn}}
\put(48,15){\begin{turn}{90} $\Re(\hat{T}_{01})$ \end{turn}}
\end{overpic}
\begin{overpic}[width=0.96 \textwidth]{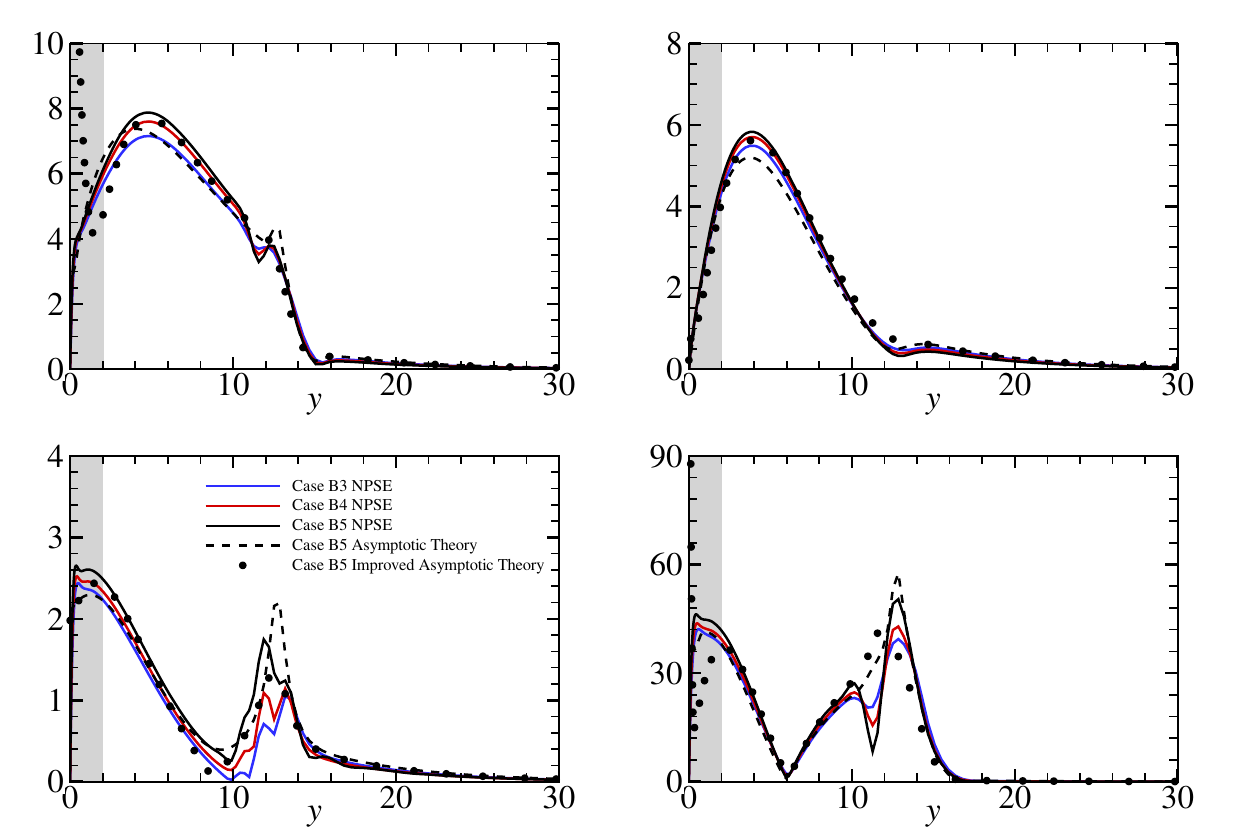}
\put(-1,62){(e)}
\put(48,62){(f)}
\put(-1,29){(g)}
\put(48,29){(h)}
\put(-1,48){\begin{turn}{90} $|\hat{u}_{11}|$ \end{turn}}
\put(48,48){\begin{turn}{90} $ |\hat{v}_{11}|$ \end{turn}}
\put(-1,15){\begin{turn}{90} $|\hat{w}_{11}|$ \end{turn}}
\put(48,15){\begin{turn}{90} $ |\hat{T}_{11}|$ \end{turn}}
\end{overpic}
\caption{ Perturbation profiles of (0,1) and (1,1) at $x=1000$ for case B. Solid lines: NPSE results; dashed lines: asymptotic predictions; symbols: improved asymptotic predictions. The profiles are normalized by the maximum of $\Re(\hat{v}_{01})$ .   } 
\label{fig:egf_compare_2000-10000_dengwen}
\end{figure}

Figures~\ref{fig:egf_compare_2000-10000} and  \ref{fig:egf_compare_2000-10000_dengwen} compare the perturbation profiles of the streak mode (0,1) and the 3D traveling mode (1,1) obtained by the NPSE calculations and {\color{red}the main-layer asymptotic predictions (\ref{eq:11}) and (\ref{eq:01})} for cases A and B. According to the system~(\ref{eq:01}), $\hat{u}_{01}$, $\hat{v}_{01}$ and $\hat{T}_{01}$ are almost real, but $\hat{w}_{01}$ is almost pure imaginary, and therefore, we display only $\Re(\hat{u}_{01})$, $\Re{\hat{v}_{01}}$, $\Im (\hat{w}_{01})$ and $\Re(\hat{T}_{01})$. 
{\color{red}Note that the asymptotic predictions in these figures are only from the main-layer equations (\ref{eq:11}) and (\ref{eq:01}), and the solutions in the wall layer shown in $\S$\ref{sec:wall_layer} are not included. In principle, if we construct a composite solution based on both the main-layer and wall-layer solutions, then the agreement to the NPSE calculations should be throughout the whole boundary layer. However, as mentioned before, because the main-layer solution undergoes a logarithmic singularity as the wall is approached, a consistent composite solution requires consideration of the wall-layer solutions up to the second order, which requires more complicated mathematical processes but yields the same accuracy on predicting the growth rate. Thus, we only probe the leading-order wall-layer solution, which is sufficient to construct the improved asymptotic approach as shown in $\S$\ref{sec:improved}. Therefore, in figures~\ref{fig:egf_compare_2000-10000} and  \ref{fig:egf_compare_2000-10000_dengwen},  we only consider the comparison in the main layer ($y>2$).}
The original asymptotic results shown by the dashed lines are not sufficient to predict the NPSE calculations, even when $R$ is as large as $1 \times 10^{4}$. The discrepancy is greater in the region of $y<10$. However, when the wall-layer-induced correction is considered, the accuracy of the asymptotic predictions is remarkably improved, as shown by the symbols.  In the region around the critical layer, the viscous effect needs to be taken into account to achieve a better agreement. A similar observation is obtained for the comparison of $|\hat{\phi}_{11}|$, as shown in panels (e)-(h).

In the main layer, the magnitude of $\hat{u}_{01}$ is much greater than those of $\hat{v}_{01}$ and $\hat{w}_{01}$, showing a longitudinal streak nature, which agrees with the scaling relations in (\ref{eq:v01_w01}). From the NPSE calculations, as $R$ increases, the peak of $\Re (\hat{u}_{01})$ increases monotonically, with its location moving towards to wall, agreeing with the improved asymptotic predictions. The main-layer solution of $\hat{v}_{01}$ and $\hat{w}_{01}$ are almost independent of $R$, which also agrees with the improved asymptotic predictions. For the 3D travelling mode $\hat{\phi}_{11}$, the discrepancy between the improved solution and the asymptotic is even smaller than that for $\hat{\phi}_{01}$, and both methods can provide a satisfied prediction in comparison with the NPSE calculations.

\section{\label{sec:compare_SIA} Verification of the asymptotic theory by the SIA approach for large $R$ values}
\begin{table}
  \begin{center}
\def~{\hphantom{0}}
  \begin{tabular}{cccccccccccccccc}
     Case& &$M$&&$T_w/T_{ad}$ & & $R$ &  &$\omega_0$&& $\beta_{0}$& &   $\bar \epsilon_{10}$  \\ [10pt]
      Case A6 & &5.92&&1   & &$1\times 10^{4}$& &0.11&&0.1 & &0.015  \\
      Case A7 & &5.92&&1   & &$1\times 10^{5}$& &0.11&&0.1& &0.015    \\
      Case A8 & &5.92&&1   & &$1\times 10^{6}$& &0.11&&0.1& &0.015   \\
      Case A9 & &5.92&&1   & &$1\times 10^{7}$& &0.11&&0.1& &0.015   \\
      Case A10 & &5.92&&1   & &$1\times 10^{8}$& &0.11&&0.1& &0.015    \\
      Case B6 & &5.92&&0.5   & &$1\times 10^{4}$& &0.125&&0.1& &0.015   \\
      Case B7 & &5.92&&0.5   & &$1\times 10^{5}$& &0.125&&0.1& &0.015   \\
      Case B8 & &5.92&&0.5   & &$1\times 10^{6}$& &0.125&&0.1& &0.015   \\
      Case B9 & &5.92&&0.5   & &$1\times 10^{7}$& &0.125&&0.1& &0.015    \\
      Case B10 & &5.92&&0.5   & &$1\times 10^{8}$& &0.125&&0.1& &0.015   \\
  \end{tabular}
  \caption{Parameters for the FR calculations at higher $R$ values. }
  \label{tab:case_R_B}
  \end{center}
\end{table}
In this section, we verify the asymptotic predictions for large $R$ \textcolor{red}{values}. Because the NPSE calculations for higher Reynolds numbers are not stable numerically, we verify our asymptotic predictions for higher $R$ values by the SIA approach in this section, which is confirmed to be accurate at moderate $R$ values shown in figures~\ref{fig:gr_compare_SIA_NPSE} and \ref{fig:01_11_SIA_NPSE}. The comparison is made based on a wavy base flow, which consists of a laminar Blasius solution and a fundamental mode (1,0) with a finite amplitude. The latter is obtained by solving the O-S equation (\ref{eq:OS}).
Admittedly, this base flow is a bit artificial, but it is easy to demonstrate the fundamental resonance for sufficiently high Reynolds numbers.
The parameters of the case studies are listed in table~\ref{tab:case_R_B}.


\begin{figure}
\centering
\begin{overpic}[width=0.96 \textwidth]{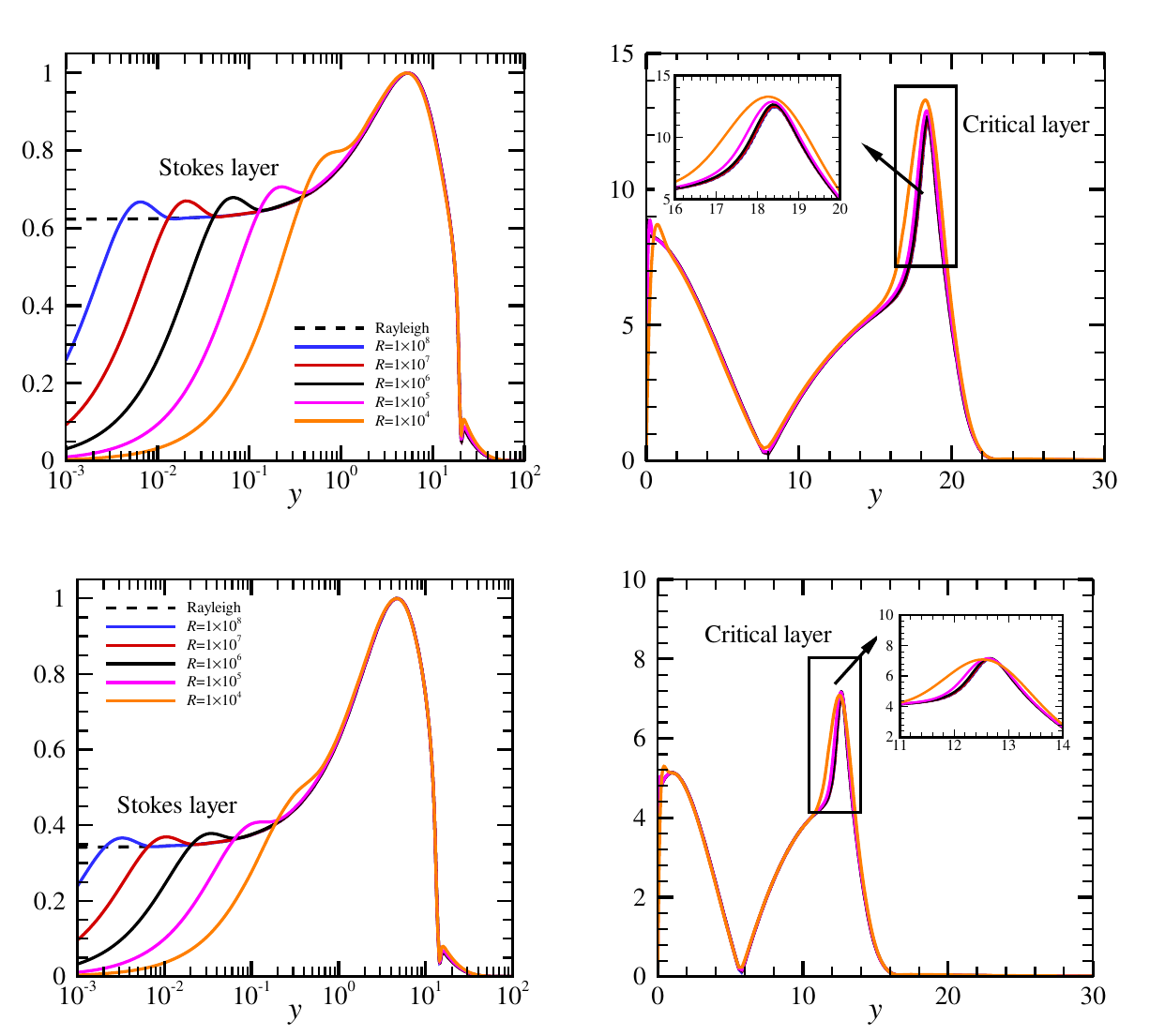}
\put(-1,83){(a)}
\put(48,83){(b)}
\put(-1,38){(c)}
\put(48,38){(d)}
\put(-1,65){\begin{turn}{90} $|\hat{u}_{10}|$ \end{turn}}
\put(48,65){\begin{turn}{90} $|\hat{T}_{10}|$ \end{turn}}
\put(-1,20){\begin{turn}{90} $|\hat{u}_{10}|$ \end{turn}}
\put(48,20){\begin{turn}{90} $|\hat{T}_{10}|$ \end{turn}}
\end{overpic}
\caption{ Perturbation profiles of the fundamental mode (1,0) for the cases listed in table~\ref{tab:case_R_B}. Left column:  streamwise perturbation velocity; right column: perturbation temperature. Top row: cases A6 to A10; bottom row: cases B6 to B10. } 
\label{fig:10_compare}
\end{figure}
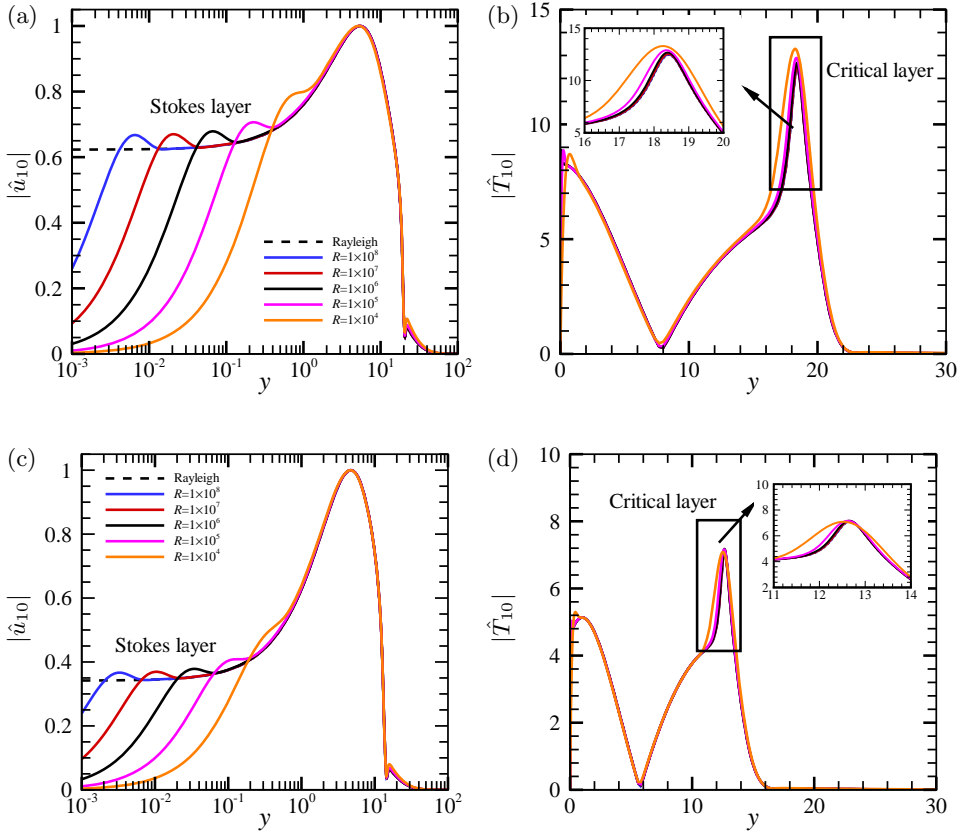

 Figure~\ref{fig:10_compare} shows the perturbation profiles of the fundamental mode (1,0) for cases A and B with different $R$ values. The results of the Rayleigh equation (\ref{eq:Rayleigh}) are also plotted by the dashed lines for comparison. Obviously, the planar Mack second mode (1,0) shows a double-deck structure, \textcolor{magenta}{a main layer in the major part of the boundary layer}, where the O-S and Rayleigh solutions agree, and a Stokes layer in the near-wall region, where the streamwise velocity and temperature damp to satisfy the no-slip and isothermal conditions. An overshoot is observed at the edge of the Stokes layer for each curve, which can be predicted by the Stokes-layer solution. The thickness of the Stokes layer $\delta_S$ decreases as $R$ increases, and a scaling of $\delta_S \sim R^{-1/2}$ is evident. The eigenfunction $|\hat{T}_{10}|$ shows a sharp peak at the critical layer but $|\hat{u}_{10}|$ does not. This is because for a 2D case, the streamwise velocity $|\hat{u}_{10}|$ shows only a logarithmic singularity, much weaker than the fist-order singularity of $|\hat{T}_{10}|$. It is found that as $R$ increases, the O-S solutions approach the Rayleigh solution, except in the near-wall region, confirming the inviscid nature of the (1,0) component. 
 

\begin{figure}
\centering
\begin{overpic}[width=0.45 \textwidth]{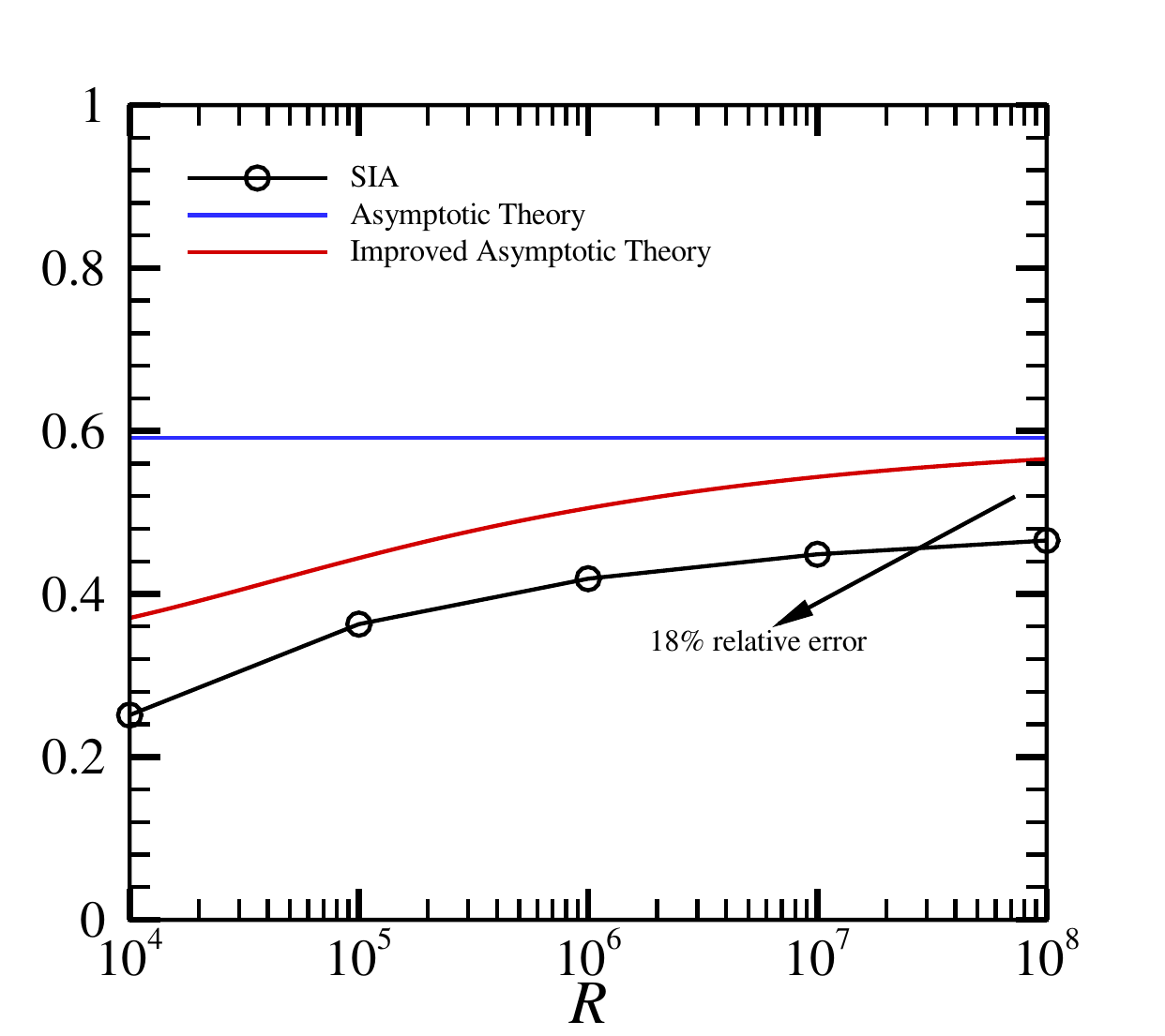}
\put(-2,78){(a)}
\put(-2,45){\begin{turn}{90} $\bar \sigma$ \end{turn}}
\end{overpic}
\begin{overpic}[width=0.45 \textwidth]{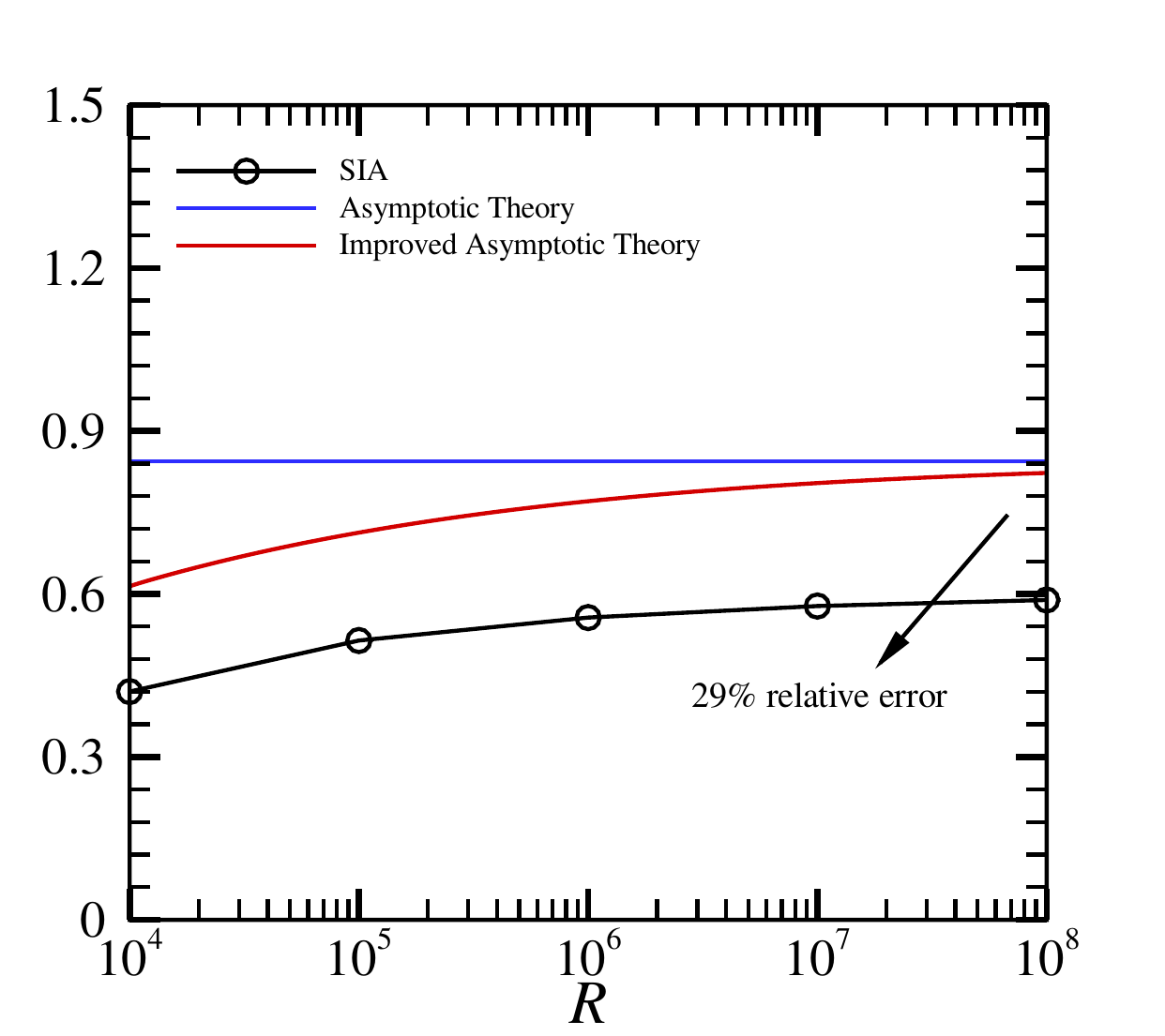}
\put(-5,78){(b)}
\put(-5,45){\begin{turn}{90} $\bar \sigma$ \end{turn}}
\end{overpic}
\caption{ Dependence on $R$ of the normalised growth rate $\bar \sigma$ obtained by the SIA approach, the asymptotic theory and the improved asymptotic theory. (a): cases A6 to A10; (b): cases B6 to B10.      }
\label{fig:large_R_validate_juere_gr}
\end{figure}
In figure~\ref{fig:large_R_validate_juere_gr}, we compare the growth rate $\bar \sigma$ of the secondary mode obtained by the SIA approach, the asymptotic theory and the improved asymptotic theory. The amplitude of the fundamental mode (1,0) is chosen to be $\bar \epsilon_{10}=0.015$. The growth rate $\bar \sigma$ obtained by the SIA increases with $R$ monotonically and approaches a constant as $R \to \infty$, indicating that the viscosity plays a stable role. In the large-$R$ limit, the relative errors between the SIA solutions and the asymptotic predictions are 18\% and 28\% for cases A and B, respectively. Although there is still a gap between the SIA solution and the asymptotic prediction when $R$ is as high as $10^{8}$, the convergent trend is clearly seen. Including the wall-layer correction to the asymptotic theory, the improved asymptotic theory leads to an increase of the accuracy at moderate $R$ values. The monotonic increase of $\bar \sigma$ with $R$ for SIA is also reproduced by the improved asymptotic theory.
\textcolor{red}{Actually, in our asymptotic analysis, there exist two small parameters, namely, $R^{-1}$ and $\bar\epsilon_{10}$, and the $O(R^{-1/2})$ and $O(\bar\epsilon_{01}^2)$ terms are neglected in the governing equations. In  figure~\ref{fig:large_R_validate_juere_gr}, although we have probed the results for large $R$ values, 
the amplitude of mode (1,0) $\bar\epsilon_{01}$ is kept unchanged, which is the reason why the there still exists a discrepancy even when $R=10^8$.} 
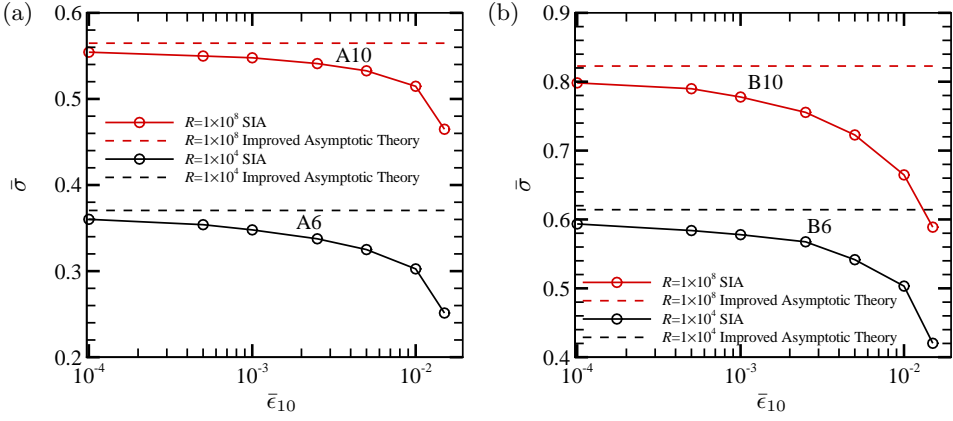
\begin{figure}
\centering
\begin{overpic}[width=0.96 \textwidth]{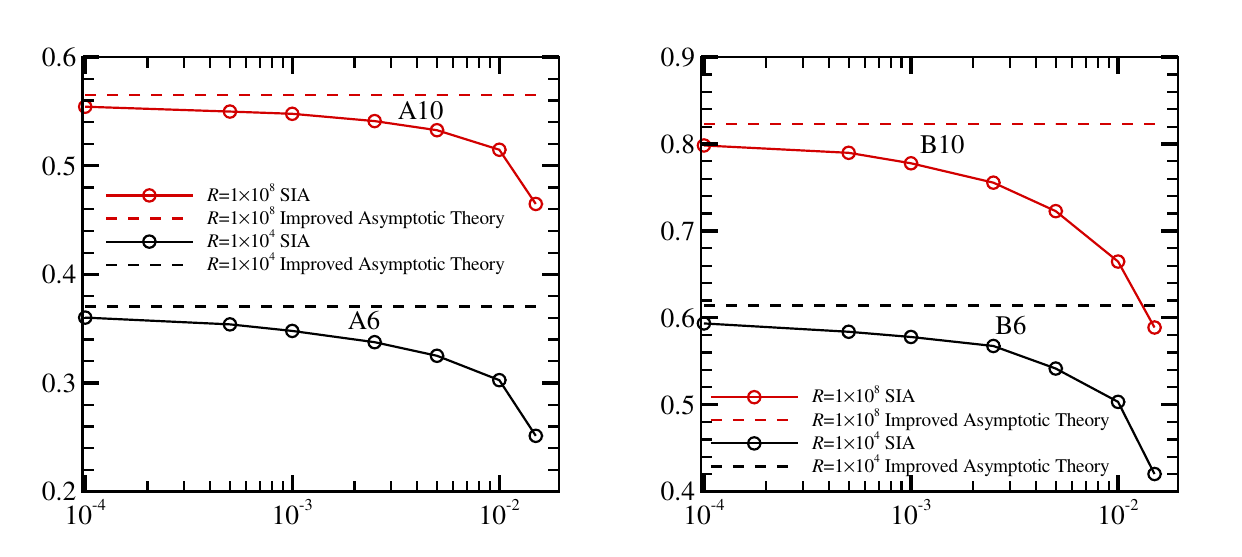}
 \put(-2,39.5){(a)}
\put(47.5,39.5){(b)}
\put(-1,22){\begin{turn}{90}${\bar \sigma}$\end{turn}}
\put(50.5,22){\begin{turn}{90}${\bar \sigma}$\end{turn}}
 \put(25,0){${\bar \epsilon}_{10}$}
\put(75,0){${\bar \epsilon}_{10}$}
\end{overpic}
\caption{ Dependence on ${\bar \epsilon}_{10}$ of the normalised growth rate $\bar \sigma$ obtained by the SIA and the improved asymptotic prediction. (a): cases A6 and A10; (b): cases B6 and B10.}
\label{fig:epsilon10_sigma}
\end{figure}
\textcolor{red}{To confirm this argument, we perform calculations of the SIA by changing the amplitude of the fundamental mode $\bar\epsilon_{01}$ with fixed Reynolds numbers, as shown in  figure~\ref{fig:epsilon10_sigma}. For each Reynolds number, the normalised growth rate obtained by SIA increases with decrease of $\bar\epsilon_{01}$ and approaches a constant when $\bar\epsilon_{01}$ is sufficiently small. The constant agrees well with the normalised growth rate predicted by the improved asymptotic theory.}


\begin{figure}
\centering
\begin{overpic}[width=0.96 \textwidth]{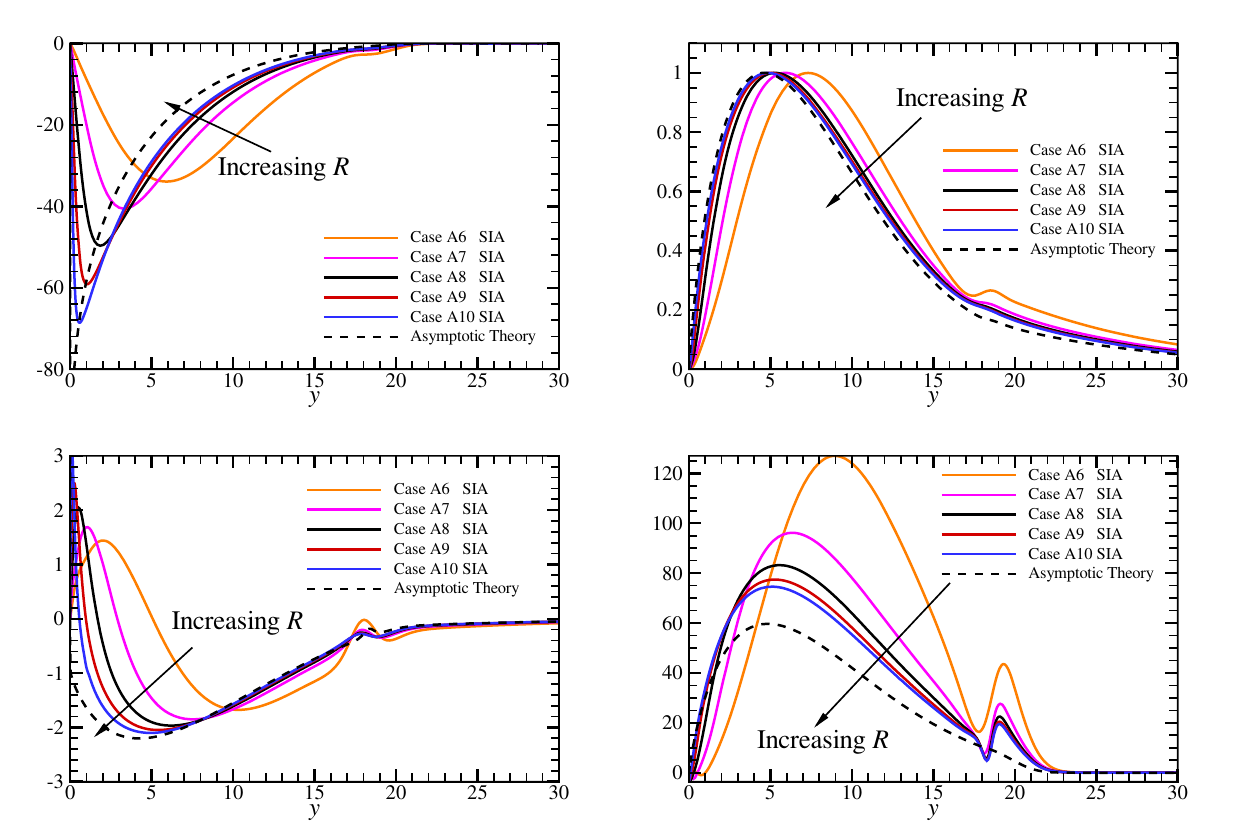}
\put(-1,62){(a)}
\put(48,62){(b)}
\put(-1,29){(c)}
\put(48,29){(d)}
\put(-1,48){\begin{turn}{90} $\Re(\hat{u}_{01})$ \end{turn}}
\put(48,48){\begin{turn}{90} $\Re(\hat{v}_{01})$ \end{turn}}
\put(-1,15){\begin{turn}{90} $\Im (\hat{w}_{01})$ \end{turn}}
\put(48,15){\begin{turn}{90} $\Re(\hat{T}_{01})$ \end{turn}}
\end{overpic}
\caption{ Comparison of the perturbation profiles for streak mode (0,1) obtained by SIA (for cases A6 to A10) and asymptotic theory. The eigenfunctions are normalised by the maximum of $\Re({\hat v}_{01})$. } 
\label{fig:large_R_validate_juere_egf}
\end{figure}

\begin{figure}
\centering
\begin{overpic}[width=0.96 \textwidth]{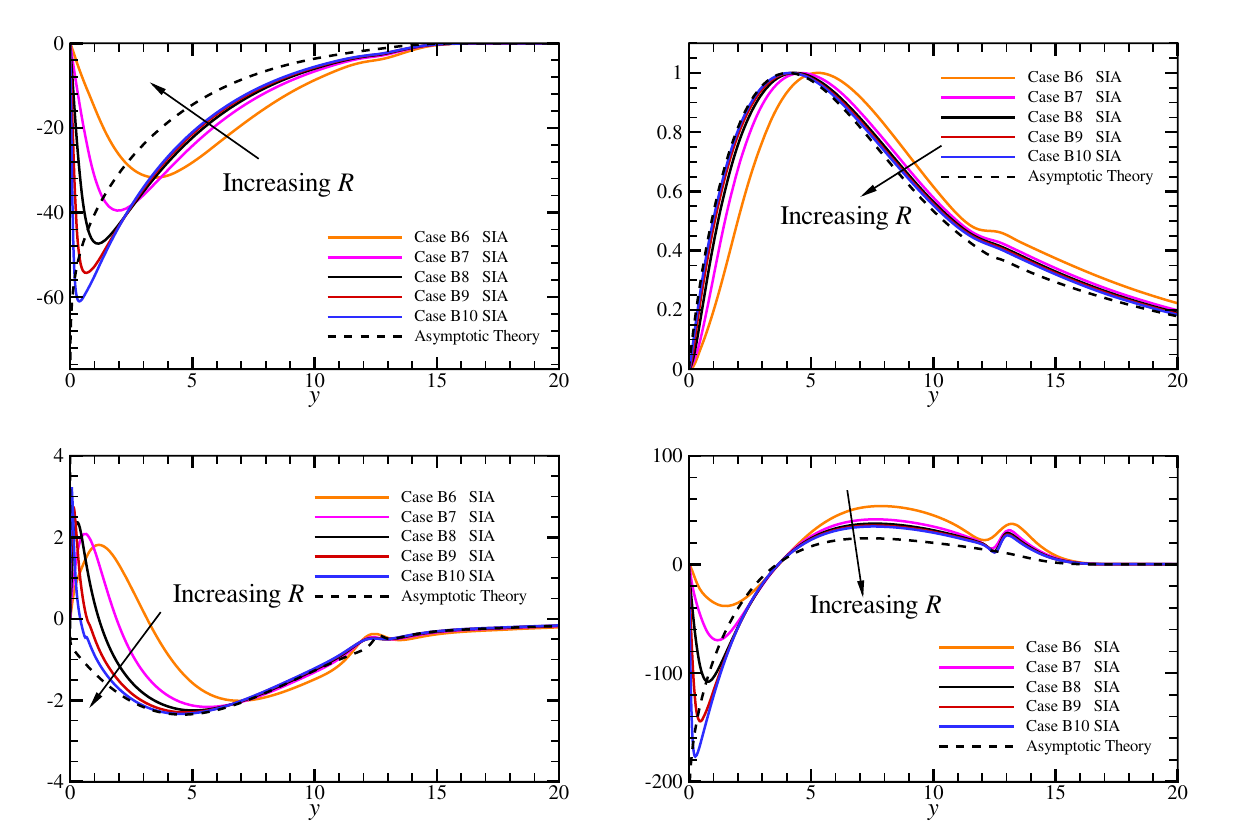}
\put(-1,62){(a)}
\put(48,62){(b)}
\put(-1,29){(c)}
\put(48,29){(d)}
\put(-1,48){\begin{turn}{90} $\Re(\hat{u}_{01})$ \end{turn}}
\put(48,48){\begin{turn}{90} $\Re(\hat{v}_{01})$ \end{turn}}
\put(-1,15){\begin{turn}{90} $\Im (\hat{w}_{01})$ \end{turn}}
\put(48,15){\begin{turn}{90} $\Re(\hat{T}_{01})$ \end{turn}}
\end{overpic}
\caption{ Comparison of the perturbation profiles for streak mode (0,1) obtained by SIA (for cases B6 to B10) and asymptotic theory. The eigenfunctions are normalised by the maximum of $\Re({\hat v}_{01})$. } 
\label{fig:large_R_validate_dengwen_egf}
\end{figure}
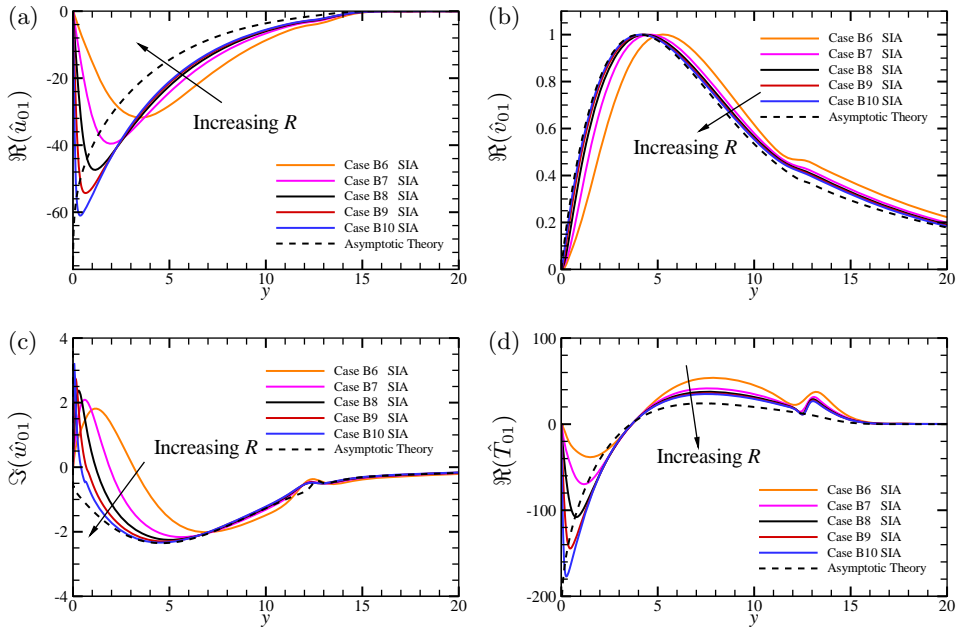

Figure~\ref{fig:large_R_validate_juere_egf} compares the perturbation profiles of the streak mode (0,1) obtained by the SIA and the asymptotic predictions for different $R$ values for cases A6 to A10. As $R$ increases, the SIA solutions approach consistently the asymptotic predictions for all the perturbation quantities. Similar observations can be found for cases B6 to B10, as shown in figure~\ref{fig:large_R_validate_dengwen_egf}. The agreement of both the growth rate and the perturbation profiles between the SIA and the improved asymptotic predictions in the large-$R$ limit confirms the accuracy of our asymptotic analysis.

\section{\label{sec:Conclusion} Concluding remarks and discussion}
In this paper, we focus on the fundamental resonance in hypersonic boundary layers, appearing when the planar Mack second mode dominates the nonlinear phase, and the infinitesimal oblique travelling waves and the streak mode amplify with a rather large rate. This regime is frequently observed in the laminar-turbulent transition in \textcolor{red}{2D or axisymmetric}  hypersonic boundary layers since the planar Mack mode is always the most unstable linear instability mode in its laminar phase.

For a Mach 5.92 flat-plate hypersonic boundary layer with different wall temperatures and Reynolds numbers, we calculate the FR process using the NPSE approach, {and show the amplitude evolution of representative Fourier components and their perturbation profiles.} It is found that when the fundamental 2D mode reaches a finite amplitude, a series of infinitesimal Fourier components with the same spanwise wavenumbers grow with the same rate, much greater than that of the fundamental mode, and the streak mode attains the greatest amplitude among these components. Such a phenomenon can be predicted quantitatively by the SIA based on a base flow consisting of the time- and spanwise-averaged mean flow and the fundamental mode. However, the SIA is not sufficient to reveal the underlying mechanism determining the energy transfer among different Fourier components and to explain the stronger amplification of the streak mode.

Therefore, a large-$R$ asymptotic theory is developed, based on the weakly nonlinear framework. The asymptotic analysis indicates that the FR is in principle a triad resonance system appearing among a dominant planar fundamental mode,  an oblique traveling  mode with the same frequency as the fundamental mode, and a streak mode with the same spanwise wavenumber as the oblique mode. {\color{red}Remarkably, the amplitude of the streamwise velocity component (streak component) of the streak mode is much greater than those of the transverse and lateral velocity components (roll components), implying that these components may be driven by different mechanisms. The triad resonance system  sketched in figure \ref{fig:sketch_mechanism} shows that} \textcolor{magenta}{in the major part of the boundary layer,} the interaction of the fundamental mode and the streak mode seeds for the growth of the oblique mode, whereas the nonlinear interaction of the fundamental mode and the oblique mode drives the roll component of the streak mode, which further encourages a stronger amplification of the streamwise component of the streak mode due to the linear lift-up mechanism. {\color{magenta}The   triad resonance system appears when (1) the dimensionless growth rates of the streak mode and the oblique mode are of the same order of the dimensionless amplitude of the fundamental mode $\bar\epsilon_{10}$, and (2) the amplitude of the streak mode is $O(\bar\epsilon_{10}^{-1})$ greater than the oblique mode.} These observations indicate that the present asymptotic theory is superior to the SIA by providing an in-depth understanding of the FR mechanism.

The asymptotic analysis also reveals the multi-layered structure of the perturbations in the FR regime as sketched in figure~\ref{fig:sketch_structure}. The main-layer solutions of the streamwise velocity, spanwise velocity and temperature of both the streak mode and the oblique mode become singular as the wall is approached, and hence a viscous wall layer needs to be taken into account. {It is found that the wall layer produces an outflux velocity of $O(\epsilon \ln \epsilon)$ to the main layer,  inclusion of which leads to an improved asymptotic theory. }Comparing with the NPSE calculations for moderate Reynolds numbers and the SIA for sufficiently large Reynolds numbers, it is found that the asymptotic theory can predict both the overall growth rate and the main-layer profiles of the infinitesimal perturbations, and the improved asymptotic theory could increase the accuracy of the growth-rate predictions remarkably. {\color{red} Additionally, for a moderate amplitude of the fundamental mode $\bar\epsilon_{01}$, the error of the improved asymptotic prediction does not vanish even when $R$ is sufficiently high, this is because the $O(\bar\epsilon_{01})$ terms are also neglected in the asymptotic analysis. Further decreasing $\bar\epsilon_{01}$ leads to a remarkable reduction of the error, confirming the accuracy of the asymptotic analysis in this paper.}







 \backsection[Acknowledgements]{
This work is supported by National Science Foundation of China (Grant Nos.  11988102,  U20B2003, 12002235) and CAS project for Young Scientists in Basic Research (YSBR-087).}
 \backsection[Declaration of interests]{The authors report no conflict of interest.}



\appendix

\section{\label{sec:Appb}The coefficient matrices and the inhomogeneous forcing in  equation (\ref{eq:dis_e})}
In (\ref{eq:dis_e}), ${ \bf{G} }$, $\bf{A}$, $\bf{B}$, $\bf{C}$, $\bf{D}$, ${\bf{V}}_{xx} $, ${\bf{V}}_{yy} $, ${\bf{V}}_{z z} $, ${\bf{V}}_{xy} $, ${\bf{V}}_{y z} $ and ${\bf{V}}_{x z} $ are all $5 \times 5$-order matrices, whose non-zero elements
are
 \begin{equation}
     \begin{split}
&{G _{11}} = 1, {\kern 2pt}{G _{22}}={G _{33}} ={G _{44}} = \rho_B, {\kern 2pt}{G _{51}} =  - \frac{({\gamma  - 1}){T_B}}{\gamma }, {\kern 2pt}{G _{55}} = \frac{{{\rho _B}}}{\gamma }, \\
 & {A_{11}} = {U_B},{\kern 1pt} {\kern 1pt} {A_{12}} = {\rho _B},{\kern 1pt} {\kern 1pt} {A_{21}} = \frac{{{T_B}}}{{\gamma {M^2}}},{\kern 1pt} {\kern 1pt}{A_{22}} = {\rho _B}{U_B} - \frac{{4 { \mu _{B,x}} }}{{3R}},{\kern 1pt} {\kern 1pt} {A_{23}} =  - \frac{{ {  \tau {T_{B,y}}} }}{R}, {\kern 1pt} {\kern 1pt} \\
  & {A_{25}} = \frac{{{\rho _B}}}{{\gamma M{^2}}} + \frac{{2\tau }}{R}\left( {\frac{1}{3}\nabla  \cdot {{\bf U}} - {S_{11}}} \right),{\kern 1pt} {\kern 1pt}  {A_{32}} = \frac{{2 \mu_{B,y} }}{{3R}},{\kern 1pt} {\kern 1pt} {A_{33}} = {\rho _B}{U_B} - \frac{{ {\mu_{B,x} } }}{R},{\kern 1pt} {\kern 1pt} \\
\nonumber
    \end{split}
\end{equation}

\begin{equation}
    \begin{split}
      & {A_{35}} = \frac{{ - 2\tau {S_{12}}}}{R},{\kern 1pt} {\kern 1pt} {A_{44}} = {\rho _B}{U_B} - \frac{{ { \mu_{B,x}} }}{R},{\kern 1pt} {\kern 1pt} {A_{51}} =  - \frac{{\gamma  - 1}}{\gamma }{T_B}{U_B},\\
&{A_{52}} =  - \frac{{4\left( {\gamma  - 1}\right) {M^2}}}{R}{\mu _B}\left( {{S_{11}} - \frac{1}{3}\nabla  \cdot {{\bf U}}} \right),{\kern 1pt} {\kern 1pt}{A_{53}} =  - \frac{{4\left( {\gamma  - 1} \right){M^2}}}{R}{\mu _B}{S_{12}}, {\kern 1pt} {\kern 1pt} \\
 &  {A_{55}} = \frac{{{\rho _B}{U_B}}}{\gamma } - \frac{{ {2\tau {T_{B,x}} } }}{{RPr}},\\
    & {B_{11}} = {V_B},{\kern 1pt} {\kern 1pt} {B_{13}} = {\rho _B},{\kern 1pt} {\kern 1pt} {B_{22}} = {\rho _B}{ V_B} - \frac{{ {  {\mu _{B,y}}} }}{R},{\kern 1pt} {\kern 1pt} {B_{23}} = \frac{{2{\mu _{B,x}}}}{{3R}},{\kern 1pt} {\kern 1pt} {B_{25}} = \frac{{ - 2\tau {S_{21}}}}{R},{\kern 1pt} {\kern 1pt} \\
&{B_{31}} = \frac{{{T_B}}}{{\gamma {M^2}}},{\kern 1pt} {\kern 1pt} {B_{32}} =  - \frac{{{\mu _{B,x}} }}{R}  ,{\kern 1pt} {\kern 1pt}{B_{33}} = {\rho _B}{ V_B} - \frac{4 \mu _{B,y}}{{3R}}  ,{B_{35}} = \frac{{{\rho _B}}}{{\gamma {M^2}}} + \frac{{2 \tau}}{R}\left( {\frac{1}{3}\nabla  \cdot {{\bf U}} - {S_{22}}} \right),{\kern 1pt} {\kern 1pt}\\
&  {B_{44}} = {\rho _B}{ V_B} - \frac{{{ {\mu _{B,y}}} }}{R},{\kern 1pt} {\kern 1pt}{B_{51}} =  - \frac{{\gamma  - 1}}{\gamma }{T_B}{ V_B},{\kern 1pt} {\kern 1pt}{B_{52}} =  - \frac{{4\left( {\gamma  - 1} \right){M^2}}}{R}{\mu _B}{S_{21}},{\kern 1pt} {\kern 1pt}\\
 & {B_{53}} =  - \frac{{4\left( {\gamma  - 1} \right){M^2}}}{R}{\mu _B}\left( {{S_{22}} - \frac{1}{3}\nabla  \cdot {{ {\bf U}}}} \right), {\kern 1pt} {\kern 1pt}  {B_{55}} = \frac{{{\rho _B}{ V_B}}}{\gamma } - \frac{{2\tau T_{B,y} }}{{R\Pr }},\\
     &{C_{14}} = \rho _B,{\kern 1pt} {\kern 1pt}  {C_{24}} =  \frac{{2{\mu _{B,x}}}}{{3R}}, {\kern 1pt} {\kern 1pt} {C_{34}} = \frac{{ {2{\mu _{B,y}} } }}{{3R}},{\kern 1pt} {\kern 1pt}{C_{41}} = \frac{{{T_B}}}{{\gamma M{^2}}},{\kern 1pt} {\kern 1pt} {C_{42}} =  - \frac{{ {\mu _{B,x}}}}{{R}}  ,{\kern 1pt} {\kern 1pt} {C_{43}} =  - \frac{{ {\mu _{B,y}}}}{{R}}  ,{\kern 1pt} {\kern 1pt} \\
&  {C_{45}} = \frac{{{\rho _B}}}{{\gamma M{^2}}} + \frac{{2\tau {\nabla  \cdot {{\bf U}} } }}{{3R}} ,{\kern 1pt} {\kern 1pt}   {C_{54}} =   \frac{{4\left( {\gamma  - 1} \right){M^2}}{\mu _B}{ \nabla  \cdot {{\bf U}}} }{{3R}} ,  {\kern 1pt} {\kern 1pt}   \\ 
    &{D_{11}} = \nabla  \cdot {{\bf U}},{\kern 1pt} {\kern 1pt} {D_{12}} = {\rho _{B,x}} ,{\kern 1pt} {\kern 1pt} {D_{13}} = {\rho _{B,y}},{\kern 1pt} {\kern 1pt} \\
        &{D_{21}} = \left( {{U_B}{U_{B,x}} + { V_B}{U_{B,y}} } \right) + \frac{{{T_{B,x}}}}{{\gamma {M^2}}},{\kern 1pt} {\kern 1pt} {D_{22}} = {\rho _B}{U_{B,x}} ,{\kern 1pt} {\kern 1pt}  {D_{23}} = {\rho _B}{U_{B,y}} ,{\kern 1pt} {\kern 1pt} \\
    &{D_{25}} = \frac{{{\rho _{B,x}}}}{{\gamma {M^2}}} 
 - \frac{1}{R}\left\{ {\tau \left[ {{{\left( {2{S_{11}} - \frac{2}{3}\nabla  \cdot {\bf{U}} } \right)}_x} + 2{S_{21,x}} } \right]  + \left( {2{S_{11}} - \frac{2}{3}\nabla  \cdot {\bf{U}} } \right){\tau _x} + 2{S_{21}}{\tau _y}} \right\},\\ 
 &{D_{31}} = \left( {{U_B}{ V_{B,x}} + { V_B}{ V_{B,y}} } \right) + \frac{{{T_{B,y}}}}{{\gamma {M^2}}},{\kern 1pt} {\kern 1pt}{D_{32}} = {\rho _B}{ V_{B,x}},{\kern 1pt} {\kern 1pt} {D_{33}} = {\rho _B}{ V_{B,y}} ,{\kern 1pt} {\kern 1pt} \\ 
&{D_{35}} = \frac{{{\rho _{B,y}}}}{{\gamma {M^2}}} -
\frac{1}{R}\left\{ \tau \left[ {2{S_{12,x}} + {{\left( {2{S_{22}} - \frac{2}{3}\nabla  \cdot  {\bf{U}} } \right)}_y} }  \right] { + 2{S_{12}}{\tau _x} + \left( {2{S_{22}} - \frac{2}{3}\nabla  \cdot  {\bf{U}} } \right){\tau _y}} \right\},\\
& {D_{51}} = \frac{{{{\bf U}} \cdot \nabla {T_B}}}{\gamma },{\kern 1pt} {\kern 1pt} {D_{52}} = \frac{{{\rho _B}{T_{B,x}}}}{\gamma }   - \frac{{\left( {\gamma  - 1} \right){T_B}{\rho _{B,x}}}}{\gamma }, {D_{53}} = \frac{{{\rho _B}{T_{B,y}}}}{\gamma }   - \frac{{\left( {\gamma  - 1} \right){T_B}{\rho _{B,y}}}}{\gamma },\\ 
& {D_{55}} =   - \frac{{2\left( {\gamma  - 1} \right){M^2}\tau }}{R}\left[ {{\bf{S}}_{B}:{\bf{S}}_{B} - \frac{1}{3}{{\left( {\nabla  \cdot {{\bf U}}} \right)}^2}} \right]  - \frac{{\nabla  \cdot \left( {{\mu _B}\nabla {T_B}} \right)}}{{RPr}} - \frac{{\gamma  - 1}}{\gamma }\left( {{{\bf U}} \cdot \nabla {\rho _B}} \right),\\
& V_{xx,22}=-\frac{4\mu_{B}}{3R},{\kern 2pt} V_{xx,33}=-\frac{\mu_{B}}{R},{\kern 2pt}V_{xx,44}=-\frac{\mu_{B}}{R},{\kern 2pt} V_{xx,55}=-\frac{\mu_{B}}{R Pr},{\kern 2pt}V_{yy,22}=-\frac{\mu_{B}}{R},{\kern 2pt} \\
& V_{yy,33}=-\frac{4\mu_{B}}{3R}, {\kern 2pt}V_{yy,44}=-\frac{\mu_{B}}{R}, {\kern 2pt} V_{yy,55}=-\frac{\mu_{B}}{R Pr},V_{z z ,22}=-\frac{\mu_{B}}{R},{\kern 2pt}V_{z z ,33}=-\frac{\mu_{B}}{R},{\kern 2pt} \\
& V_{z z ,44}=-\frac{4\mu_{B}}{3R},{\kern 2pt}V_{\varphi \varphi,55}=-\frac{\mu_{B}}{R Pr},{\kern 2pt} V_{xy,23}=V_{xy,32}=-\frac{\mu_{B}}{3R},{\kern 2pt} V_{x z,24}=V_{xz,42}=-\frac{\mu_{B}}{3R},{\kern 2pt}\\
&V_{y z,34}=V_{yz,43}=-\frac{\mu_{B}}{3R}, \nonumber
    \end{split}
\end{equation}
with $\tau = {d \mu_B}/{d T_{B}}$, ${{\bf U}} = {\left[ {{U_B},{\kern 1pt} {\kern 1pt} { V_B},{\kern 1pt} {\kern 1pt} 0} \right]^T}$, $\nabla  \cdot {{\bf U}} = {U_{B,x}} + { V_{B,y}} $ and ${\bf{S}}_{B}$ the rate of strain tensor of the base flow, whose components are  
\begin{equation}
    \begin{split}
    &{S_{11}} = {U_{B,x}},{\kern 2pt} {S_{22}} = { V_{B,y}},{\kern 2pt} , {S_{12}} ={S_{21}} = \left( {{ V_{B,x}} + {U_{B,y}}} \right)/2. \nonumber
    \end{split} 
\end{equation}

$\bf{F}$ is a five-dimensional vector, whose elements are 
\begin{equation}
    \begin{split}
    {F^{(1)}} = & - \left( {\tilde \rho \nabla  \cdot {{\bf u}} + { {\bf u}} \cdot \nabla \tilde \rho } \right),  \\
    {F^{(2)}} = & - \tilde \rho {{\tilde u}_t} - \tilde \rho \left( {{U_B}{{\tilde u}_x} + {\tilde V_B}{{\tilde u}_y} +  \tilde u{U_{B,x}} + \tilde v{U_{B,x}} } \right)   \\
&-\left( {\tilde \rho  + {\rho _B}} \right)\left( {\tilde u{{\tilde u}_x} + \tilde v{{\tilde u}_y} + {\tilde w{{\tilde u}_z }}} \right) - \frac{{{{\left( {\tilde \rho \tilde T} \right)}_x}}}{{\gamma {M^2}}} \\
 &- \frac{2}{{3R}}{\left( {\tilde \mu \nabla  \cdot { {\bf u}}} \right)_x}   + 
\frac{2}{R}\left[ {{{\left( {\tilde \mu {{\tilde S}_{11}}} \right)}_x} + {{\left( {\tilde \mu {{\tilde S}_{21}}} \right)}_y} +{{\left( {\tilde \mu {{\tilde S}_{31}}} \right)}_z}  }\right] , \\
    {F^{(3)}}  = & - \tilde \rho {{\tilde v}_t} - \tilde \rho \left( {{U_B}{{\tilde v}_x} + {\tilde V_B}{{\tilde v}_y}  + \tilde u{\tilde V_{B,x}} + \tilde v{\tilde V_{B,y}}} \right)\\
 &- \left( {\tilde \rho  + {\rho _B}} \right)\left( {\tilde u{{\tilde v}_x} + \tilde v{{\tilde v}_y} + \tilde w{{\tilde v}_z }  } \right) - \frac{{{{\left( {\tilde \rho \tilde T} \right)}_y}}}{{\gamma {M^2}}} \\
 & - \frac{2}{3R}{\left( {\tilde \mu \nabla  \cdot { {\bf u}}} \right)_y} + 
\frac{2}{R}\left[ {{{\left( {\tilde \mu {{\tilde S}_{12}}} \right)}_x} + {{\left( {\tilde \mu {{\tilde S}_{22}}} \right)}_y} +{{\left( {\tilde \mu {{\tilde S}_{32}}} \right)}_z} }  \right],   \\
  {F^{(4)}} = & - \tilde \rho {{\tilde w}_t} - \tilde \rho \left[ {{U_B}{{\tilde w}_x} + {\tilde V_B}{{\tilde w}_y} } \right] - \left( {\tilde \rho  + {\rho _B}} \right)\left( {\tilde u{{\tilde w}_x} + \tilde v{{\tilde w}_y} + {\tilde w{{\tilde w}_z }} } \right) \\
& - \frac{{{{\left( {\tilde \rho \tilde T} \right)}_z }}}{{\gamma {M^2}}}- \frac{2}{{3R}}{\left( {\tilde \mu \nabla  \cdot { {\bf u}}} \right)_z } + 
\frac{2}{R}\left[ {{{\left( {\tilde \mu {{\tilde S}_{13}}} \right)}_x} + {{\left( {\tilde \mu {{\tilde S}_{23}}} \right)}_y}+{{\left( {\tilde \mu {{\tilde S}_{33}}} \right)}_z}  }\right], \nonumber \\
    \end{split}
\end{equation}

\begin{equation}
    \begin{split}
   {F^{(5)}} = & - \frac{{\tilde \rho }}{\gamma }\left( {{{\tilde T}_t} + { {\bf u}} \cdot \nabla {T_B} + { {\bf U}} \cdot \nabla \tilde T + { {\bf u}} \cdot \nabla \tilde T} \right) + \frac{{\gamma  - 1}}{\gamma }{T_B}\left( {{ {\bf u}} \cdot \nabla \tilde \rho } \right) \\
&  + \left( {\gamma  - 1} \right){M^2}\left\{ {2\tilde \mu {\bf{\tilde S}}:{\bf{\tilde S}} - \frac{2}{3}\tilde \mu {{\left( {\nabla  \cdot { {\bf u}}} \right)}^2}} \right\}- \frac{{{\rho _B}}}{\gamma }\left( {{ {\bf u}} \cdot \nabla \tilde T} \right) \\
& + \frac{{\gamma  - 1}}{\gamma }\tilde T\left( {{{\tilde \rho }_t} + { {\bf u}} \cdot \nabla {\rho _B} + { {\bf U}} \cdot \nabla \tilde \rho  + { {\bf u}} \cdot \nabla \tilde \rho } \right) + \nabla  \cdot \left( {\tilde \mu \nabla \tilde T} \right)\\
 &+ \left( {\gamma  - 1} \right){M^2}\left[ {2{\mu _B}{\bf{\tilde S}}:{\bf{\tilde S}} - \frac{2}{3}{\mu _B}{{\left( {\nabla  \cdot   {\bf{u}} } \right)}^2}}+ 4\tilde \mu {\bf{\tilde S}}:{\bf{\tilde S}}  {   - \frac{4}{3}\tilde \mu \left( {\nabla  \cdot   {\bf{U}} } \right)\left( {\nabla  \cdot   {\bf{u}} } \right)} \right] , \nonumber
    \end{split}
\end{equation}
where ${ {\bf u}} = {\left[ {\tilde u,\tilde v,\tilde w} \right]^T}$, $\nabla  \cdot { {\bf u}} = \left( {{{\tilde u}_x} + {{\tilde v}_y} + {{\tilde w}_z}} \right)$, and $\tilde { \bf{S}}$ is the rate of strain tensor of disturbance with
\begin{equation}
    \begin{split}
    &{\tilde S_{11}} = {\tilde u_{x}},{\kern 2pt} {\tilde S_{22}} = {\tilde v_{y}},{\kern 2pt} {\tilde S_{33}} = {\tilde w_{z}},{\kern 2pt} \tilde S_{12} ={\tilde S_{21}} = \left( {{\tilde v_{x}} + {\tilde u_{y}}} \right)/2,\\
  &  {\tilde S_{13}}={\tilde S_{31}} = \left( {{\tilde w_{x}} + {\tilde u_{z}} } \right)/2,  {\kern 2pt}{\tilde S_{23}} ={\tilde S_{32}}= \left( {{\tilde w_{y}}+ {\tilde v_{z}}  } \right)/2. \nonumber
    \end{split} 
\end{equation}

\bibliographystyle{jfm}

\bibliography{jfm-instructions}

\end{document}